\documentclass{article}  

\usepackage{graphicx}

\usepackage{geometry}
\usepackage{graphicx}
\usepackage[round]{natbib}
\usepackage{color}
\usepackage{amssymb}
\usepackage{amsmath}
\usepackage{lineno}
\usepackage{txfonts}
\usepackage{amstext}
\usepackage{mathabx}
\usepackage{multirow}
\usepackage{ulem}
\usepackage{lscape}
\usepackage{txfonts}
\usepackage{subfigure}
\usepackage{float}
\usepackage{chngpage}
\usepackage{hyperref}
\usepackage{longtable}

\begin{document} 


\Large
\begin{center}\textbf{A 3D model simulation of hydrogen chloride photochemistry on Mars: Comparison with satellite data}\end{center}\normalsize

\vspace{0.2cm}
\large\noindent Benjamin Benne$^{1,2}$, Paul I. Palmer$^{1,2}$, Benjamin M. Taysum$^{3}$, Kevin S. Olsen$^{4,5}$, Franck Lef\`evre$^{6}$ \normalsize\\
\vspace{0.2cm}

\noindent$^1$The University of Edinburgh, School of GeoSciences, Edinburgh, United Kingdom (ORCID: 0009-0003-8615-5002, email: benjamin.benne@ed.ac.uk; ORCID: 0000-0002-1487-0969, email: pip@ed.ac.uk)\\ 
$^2$Centre for Exoplanet Science, University of Edinburgh, Edinburgh, United Kingdom\\
$^3$Institut für Planetenforschung (PF), Deutsches Zentrum für Luft- und Raumfahrt (DLR), Rutherfordstraße 2, 12489 Berlin, Germany (ORCID: 0000-0002-0856-4340)\\
$^4$Department of Physics, University of Oxford, Oxford, United Kingdom (ORCID: 0000-0002-2173-9889)\\
$^5$School of Physical Sciences, The Open University, Milton Keynes, United Kingdom\\
$^6$Laboratoire Atmosph\`eres, Milieux, Observations Spatiales (LATMOS/CNRS), Paris, France (ORCID: 0000-0001-5294-5426)\\

\vspace{0.2cm}
\noindent\textbf{Received:} 23 January 2025\\
\noindent\textbf{Accepted:} 26 May 2025\\
\vspace{0.2cm}

\noindent\textbf{DOI:} https://doi.org/10.1051/0004-6361/202553872\\
\vspace{0.5cm}

\section*{Abstract}
\textit{Context } Hydrogen chloride (HCl) was independently detected in the Martian atmosphere by the Nadir and Occultation for MArs Discovery (NOMAD) and Atmospheric Chemistry Suite (ACS) spectrometers aboard the ExoMars Trace Gas Orbiter (TGO). Photochemical models show that using gas-phase chemistry alone is insufficient to reproduce these data. Recent work has developed a heterogeneous chemical network within a 1D photochemistry model, guided by the seasonal variability in HCl. This variability includes detection almost exclusively during the dust season, a positive correlation with water vapour, and an anticorrelation with water ice.

\noindent \textit{Aims } The aim of this work is to show that incorporating heterogeneous chlorine chemistry into a global 3D model of Martian photochemistry with conventional gas-phase chemistry can reproduce spatial and temporal changes in hydrogen chloride on Mars, as observed by instruments aboard the TGO.

\noindent \textit{Methods } We incorporated this heterogeneous chlorine scheme into the Mars Planetary Climate Model (MPCM). After some refinements to the scheme, mainly associated with it being employed in a 3D model, we used it to model chlorine photochemistry during Mars Years (MYs) 34 and 35. These two years provide contrasting dust scenarios, with MY 34 featuring a global dust storm. We also examined correlations in the model results between HCl and other key atmospheric quantities, as well as production and loss processes, to understand the impact of different factors driving changes in HCl.

\noindent \textit{Results } We find that the 3D model of Martian photochemistry using the proposed heterogeneous chemistry is consistent with the changes in HCl observed by ACS in MY 34 and MY 35, including detections and 70\% of non-detections. For the remaining 30\% of non-detections, model HCl is higher than the ACS detection limit due to biases associated with water vapour, dust, or water ice content at these locations.  As with previous 1D model calculations, we find that heterogeneous chemistry is required to describe the loss of HCl, resulting in a lifetime of a few sols that is consistent with the observed seasonal variation in HCl. As a result of this proposed chemistry, modelled HCl is correlated with water vapour, airborne dust, and temperature, and anticorrelated with water ice. 
Our work shows that this chemical scheme enables the reproduction of aphelion detections in MY 35.

\section{Introduction}
\label{sect_intro}

The ExoMars Trace Gas Orbiter (TGO) was launched in 2016 and, after a period of aerobraking to reach its final orbital altitude of $\simeq$400~km, the onboard instruments began collecting data in early 2018. The TGO comprises four instruments, two of which are focused on studying the composition of the Martian atmosphere: the Atmospheric Chemistry Suite (ACS) and the Nadir and Occultation for MArs Discovery (NOMAD). 
These instruments were designed to detect new trace gases that could indicate active geological and/or biological activity in the Martian atmosphere \citep{vago_esa_2015,korablev_atmospheric_2018,vandaele_nomad_2018}. 
To date, the only new trace gas discovery is hydrogen chloride (HCl), which was first observed during Mars Year (MY) 34 by the ACS mid-infrared (MIR) instrument \citep{korablev_transient_2021, olsen_seasonal_2021}, and corroborated by NOMAD \citep{aoki_annual_2021}. 
Detections of HCl have since been reported from MY 35 to MY 37, occurring almost exclusively during the dust season \citep{olsen_seasonal_2021, aoki_annual_2021,olsen_relationships_2024,olsen_relationships_2024-1}, and are correlated with water vapour and anticorrelated with water ice. 
Ground-based observations during MY 35 at solar longitudes L$_{\text{S}}$=273 and 306$^{\circ}$ also detected HCl, as reported by \citet{aoki_global_2024}, using the iSHELL high-resolution near-infrared echelle spectrometer \citep{rayner_ishell_2022} at the NASA Infrared Telescope Facility (IRTF).
Since the first HCl detections on Mars, photochemical models have been applied to explain the chemical cycle of chlorine in the Martian atmosphere. These models typically refined gas-phase chemistry that is classically used in models of the atmospheres of Venus and Earth, where chlorine species play an important role \citep{aoki_annual_2021, krasnopolsky_photochemistry_2022,taysum_observed_2024}. 
However, it quickly became apparent that using only gas-phase chemistry in a 1D model \citep{taysum_observed_2024} and a 3D General Circulation Model (GCM) \citep{rajendran_global_2025} is insufficient to explain the observed rapid appearance and disappearance of atmospheric HCl or the observed seasonal variations of HCl, and that heterogeneous chemistry would be needed to reproduce these observations. Subsequent calculations that included heterogeneous chlorine chemistry in 1D \citep{taysum_observed_2024} and 3D models \citep{streeter_global_2025} show significant improvement with HCl observations compared to models that considered only gas-phase chlorine chemistry.

The discovery of HCl on Mars is important because halogenated gases can play an important role in atmospheric chemistry. We currently do not have a definitive understanding of the source of chlorine in the Martian atmosphere, but the most likely
candidates are chlorinated salts (e.g. NaCl, FeCl$_3$, \citealt{krasnopolsky_photochemistry_2022}) present in the Martian soil. 
Chlorine could then be liberated from the dust when it is lofted into the atmosphere during the dust season, potentially explaining the apparent correlation between this season and HCl detections. 
The process by which chlorine is separated from the dust is still unknown, with many processes considered possible, such as dust aerosol chemistry (\citealt{korablev_transient_2021} and references therein) or electric discharges \citep{rao_production_2012,martinez-pabello_production_2019,wang_chlorine_2020}. 
An alternative source of HCl is volcanoes, but this seems unlikely because TGO did not detect trace gases typically associated with this activity, such as OCS, SO$_2$, and H$_2$S, during MY 34 and MY 35 \citep{braude_no_2022}.

To interpret ACS and NOMAD retrievals of HCl for MY 34, \citet{taysum_observed_2024} developed a chlorine chemistry network, including gas-phase and heterogeneous chemistry, and incorporated it into the 1D sub-model of the Laboratoire de Météorologie Dynamique (LMD)-UK Mars GCM. 
These researchers included heterogeneous chemistry to release atomic chlorine from airborne dust and destroy HCl through reactions with dust and water ice, prompted by TGO observations. 
They show that including heterogeneous chemistry reduces the chemical lifetime of HCl from $>$100 to $>$1000 sols, as determined by gas-phase chemistry \citep{aoki_annual_2021}, to 1--24 hours, allowing the model to reproduce the seasonal cycle of HCl observations. Baseline photochemical model calculations used physical atmospheric parameters from the Mars Climate Database (MCD) v6.1 for the appropriate MY \citep{millour_mars_2018,millour_mars_2022}. \citet{taysum_observed_2024} found that using retrieved values of aerosols and water vapour from ACS and NOMAD, they were able to improve agreement with the HCl observations for most latitudes, but found negative model biases at high northern latitudes. Their explanation for this discrepancy was the lack of horizontal transport in the model, which prevented HCl from being transported from the water-rich southern regions to the northern hemisphere. This would naturally be addressed by using a global 3D GCM. 
Based on these findings, we implemented the chlorine scheme developed by \citet{taysum_observed_2024} in the Mars Planetary Climate Model (MPCM) \citep{forget_improved_1999,pottier_unraveling_2017} to better understand the chlorine atmospheric cycle on Mars, as observed by TGO.
   
In the next section, we present the MPCM and describe how we implement the chlorine photochemistry described by \citet{taysum_observed_2024} as well as the model configuration of our numerical experiments. In Sect. \ref{sect_results}, we report results from our HCl model experiments and compare them with ACS observations, including non-detections and correlations with other atmospheric quantities. In particular, we explain the biases between model results and observations by analysing the production and loss processes that determine HCl in the Martian atmosphere. We conclude the paper in Sect. \ref{sect_conclu}.

\section{MPCM set-up}
\label{sect_MPCM_setup}

The MPCM has been used to study various aspects of the photochemistry of the Martian atmosphere (e.g. \citealt{lefevre_threedimensional_2004, lefevre_heterogeneous_2008, lefevre_observed_2009,lefevre_relationship_2021}). 
This model considers the photochemistry of CO$_2$, oxygen, hydrogen, nitrogen, and water vapour. Ion-neutral and deuterium chemistry are also included in the model but are not considered in the present work. 
Here we describe the chlorine chemical network used to simulate TGO observations of HCl, as well as the model configurations for our numerical experiments. 
For a broader description of the model, we refer the reader to more detailed papers on that subject \citep{forget_improved_1999,lefevre_threedimensional_2004,madeleine_revisiting_2011,navarro_global_2014,pottier_unraveling_2017}.

\subsection{Chemical scheme}
\label{subsect_chem_scheme}

We used the standard chemical network in the MPCM \citep{lefevre_relationship_2021} to which we added the chlorine scheme developed by \citet{taysum_observed_2024}. In this study, we did not use the methane or ethane chemistry networks described by \citet{taysum_photochemistry_2020, taysum_observed_2024}.
To describe chlorine chemistry, we added 12 chlorinated species to the baseline MPCM chemical network, seven photodissociation reactions, 50 gas-phase reactions, and five heterogeneous reactions. The added species and reactions are reported in Tables \ref{tab_app_added_species} and \ref{Tab_app_chlo_scheme}, respectively. 
Among the added heterogeneous reactions, three have a direct impact on HCl abundances:
\begin{table}[!h]\label{heter_react}
    \begin{center}
        \begin{tabular}{l l l l}
        h1 & 6 H$_2$O + dust + $h\nu$ & $\longrightarrow$ & hydrated dust + Cl + 4 O           \\
        h2 & 2 HCl + dust        & $\longrightarrow$ & dust + H$_2$O + CO$_2$  \\
        h3 & HCl + water ice   & $\longrightarrow$ & H + Cl, or no products.\end{tabular}    
    \end{center}
\end{table}

\vspace{-0.5cm}
One issue with the use of these heterogeneous reactions is that their uptake coefficients, and therefore their reaction rates, are not well constrained. We used data from the few experimental measurements available but modified some of the estimated uptake coefficients to improve the match with ACS observations based on sensitivity studies, as described below.
\newline

Reaction h1 is derived from experimental work by \citet{zhang_reaction_2022}, where hydrated dust in the form of Mg(ClO$_4$)$_2 \cdot$ 6H$_2$ O samples was placed under UV radiation. In our model, reaction h1 then considers that six water vapour molecules first react with dust to produce hydrated dust, which can then interact with UV radiation. \citet{zhang_reaction_2022} hypothesised that this reaction leads to photodissociation of ClO$_4^-$ groups and therefore to the production of one Cl atom, four oxygen atoms, and one electron. As in \citet{taysum_observed_2024}, we neglect electron production in our model.
For this reaction, \citet{taysum_observed_2024} used the following uptake coefficient: 
\begin{equation*}
    \gamma_1(z) = 6.3\times 10^{-2} \times j_{\text{ClO}}(z) \times \text{min(1,$RH(z)$)},
\end{equation*}
where $RH(z)$ is the relative humidity ratio for water vapour and $j_{\text{ClO}}(z)$ is the photodissociation rate of ClO at altitude $z$.
The scaling of $\gamma_1$ with $RH$ was chosen based on the experimental studies of \citet{jia_phase_2018} and \citet{gu_water_2017}. These studies show that Mg(ClO$_4$)$_2$ samples are quickly converted to Mg(ClO$_4$)${_2}\cdot$6H$_2$O at low $RH$ values and that the mass of Mg(ClO$_4$)${_2}\cdot$6H$_2$O samples increases approximately linearly with $RH$ values $\simeq$40-80\%. However, we found that using a factor based on $RH$ in the GCM led to strong variations in HCl volume mixing ratios (VMRs) that were inconsistent with TGO observations. As these variations were directly caused by strong variations in the values of $RH$, we removed this term and instead, based on sensitivity experiments, described the uptake coefficient of reaction h1 as: 
\begin{equation*}
\gamma_1(z) = 0.1 \times 6.3\times 10^{-2} \times j_{\text{ClO}}(z).
\end{equation*}

\noindent Reactions h2 and h3 used in \citet{taysum_observed_2024} are derived from observations made on Earth, where dust and water ice are found to be efficient sinks of HCl \citep{sullivan_mineral_2007, abbatt_interaction_1992, isakson_adsorption_1999}.
Reaction h2 is derived from the work of \citet{kelly_thermodynamics_2005}, where HCl reacts with CaCO$_3$ groups present on dust grains:
\begin{table}[!h]
    \begin{center}
        \begin{tabular}{l l l}
        2 HCl$_{\text{(g)}}$ + CaCO$_3$$_{\text{(s)}}$ & $\longrightarrow$ & CaCl$_2$$_{\text{(s)}}$ + H$_2$O$_{\text{(g)}}$ + CO$_2$$_{\text{(g)}}$.\end{tabular}    
    \end{center}
\end{table}

\noindent \citet{taysum_observed_2024} then determined the uptake coefficient by fitting experimental results from \citet{huynh_heterogeneous_2020, huynh_heterogeneous_2021}, which takes the form of the Mediated Precursor Model (MPM, \citealt{berland_surface_1997}):
\begin{equation*}
    \gamma_2(z) = \frac{0.2}{1+2.03\times 10^3 \exp{\left ( \frac{-12.30\times 10^3}{RT(z)} \right )}},
\end{equation*}
where $R$ is the ideal gas constant and $T(z)$ is the temperature at altitude $z$. In our work, we used this value of $\gamma_2(z)$, which is therefore 10 times higher than the value used in \citet{taysum_observed_2024}. They justified dividing this uptake coefficient by 10 to account for the chemical ageing of dust grains. 
Our sensitivity studies indicated that model results were inconsistent with ACS observations when using their $\gamma_2(z)$ value. 
Because these uptake coefficients are unconstrained, the use of a higher $\gamma_2(z)$ coefficient compared to \citet{taysum_observed_2024} may be a consequence of the change in $\gamma_1$. This highlights the need for better measurements of these uptake coefficients to resolve this degeneracy.
\newline

Lastly, \citet{taysum_observed_2024} derived the uptake coefficient of reaction h3 from experimental results by \citet{hynes_interaction_2001} and the work of \citet{kippenberger_trapping_2019}:
\begin{equation*}
    \gamma_3(z) = \frac{0.09}{1+8.33\times 10^8 \exp{\left ( \frac{-30.50\times 10^3}{RT(z)} \right )}} \times \left [ 1-\theta_{\text{Lang}}(z) \right ].
\end{equation*}
The first term also takes the form of the MPM, while the second accounts for the surface coverage $\theta_{\text{Lang}}$ \citep{kippenberger_trapping_2019}:
\begin{equation*}
    \theta_{\text{Lang}}(z) = \frac{K_{\text{Lang}}[\text{HCl}](z)}{1+K_{\text{Lang}}[\text{HCl}](z)},
\end{equation*}
where $K_{\text{Lang}}=9.6\times 10^{-11}$ cm$^3$.mol$^{-1}$ and [HCl]($z$) is the number density of HCl (mol.cm$^{-3}$) at altitude $z$. Following \citet{taysum_observed_2024}, the expression of $\gamma_3(z)$ includes a factor 0.1 to account for imperfections in the ice crystals.

Although we did not change the uptake coefficient of reaction h3 compared to \citet{taysum_observed_2024}, we slightly modified the way products of this reaction are handled in the model. In \citet{taysum_observed_2024}, H and Cl are produced after the heterogeneous uptake of HCl on water ice if the reaction takes place in the sunlit part of the planet, but otherwise HCl is permanently lost. 
This approach reflects the observed reactivity of HCl with the ice surface, which leads to molecular dissociation. 
As this reaction takes place on the surface of the ice grain, hydrogen and chlorine atoms can then be released in the atmosphere.
However, this release can only occur if HCl is not buried under ice layers, which happens when ice grows during the night. In that case, HCl is permanently lost. 
In our model, we improve this representation by assuming that HCl is permanently lost when the condensation rate of water vapour in the atmosphere is positive at a particular point in the atmosphere, but otherwise we release gaseous H and Cl, independent of where the reaction takes place. 
\newline

The reaction rates of heterogeneous reactions are then computed as
\begin{equation*}
    k_{\text{h}}(z) = \frac{1}{4}\times v_{\text{th}}(z) \times S(z) \times \gamma(z) \times \alpha,
\end{equation*}
where $v_{\text{th}} = \sqrt{\frac{8RT(z)}{\pi M}}$ is the thermal speed of the reactant species, $M$ is the molar mass in g.mol$^{-1}$, $S$ is the surface area of dust or water ice in cm$^{-1}$, and $\gamma$ is the uptake coefficient. 
The additional factor $\alpha$ is only used for dust uptake reactions and allows us to take into account the abundance of specific chemical groups in dust grains. Therefore, we use the abundance of MgClO$_4$ for reaction h1 ($\alpha=0.49\%$) and the abundance of CaCO$_3$ for reaction h2 ($\alpha=4\%$) \citep{taysum_observed_2024}. 
\newline

However, if considered as presented at the beginning of this section, these heterogeneous reactions would not conserve mass during the simulations. For example, H$_2$O would be permanently lost in reaction h1, while 4 oxygen atoms would be added to the atmosphere. 
Therefore, to allow for better mass conservation, we modified the products of these reactions as follows: 
\begin{table}[!h]
    \begin{center}
        \begin{tabular}{l l l l}
        h1 & 6 H$_2$O + dust + $h\nu$ & $\longrightarrow$ & Cl + 6 H$_2$O           \\
        h2 & 2 HCl + dust        & $\longrightarrow$ & 2 H  \\
        h3 & HCl + water ice   & $\longrightarrow$ & H + Cl or H.\end{tabular}    
    \end{center}
\end{table}

\begin{figure*}[h!]
   \resizebox{\hsize}{!}
            {
            \includegraphics{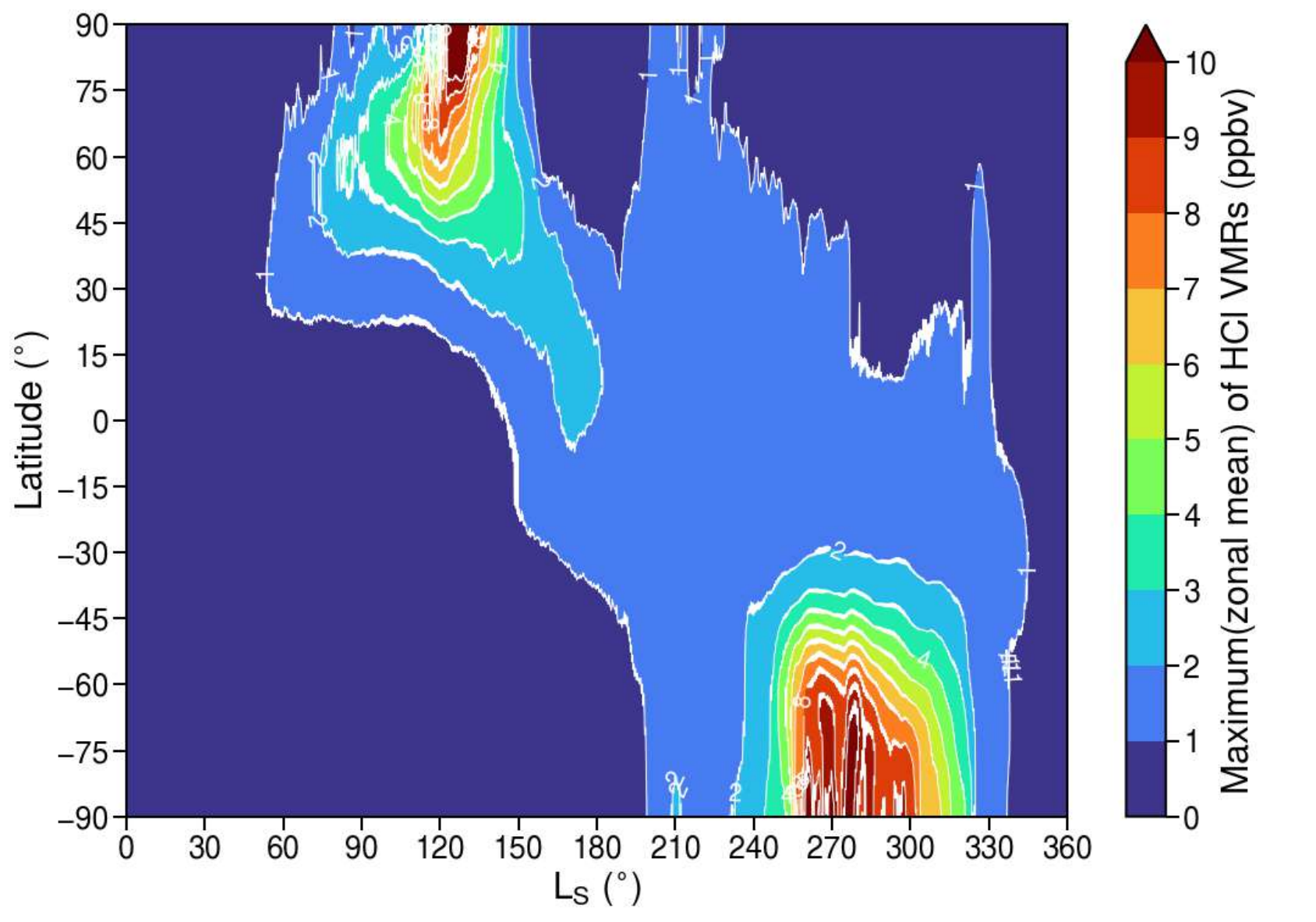}
            \includegraphics{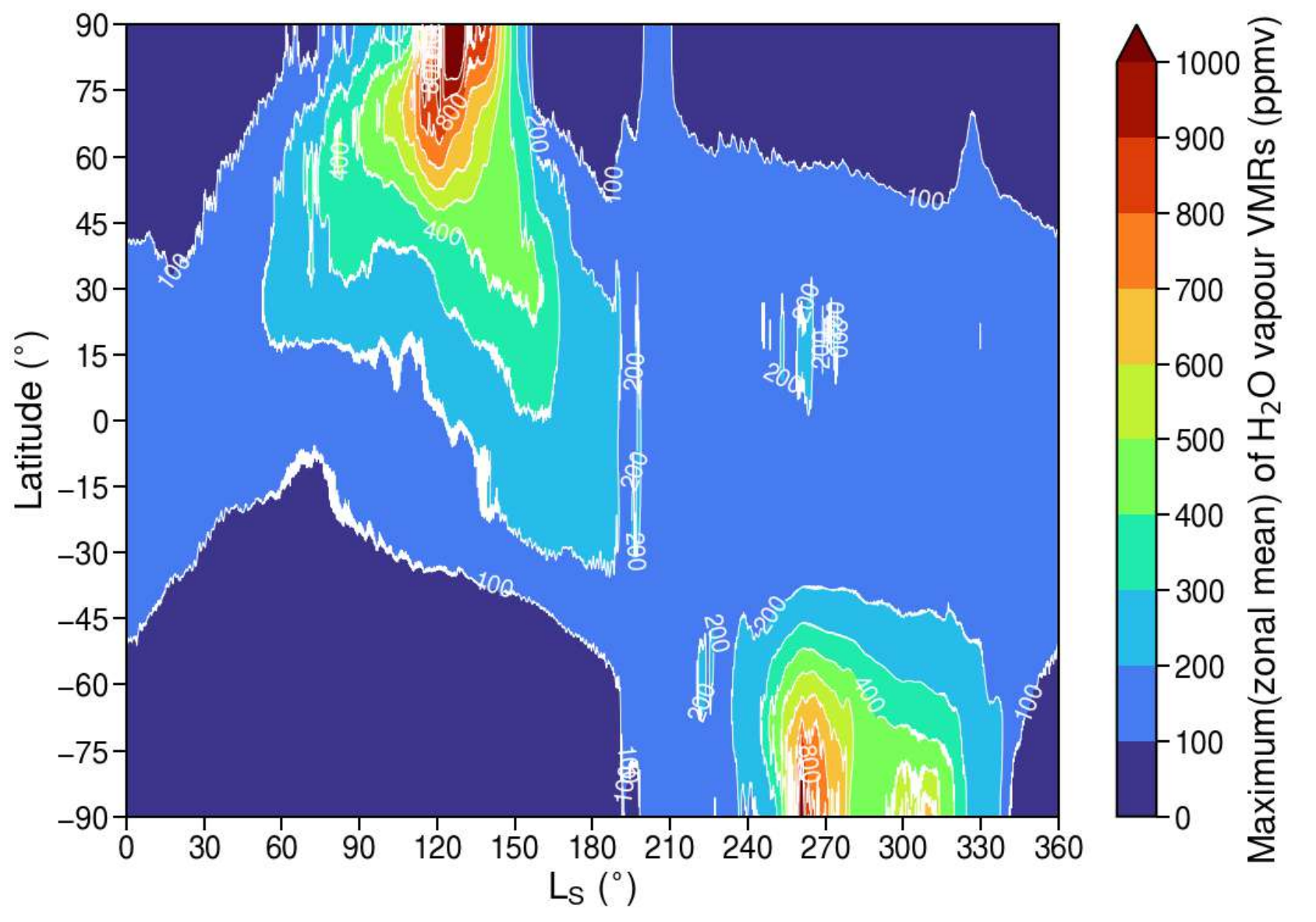}
            }
    \resizebox{\hsize}{!}
            {
            \includegraphics{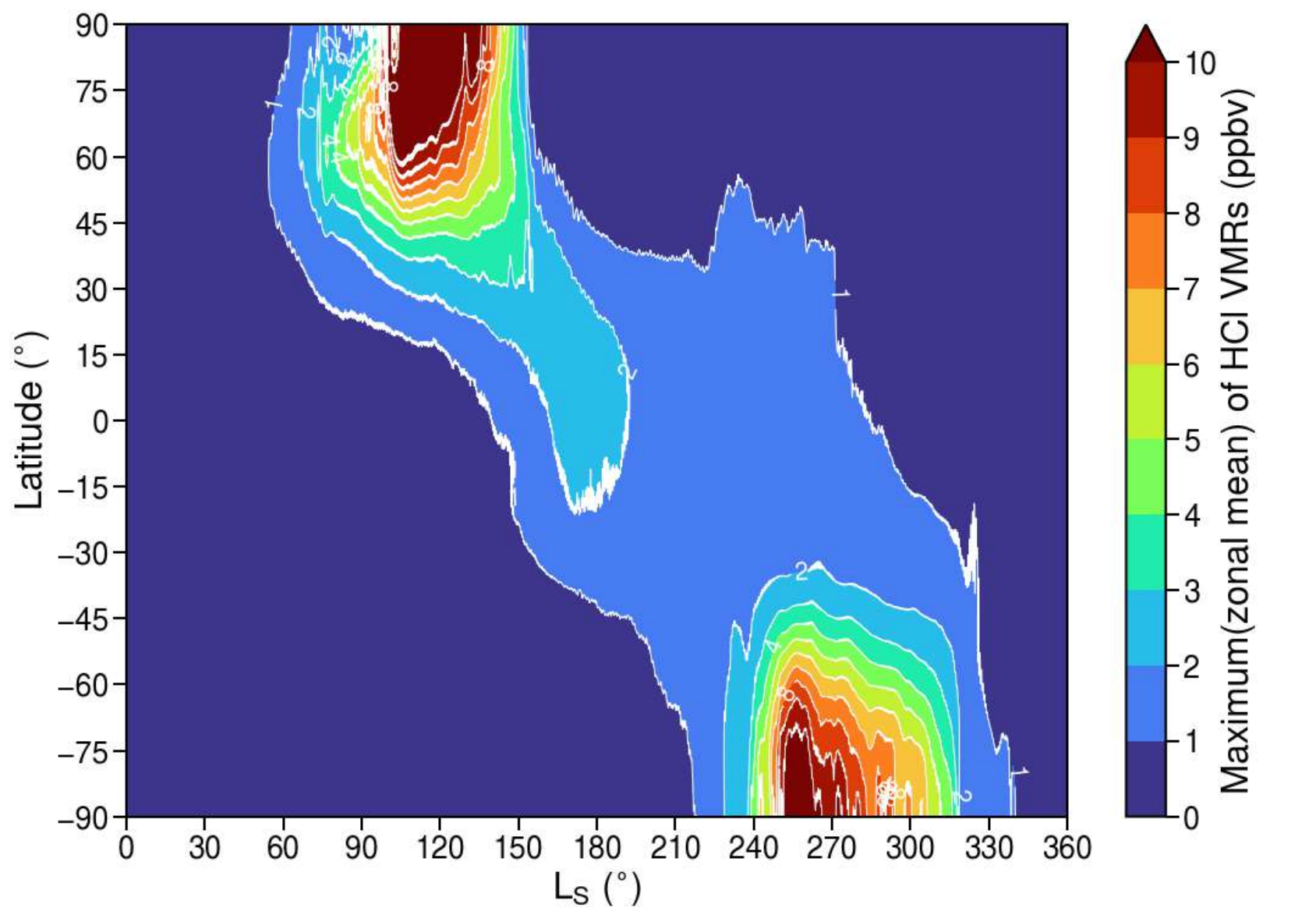}
            \includegraphics{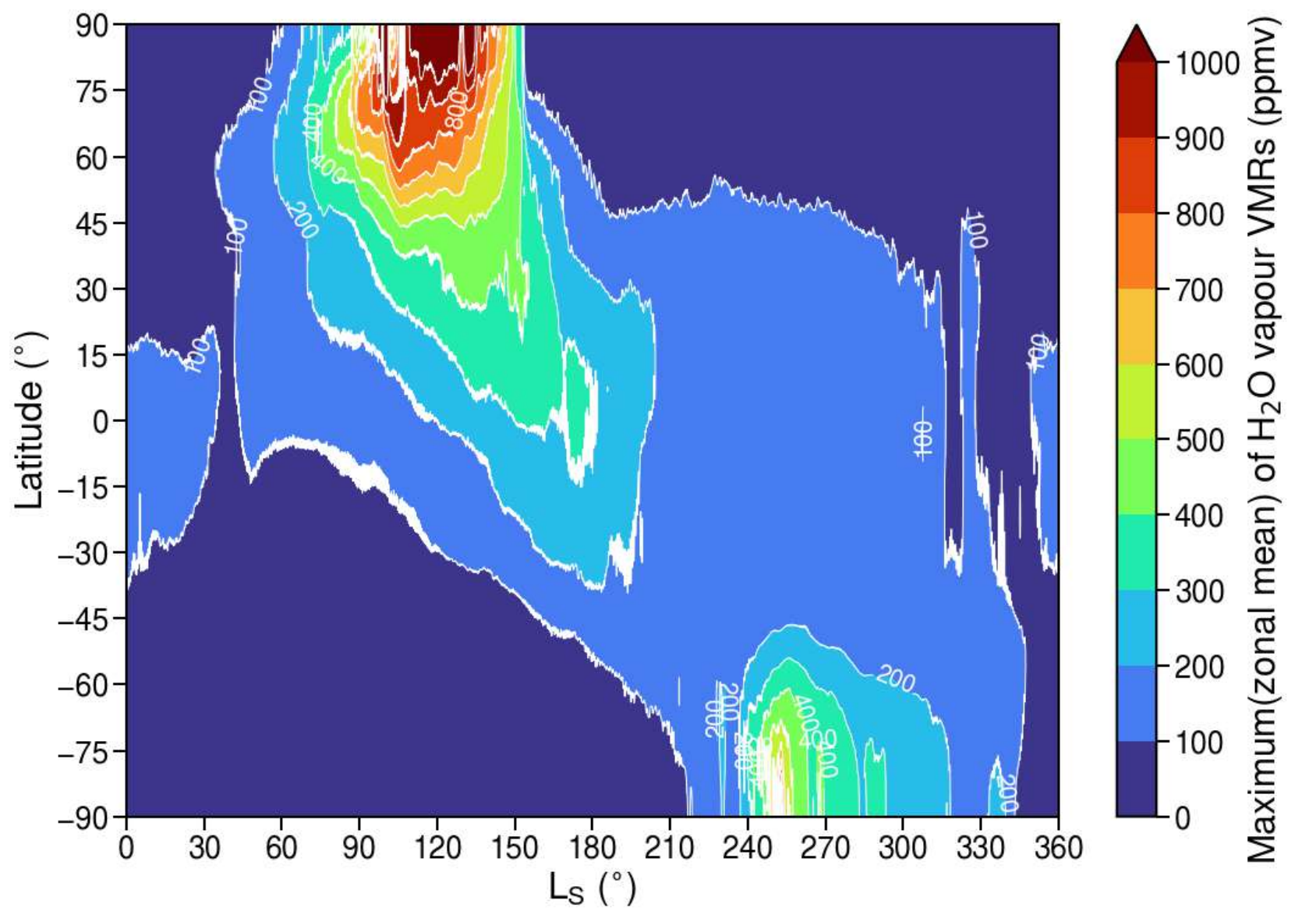}
            }
      \caption{Model column maximum of the zonal mean VMRs for (left) HCl and (right) water vapour. The top panels refer to MY 34 and the bottom panels correspond to MY 35.}
         \label{Fig_climato_hcl_vap}
\end{figure*}

\subsection{Model configuration}
\label{subsect_simu_setup}

We used the MPCM in its most recent configuration, PCM6. In particular, this allowed us to take advantage of a refined description of the water cycle from an improved parameterisation of the microphysics (Naar et al., in prep). Dust activity in the model used dust scenarios derived from Mars Climate Sounder (MCS) observations \citep{montabone_eight-year_2015, montabone_martian_2020}.

We used the standard latitude $\times$ longitude $\times$ pseudo-altitude grid used for PCM6 runs, corresponding to a 64 $\times$ 48 $\times$ 54 levels grid. Pseudo-altitude levels are spread unevenly from the surface to 110 km, ranging from a few dozens of metres near the surface to 7~km above an altitude of 89~km. Our model vertical grid excludes the thermosphere, but our study is focused on chlorine chemistry that is mainly confined to altitudes below 100~km, where HCl is observed by TGO instruments. We used a dynamical time step of 92.5 s (960 steps per sol of 88\,775 s), while the physics time step is five times longer. The model solves the photochemical equations at each physics time step. Outputs are saved 12 times per sol, allowing us to closely compare model results and observations. We also computed the mean values of model outputs over each 5$^{\circ}$L$_{\text{S}}$ interval of the orbit (with L$_{\text{S}}$ the solar longitude), allowing us to better assess the variability of the outputs in these time intervals and subsequently calculate the mean statistics of the model results.

We took the initial conditions from MCD v6.1 \citep{millour_mars_2018, millour_mars_2022}, such as physical conditions and trace gas and aerosol profiles. We focused our study on MY 34 and MY 35, which currently represent the last two years covered by this version of the MCD. These two years provide contrasting atmospheric conditions with which to test our model, with a global dust storm in MY 34. Because observations suggest a relationship between HCl and dust, reflected in our heterogeneous chemical network (see above), this could potentially have a significant effect on our ability to simulate HCl. Initially, we set all chlorine species to zero and spun up the model for 10 MYs, allowing us to reach steady-state conditions. 
HCl was only detected below 70 km by ACS MIR, so we focused our analysis on the altitude range between the Martian areoid (altitude = 0~km) and 70~km. All column values and other quantities--such as the Spearman correlation coefficients--presented in the following sections were computed for this altitude range.

\section{Results}
\label{sect_results}

\subsection{Climatology of HCl}
\label{subs_climatology}

To provide an overview of the latitudes and times of the year in which HCl VMRs are large, we computed the zonal mean HCl VMRs for every output time step of the model.
We then extracted the maximum value for each latitude level and looked at the evolution of these zonally averaged maxima depending on solar longitude L$_{\text{S}}$ for both MY 34 and MY 35, as shown in Figure \ref{Fig_climato_hcl_vap}.
In both years, we find that the mean maximum HCl VMRs are larger than one parts per billion by volume (ppbv) for most of the year, between L$_{\text{S}} \approx 50^{\circ}$ and L$_{\text{S}} \approx$ 340$^{\circ}$. HCl VMRs peak around the summer solstices in both hemispheres (L$_{\text{S}}=90^{\circ}$ for northern summer and L$_{\text{S}}=270^{\circ}$ for southern summer) at high latitudes. Therefore, our results are similar for both years, despite the global dust storm that occurred in MY 34. The main difference is the amplitude of the seasonal peak of HCl, with a larger zonal maximum for a longer period during MY 35, particularly during the northern summer months.

During MY 35, HCl VMRs larger than 10 ppbv are found only at latitudes higher than 45$^{\circ}$N or 65$^{\circ}$S, with a maximum value of 24.3 ppbv reached at 67.5$^{\circ}$N (see Fig. \ref{app_Fig_scatt_altlatlonlt_MY35}). 
We find a similar pattern during MY 34, but the largest VMR is three times higher than during MY 35, at 82.5$^{\circ}$S (see Fig. \ref{app_Fig_scatt_altlatlonlt_MY34}), which we attribute to the global dust storm. 
During winter months, the maximum zonal mean VMR is lower than 1 ppbv. Because northern winter months occur during the dust season, HCl values are typically higher than during southern winter months. 

As such, we show that large HCl VMRs are not restricted to the dust season, as suggested by the absence of regular detections in the aphelion season in the ACS MIR data \citep{korablev_transient_2021, olsen_seasonal_2021, olsen_relationships_2024} and NOMAD \citep{aoki_annual_2021}. 
Instead, seasonal changes in HCl appear to be closely related to the atmospheric water cycle, as shown in Fig. \ref{Fig_climato_hcl_vap}. 
The patterns obtained for water vapour and HCl are strikingly similar, with peak HCl abundances reached during the maximum sublimation events of the summer solstices. We explore this apparent correlation between HCl and water vapour in Sects. \ref{subsect_vert_prof_comp} and \ref{subs_correlations}. 
In MY 35, while the HCl maxima are comparable between the northern and southern summers, the water sublimation peak occurs during the northern summer, outside of the dust season. Consequently, this suggests that dust also plays an important role in seasonal changes of HCl, particularly during the southern summer months. 
HCl VMRs shown in Fig. \ref{Fig_climato_hcl_vap} are larger than 5 ppbv over a large latitude range, well within the detection limit of ACS MIR. 
In Sect. \ref{subs_non_detect} we demonstrate that our model is also consistent with non-detections of HCl over this spatial and temporal domain, providing additional confidence about the model performance. 

\subsection{Comparison of model and ACS vertical profiles of HCl}
\label{subsect_vert_prof_comp}

\begin{figure*}[h!]
   \resizebox{\hsize}{!}
            {
            \includegraphics{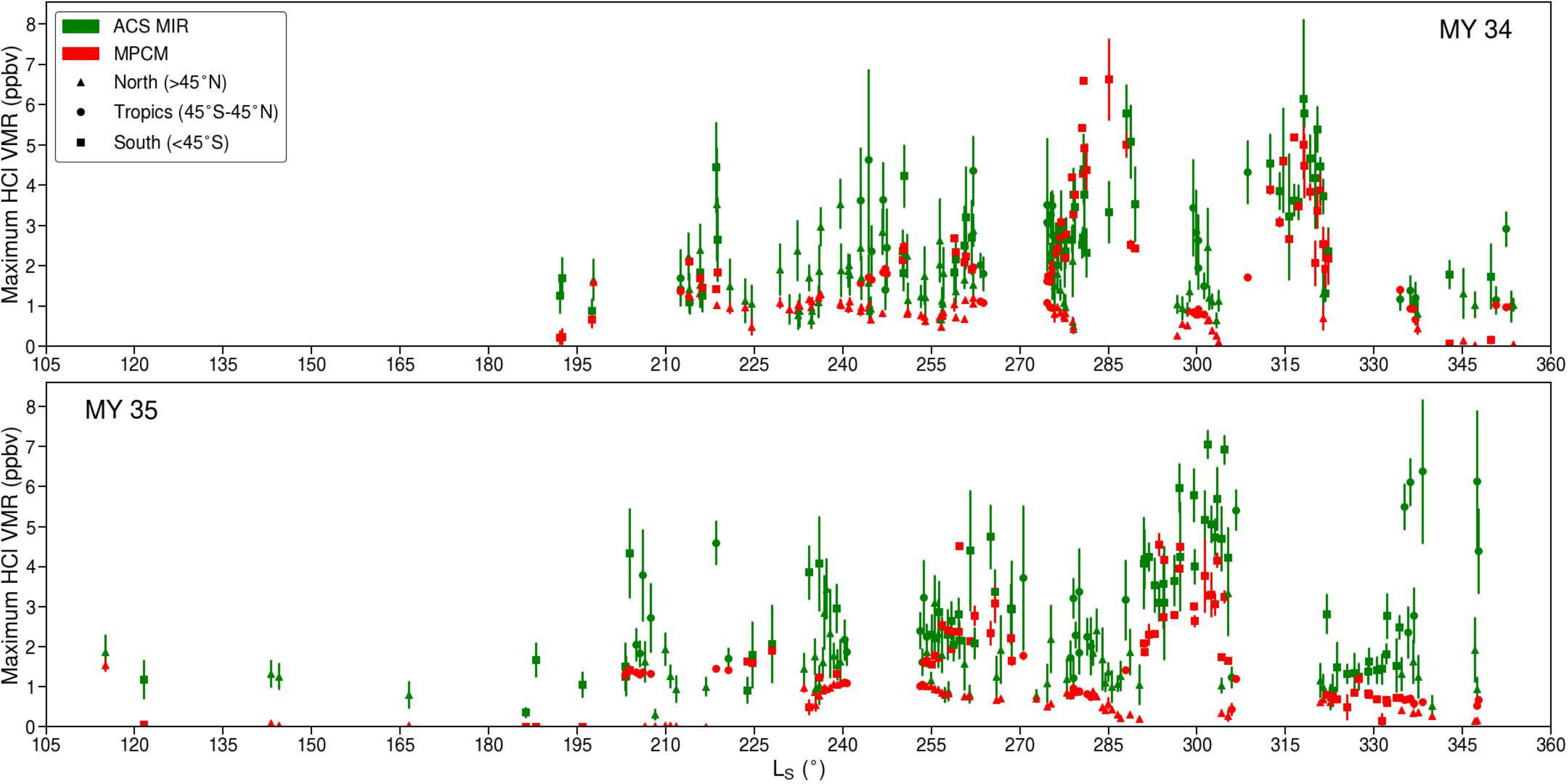}
            }
      \caption{Maximum HCl VMR from our model (red) and ACS MIR (green) at every location where HCl was retrieved by the instrument during MY 34 (top) and MY 35 (bottom). Model maxima are extracted from the altitude range over which HCl is observed by ACS MIR. Triangles correspond to observations made at northern latitudes ($>$45$^{\circ}$N), circles at tropical latitudes (45$^{\circ}$S-45$^{\circ}$N), and squares to southern latitudes ($<$45$^{\circ}$S). Error bars display the standard deviation of modelled HCl within the 5$^{\circ}$L$_{\text{S}}$ interval containing the observation and the uncertainty on the ACS value.}
    \label{Fig_MAXhcl_vs_Ls}
\end{figure*}

In Fig. \ref{Fig_MAXhcl_vs_Ls} we compare the maximum HCl VMR for each ACS MIR detection with the model maximum at the same location and time, for every detection of MY 34 and MY 35. 
We observe that our model reproduces most ACS MIR profiles if we consider both the standard deviation of the model results over the 5$^{\circ}$ L$_{\text{S}}$ interval containing the observation and the uncertainty in the ACS MIR values. 
We find that the model typically exhibits a negative HCl bias at similar times of the year, with the exception of a few observations that mostly occur in the southern hemisphere. We investigate the reasons for these biases in the following sections, after presenting our results on the main factors that influence HCl in our model. For brevity, we report representative results; the complete set of figures comparing model results and ACS MIR observations is available on \href{https://zenodo.org/records/14717873}{Zenodo}.

\subsubsection{General results}\label{subsub_general_results}

If we compute the relative difference between each point from individual ACS MIR HCl profiles and the corresponding model values, accounting for
the standard deviation of model values and the uncertainties of the ACS data, we find a median relative difference of 0\% for both MY 34 and MY 35 (see Fig. \ref{app_Fig_rel_diff_MY3435}). The corresponding mean differences are -3\% and -18\% for MY 34 and MY 35, respectively. This is a remarkable result given the large uncertainties in the heterogeneous reaction rates, as described above.

\begin{figure*}[!h]
    \centering
   \resizebox{0.7\hsize}{!}
            {
            \includegraphics{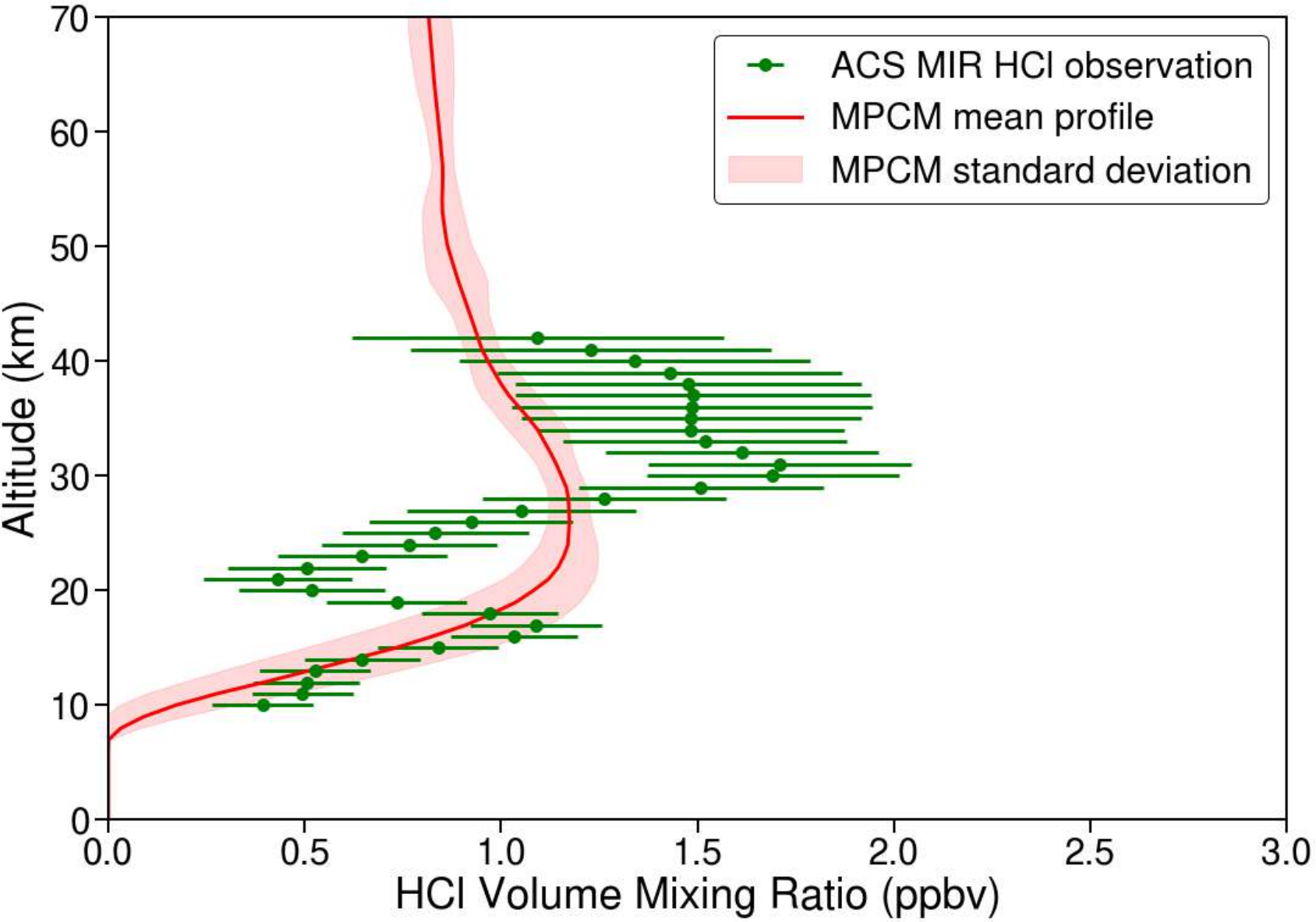}
            \includegraphics{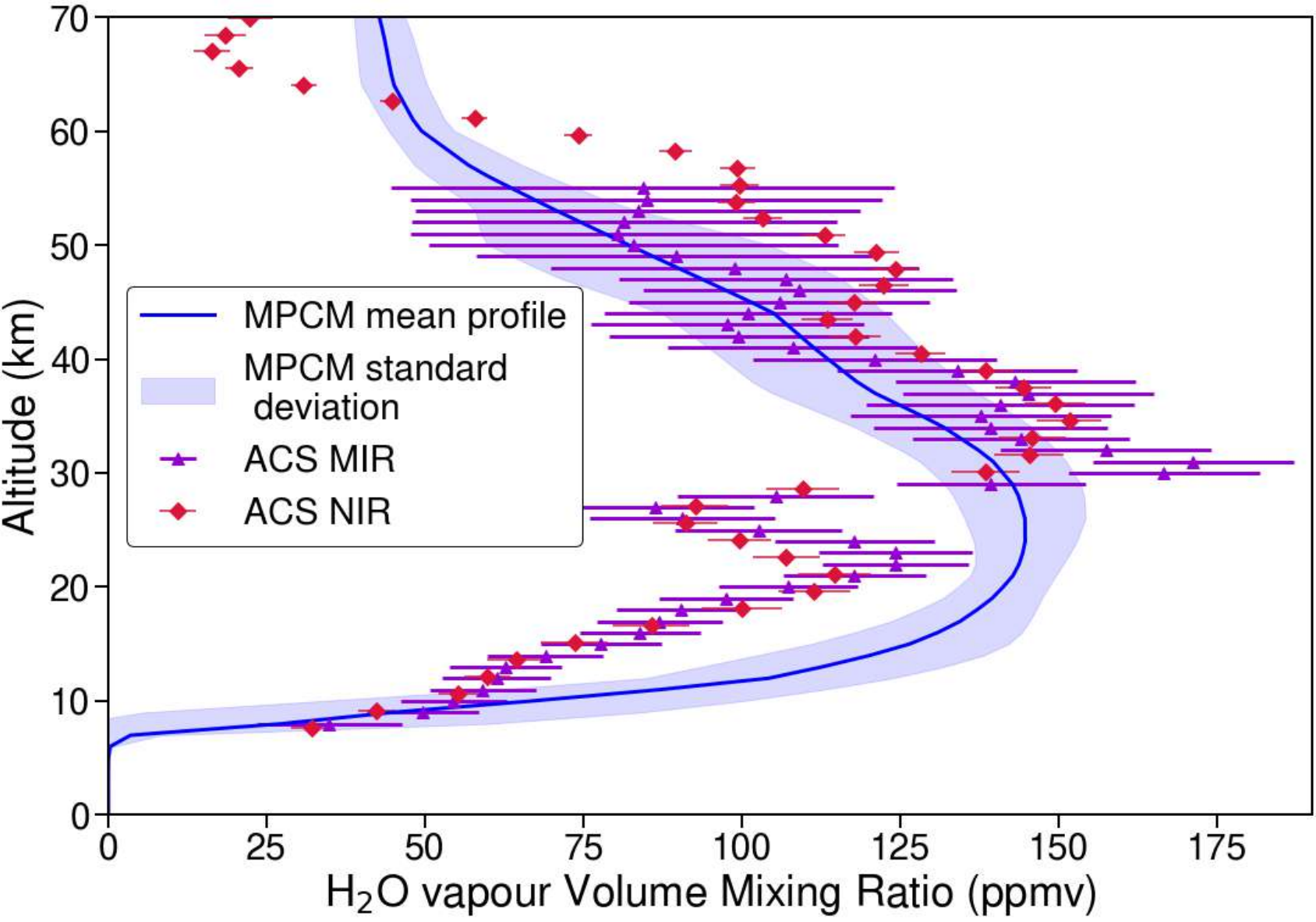}
            }
      \caption{Comparison between model and observed VMRs of HCl (left) and water vapour (right) for a representative case where the modelled HCl VMR profile is consistent with the observed profile from ACS MIR. Model profiles are averaged over the 5$^{\circ}$ L$_{\text{S}}$ interval containing the ACS observation. The standard deviation represents the variability of the computed profiles within this interval. The plots correspond to the ACS MIR occultation 003308\_N2\_E\_P1\_11\_F, located at latitude 57.14$^{\circ}$N, longitude 149.68$^{\circ}$E, 234.30$^{\circ}$ L$_{\text{S}}$, LT = 8.33 hrs in MY 34.}
         \label{Fig_HCl_vap_general_results}
\end{figure*}

\begin{figure*}[!h]
    \centering
   \resizebox{\hsize}{!}
            {
            \includegraphics{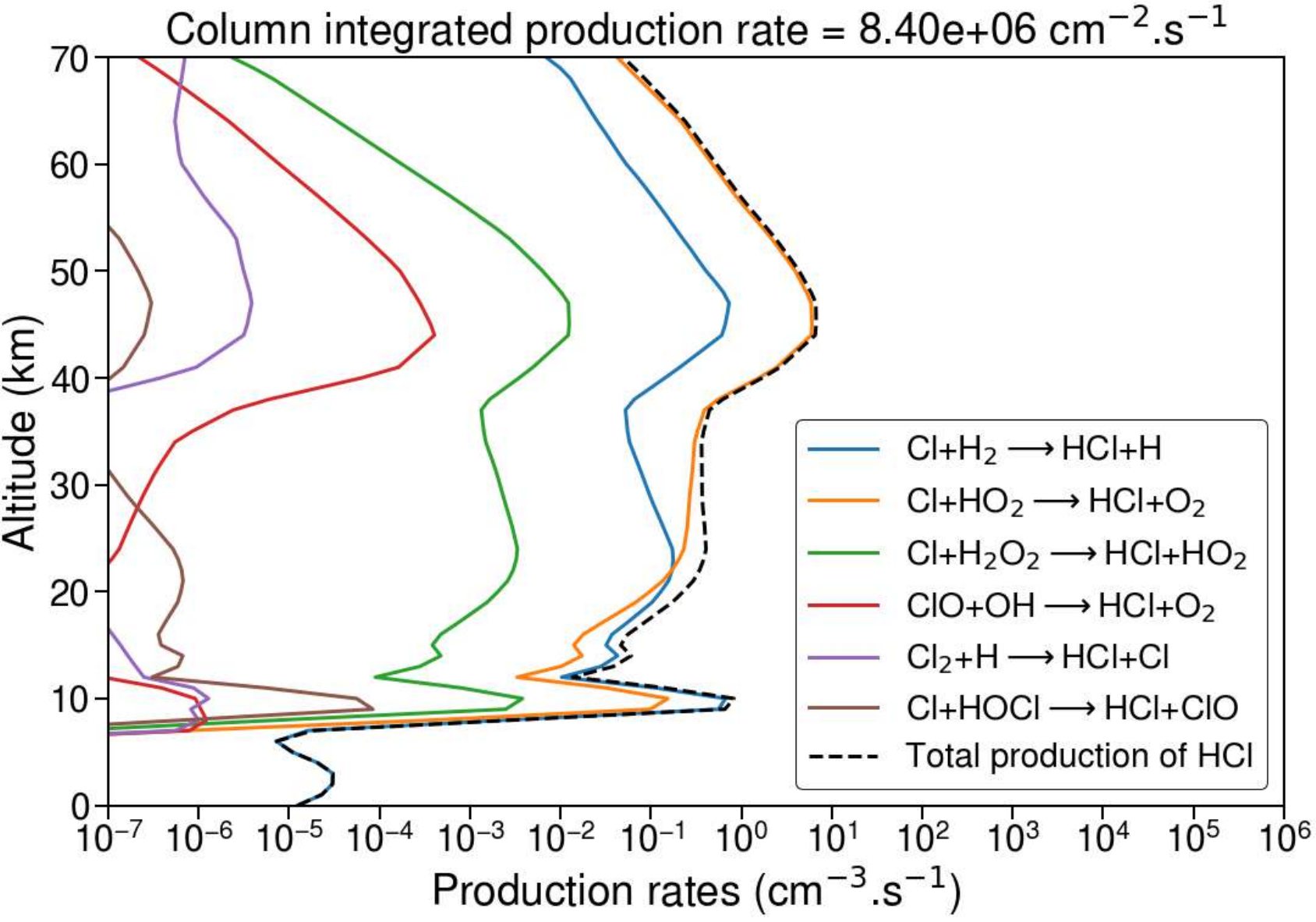}
            \includegraphics{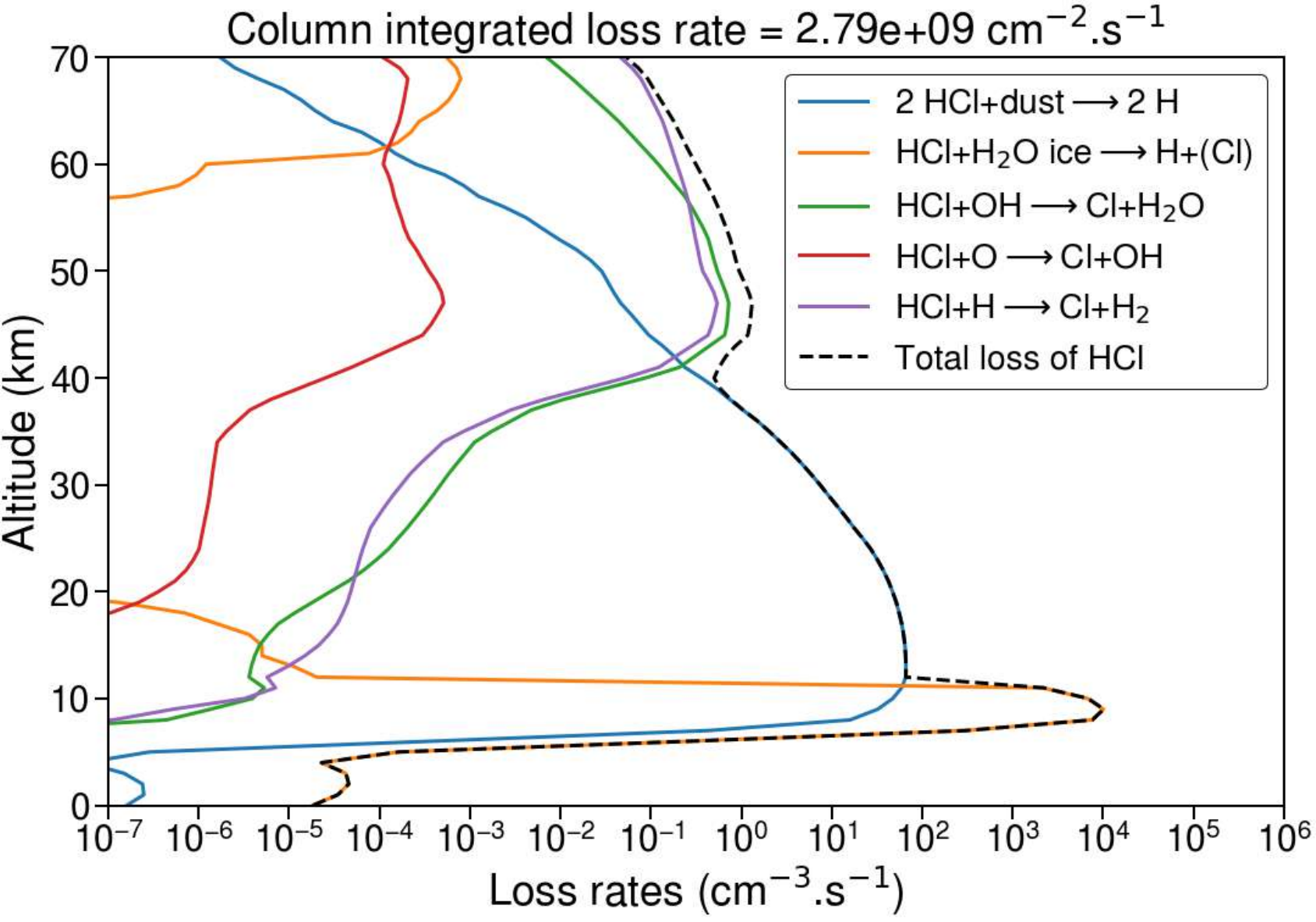}
            \includegraphics{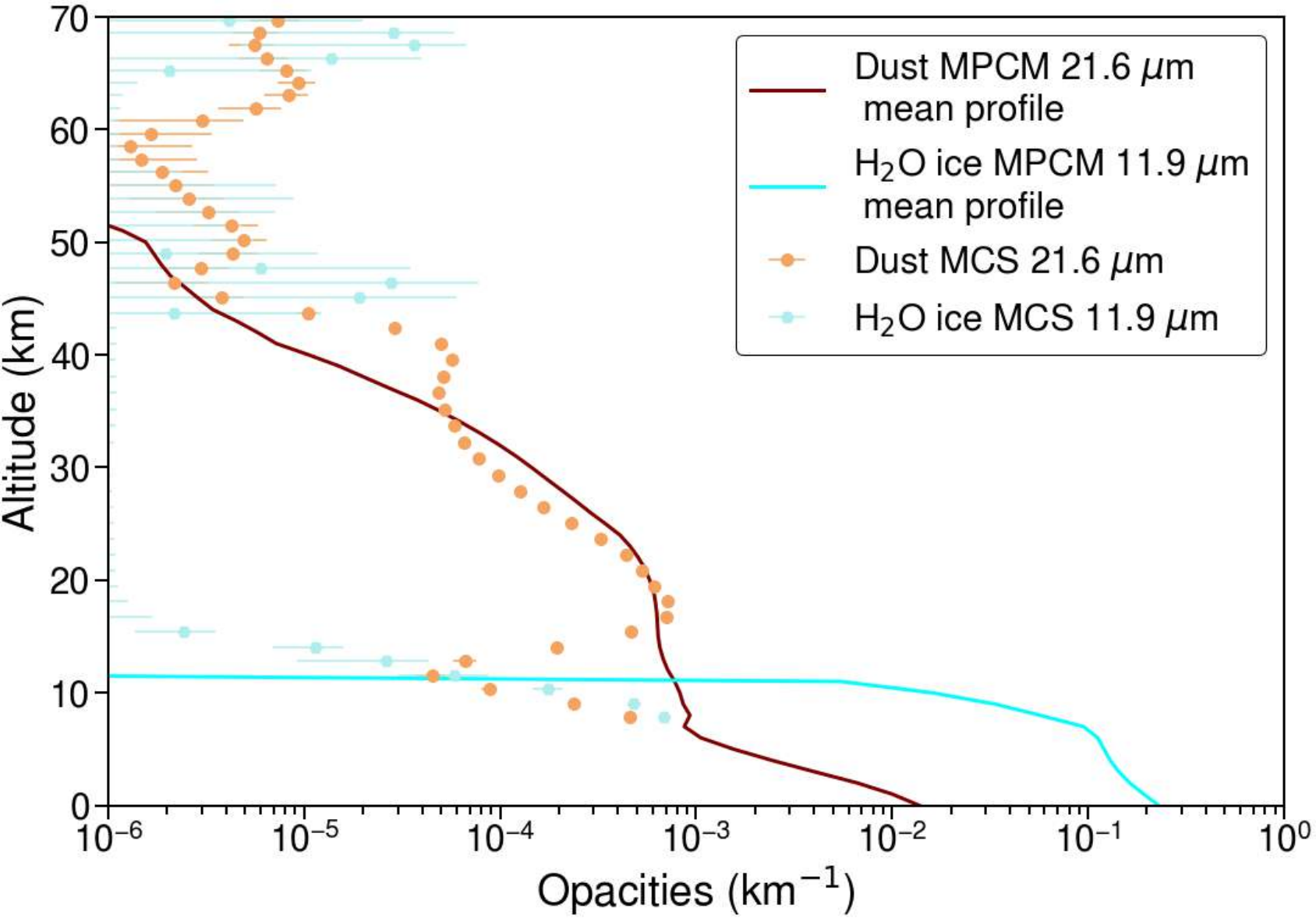}
            }
      \caption{Model production (left) and loss (middle) rates of HCl corresponding to the ACS HCl detection from Fig. \ref{Fig_HCl_vap_general_results}, where our model HCl is consistent with the ACS observation. The right panel shows modelled (solid line) and MCS (closed circles) water ice and dust opacity profiles. The MCS data are taken as close as possible in space and time to the ACS MIR HCl observation. Model profiles are averaged over the 5$^{\circ}$ L$_\text{S}$ interval containing the observation. Standard deviations are omitted here for clarity.}
         \label{Fig_prod_loss_general_results}
\end{figure*}

We find that in most cases, when the observed profile of HCl is reproduced by the model, the corresponding model water vapour profile is reasonably consistent with the ACS MIR and ACS NIR observations. An example is shown in Fig. \ref{Fig_HCl_vap_general_results}. This demonstrates that HCl and water vapour are correlated, in agreement with the observations from Sect. \ref{subs_climatology}. To better understand this correlation, we examined the model HCl chemistry associated with this profile. Figure \ref{Fig_prod_loss_general_results} shows the production and loss rates of HCl corresponding to the model and data shown in Fig. \ref{Fig_HCl_vap_general_results} 
In that case, HCl is produced mainly by: 

\begin{table}[!h]
    \label{heter_react}
      \begin{center}
         \begin{tabular}{l l l l}
         cl8 & Cl + H$_2$      & $\longrightarrow$ & HCl + H           \\
         cl9 & Cl + HO$_2$       & $\longrightarrow$ & HCl + O$_2$.\end{tabular}    
      \end{center}
\end{table}
\noindent We find this is generally true at other locations where HCl was detected by ACS MIR. Reaction cl8 dominates in the lower atmosphere ($\approx z$<$20-30$~km), while cl9 is the main source of HCl between $\approx$ 20--30 and $\approx$ 60~km. Above 60~km, both reactions contribute, with varying importance depending on the location. 
The mean contributions of these reactions to the column-integrated production of HCl depend on the latitude: cl8 contribution varies between 21\% at northern latitudes and 67\% at southern latitudes, while the contribution of cl9 ranges from 79 to 41\%. 
In the Martian atmosphere, hydrogen is produced from water photolysis \citep{haberle_atmospheric_2017}. The concentrations of H$_2$ and HO$_2$ therefore depend on the abundance of water vapour, and, by extension, so do HCl, whose production is primarily based on these two species. 
In addition, atomic chlorine is produced through reaction h1, which also involves water vapour. 
Therefore, this explains the strong correlation between HCl and water vapour.
At lower altitudes, the reaction Cl + H$_2$O$_2$ sometimes contributes a few percent to HCl production. Generally, we find that the other reactions producing HCl are negligible in our model (reactions cl13, cl28, and cl47, as described in Table \ref{Tab_app_chlo_scheme}, with contributions $\leq$ 0.02\%).

Figure \ref{Fig_prod_loss_general_results} shows that the main HCl loss processes below 40~km are heterogeneous reactions with water ice and dust (reactions h3 and h2). Generally, uptake on water ice is the main loss process wherever water ice is abundant, representing 86\% to 11\% of the total HCl loss moving from northern to southern latitudes, respectively. Uptake on dust is the second most important loss process for HCl, which we find dominates below 40--50~km when water ice is not present. Its contribution to the total loss of HCl ranges from 13\% at northern latitudes to 87\% at southern latitudes, where water ice is less abundant because it corresponds to the summer hemisphere. 
At altitudes higher than 50~km, we identify three main loss processes: photolysis (ph1), reaction with OH that produces Cl and water vapour (cl15), and the reaction with atomic hydrogen that leads to the production of Cl and H$_2$ (cl33). These reactions contribute only a small percentage to the total HCl loss, showing no clear trends except for photolysis, which increases as expected towards the south. Heterogeneous processes are therefore critical for the production of HCl, which depends on the production of atomic chlorine through h1, as well as for the loss of HCl through h2 and h3. 

As HCl and water vapour are correlated, we also studied the relative difference between water vapour VMRs from our model, and ACS MIR and NIR vertical profiles.
As for HCl, if we compute the relative difference for every point from the ACS MIR or ACS NIR profile and consider the uncertainty in these measurements and the standard deviation of our results over 5$^{\circ}$ intervals of L$_{\text{S}}$, we find a median relative difference of 0\% for both years and both instruments (see Fig. \ref{app_Fig_rel_diff_MY3435}). 
However, we find that negative model biases for water vapour are greater in MY 35. This partly explains the larger negative model biases for HCl in this year compared to MY 34 (Fig. \ref{Fig_MAXhcl_vs_Ls}). 

\subsubsection{Study of model negative biases}
\label{subs_study_neg_biases}

In Fig. \ref{Fig_MAXhcl_vs_Ls}, we find that the model typically has a negative HCl bias relative to the ACS data. The largest model biases are found in three different periods: (1) before L$_{\text{S}}$=220$^{\circ}$ in MY 35, (2) between L$_{\text{S}}$=275--310$^{\circ}$, and (3) after L$_{\text{S}}$=330$^{\circ}$. We also find that large biases are found during the last two of these periods in MYs 34 and 35. 

\begin{figure*}[!h]
    \centering
   \resizebox{\hsize}{!}
            {
            \includegraphics{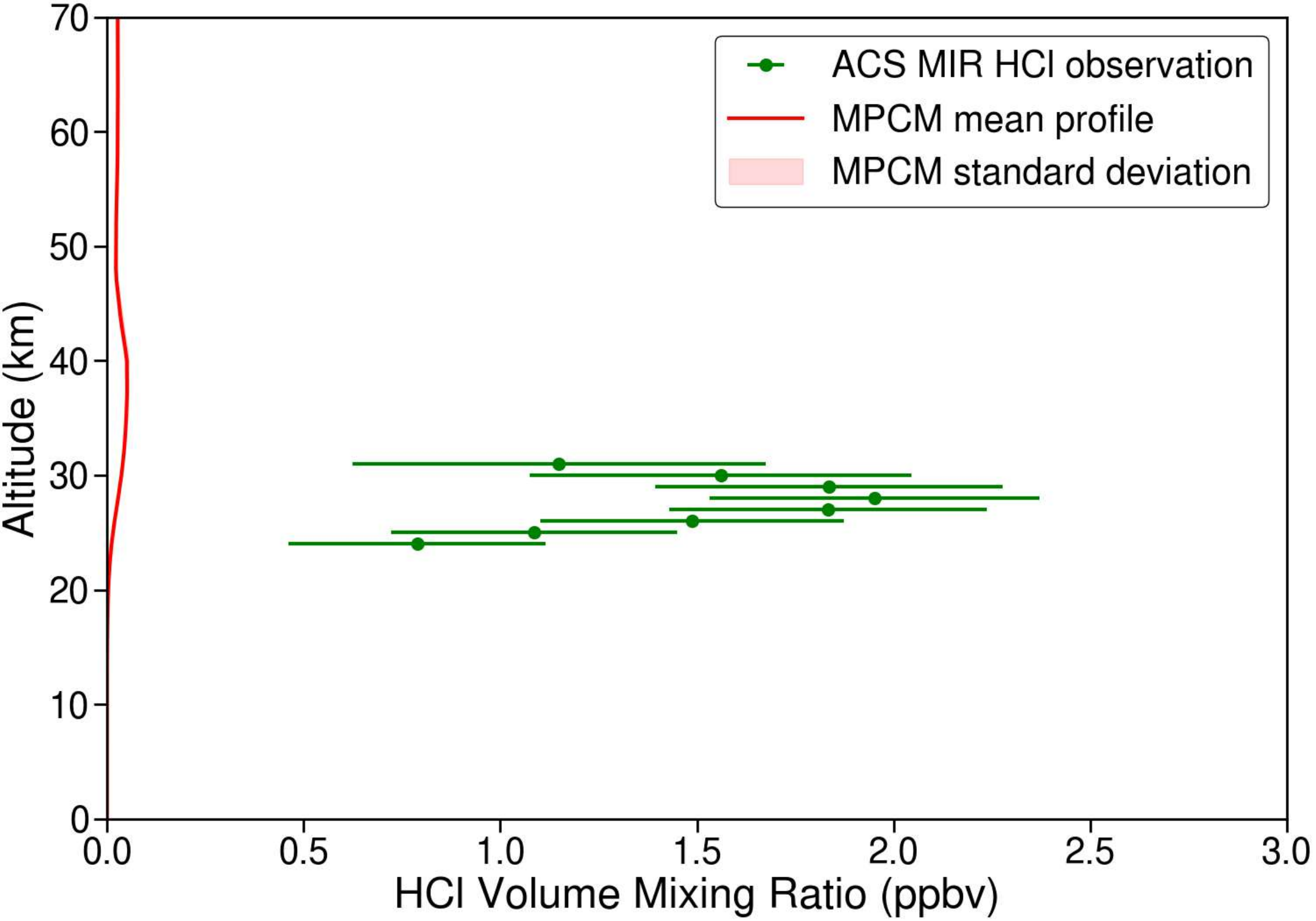}
            \includegraphics{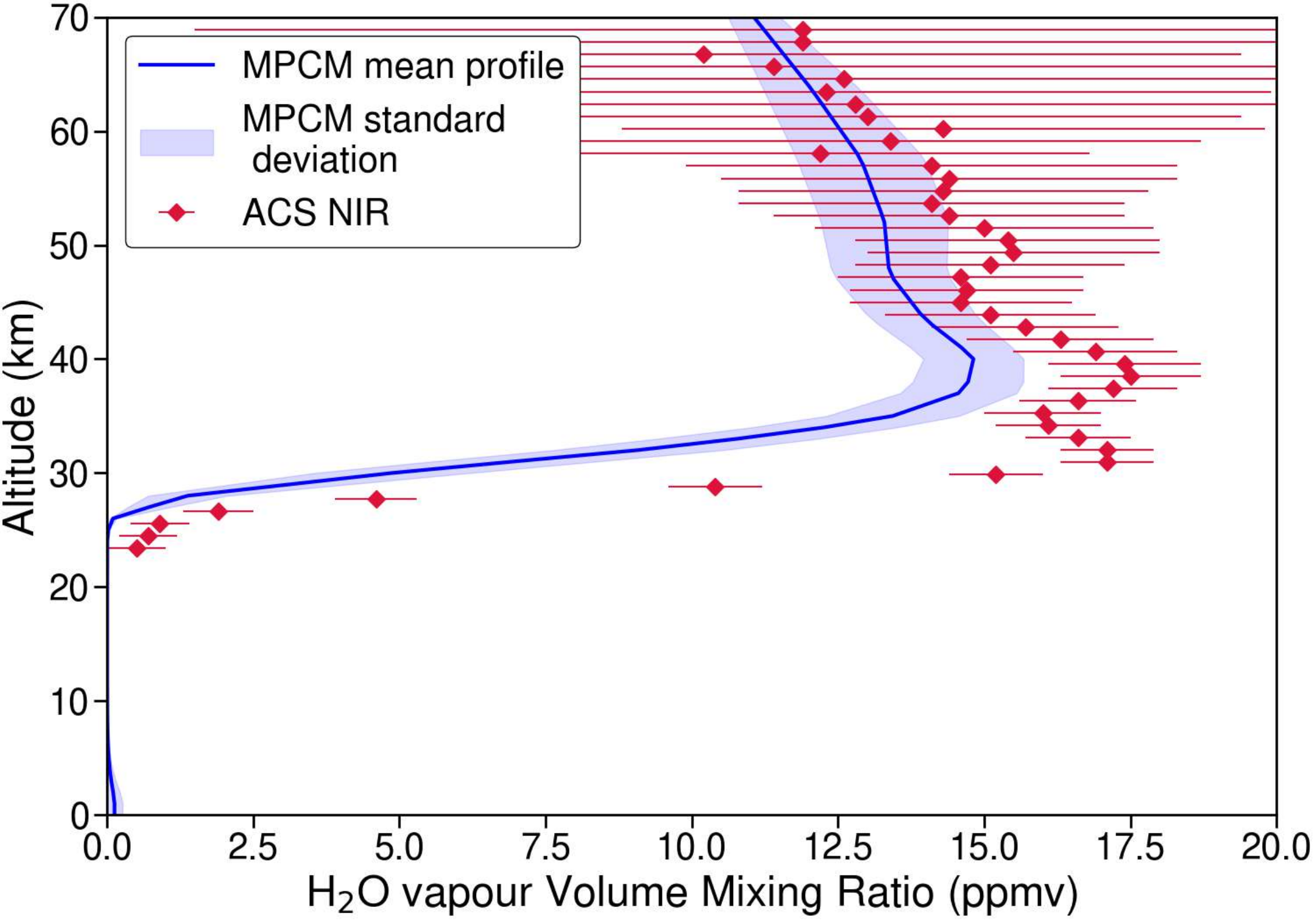}
            \includegraphics{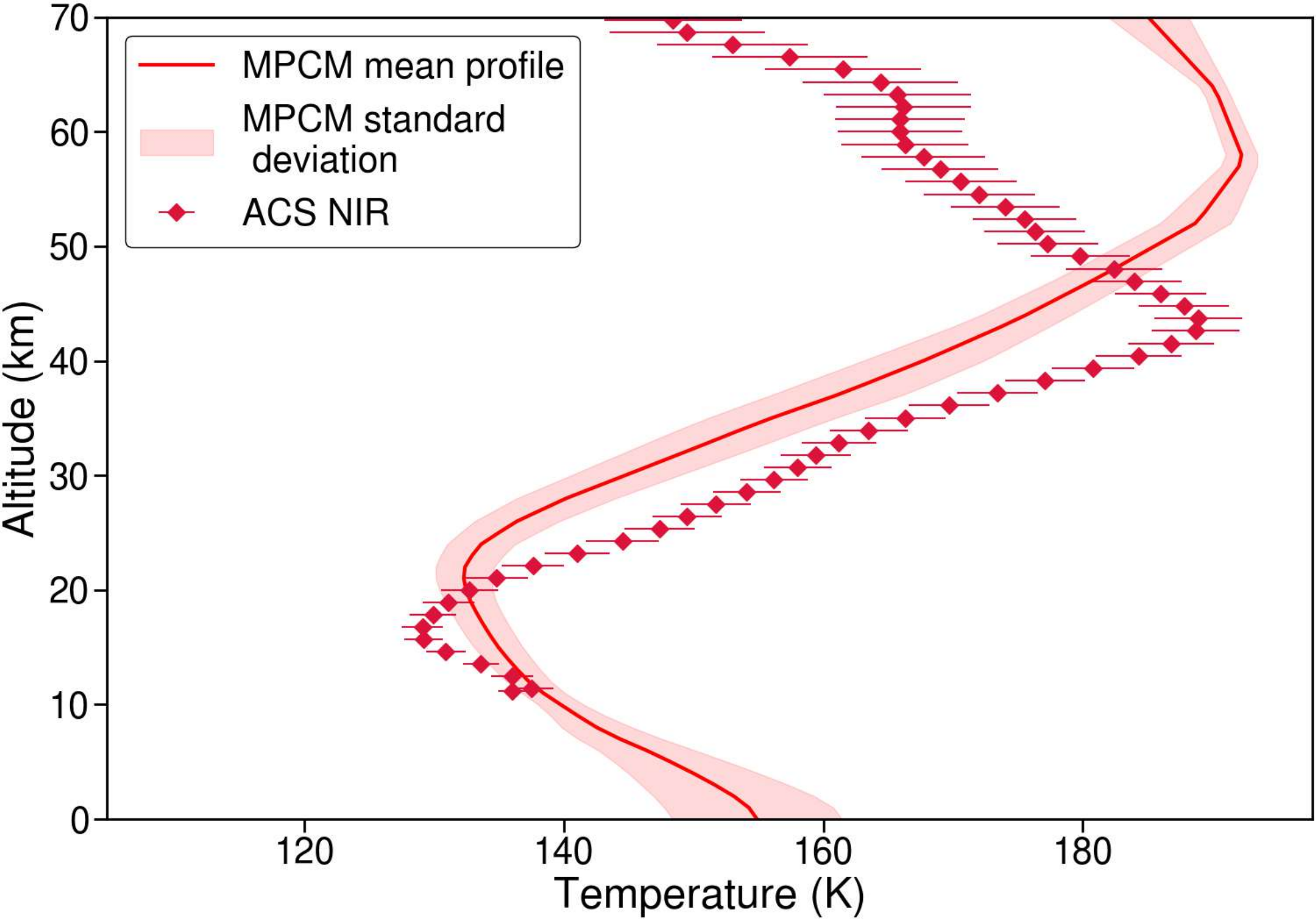}
            }
      \caption{Comparison between model and observed HCl VMRs (left), water vapour VMRs (middle), and temperature (right) for a representative case where the modelled HCl VMR profile is lower than the observed profile from ACS MIR. Model profiles are averaged over the 5$^{\circ}$ L$_{\text{S}}$ interval containing the ACS observation. The standard deviation represents the variation of the computed profiles on this interval. These plots correspond to the ACS MIR occultation 011220\_N2\_E\_P1\_11\_F, located at latitude 76.94$^{\circ}$N, longitude 56.38$^{\circ}$E, 209.92$^{\circ}$ L$_{\text{S}}$, LT = 10.67 hrs in MY 35.}
         \label{Fig_HCl_vap_biases_north}
\end{figure*}

\begin{figure*}[!h]
    \centering
   \resizebox{\hsize}{!}
            {
            \includegraphics{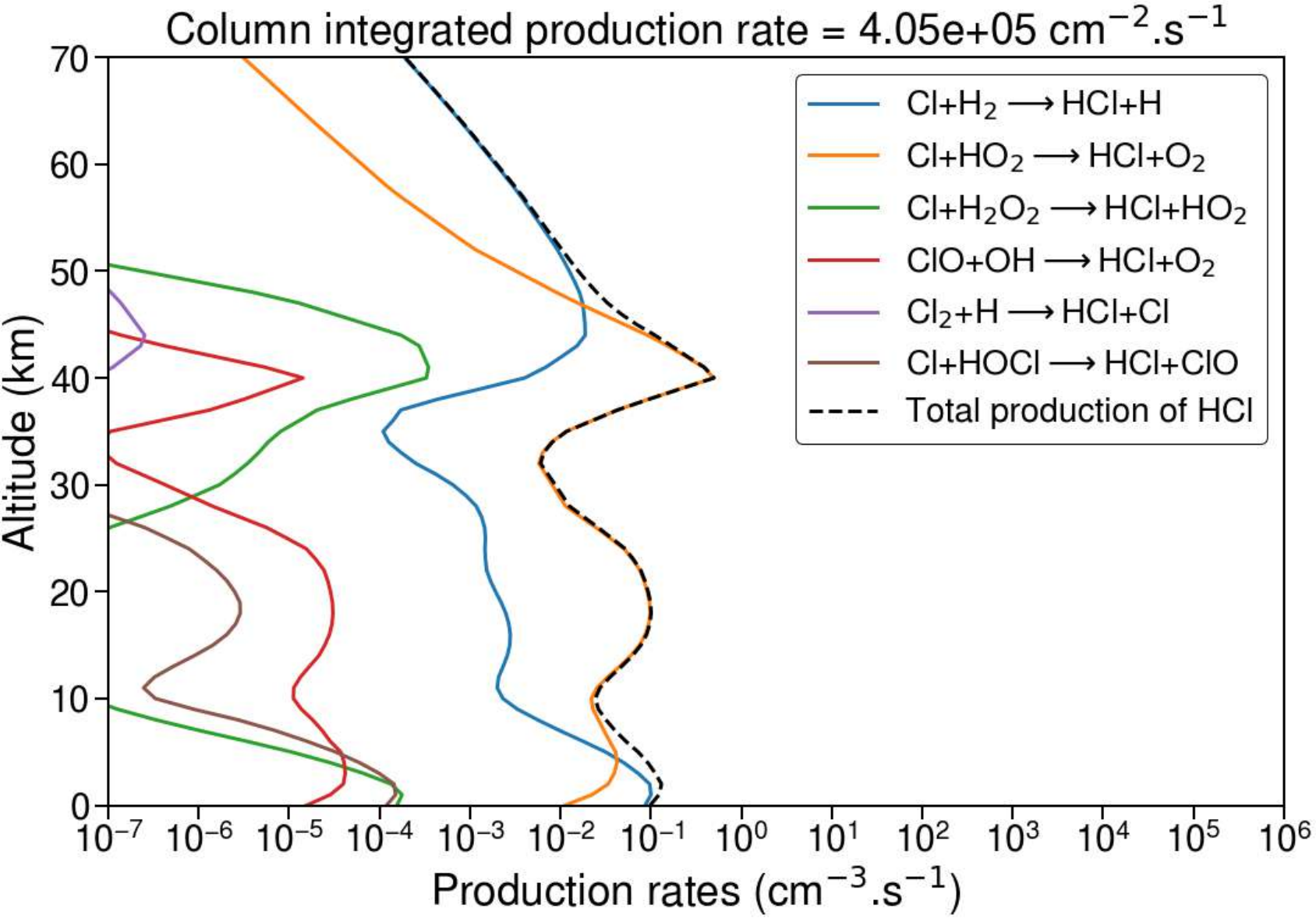}
            \includegraphics{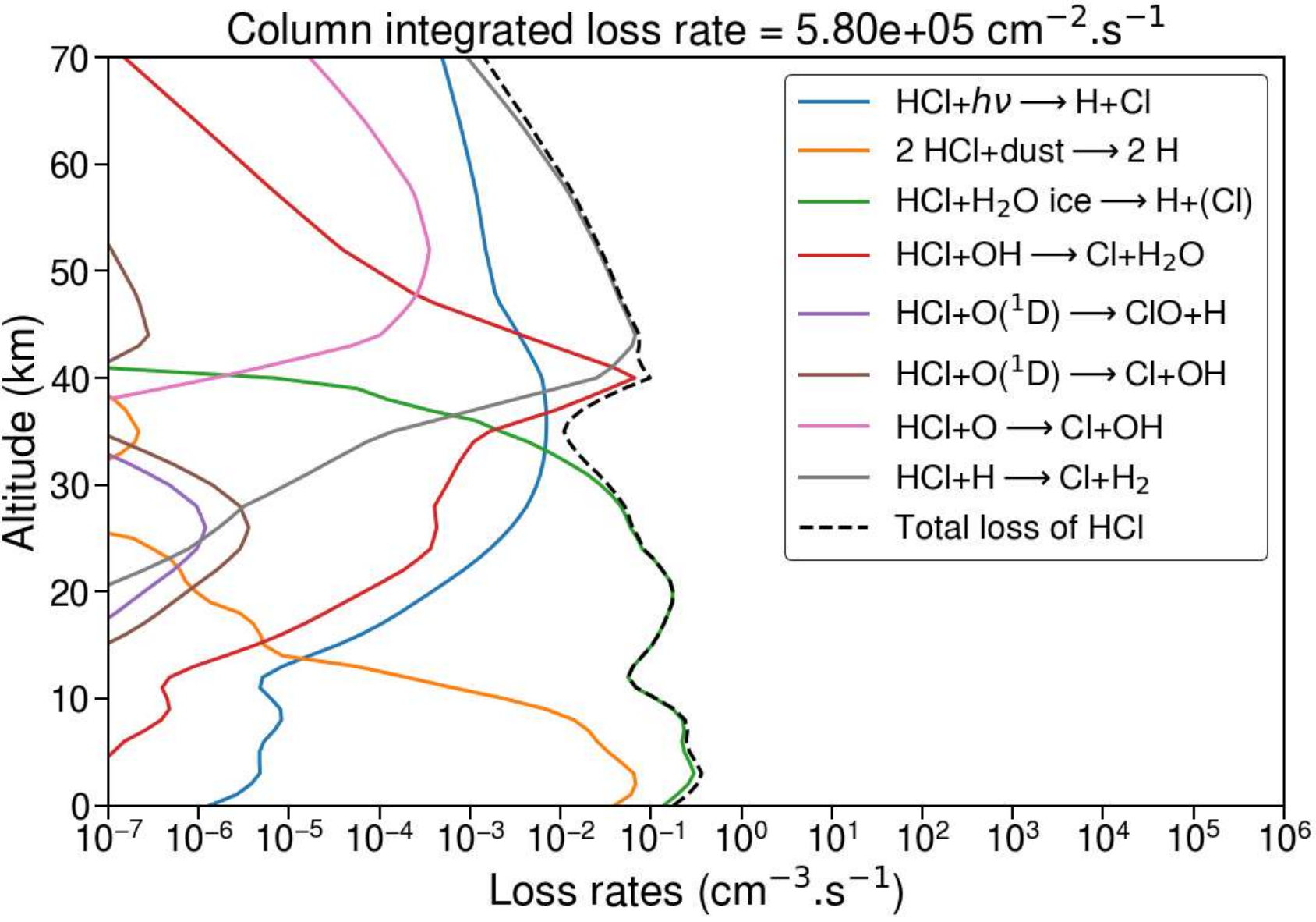}
            \includegraphics{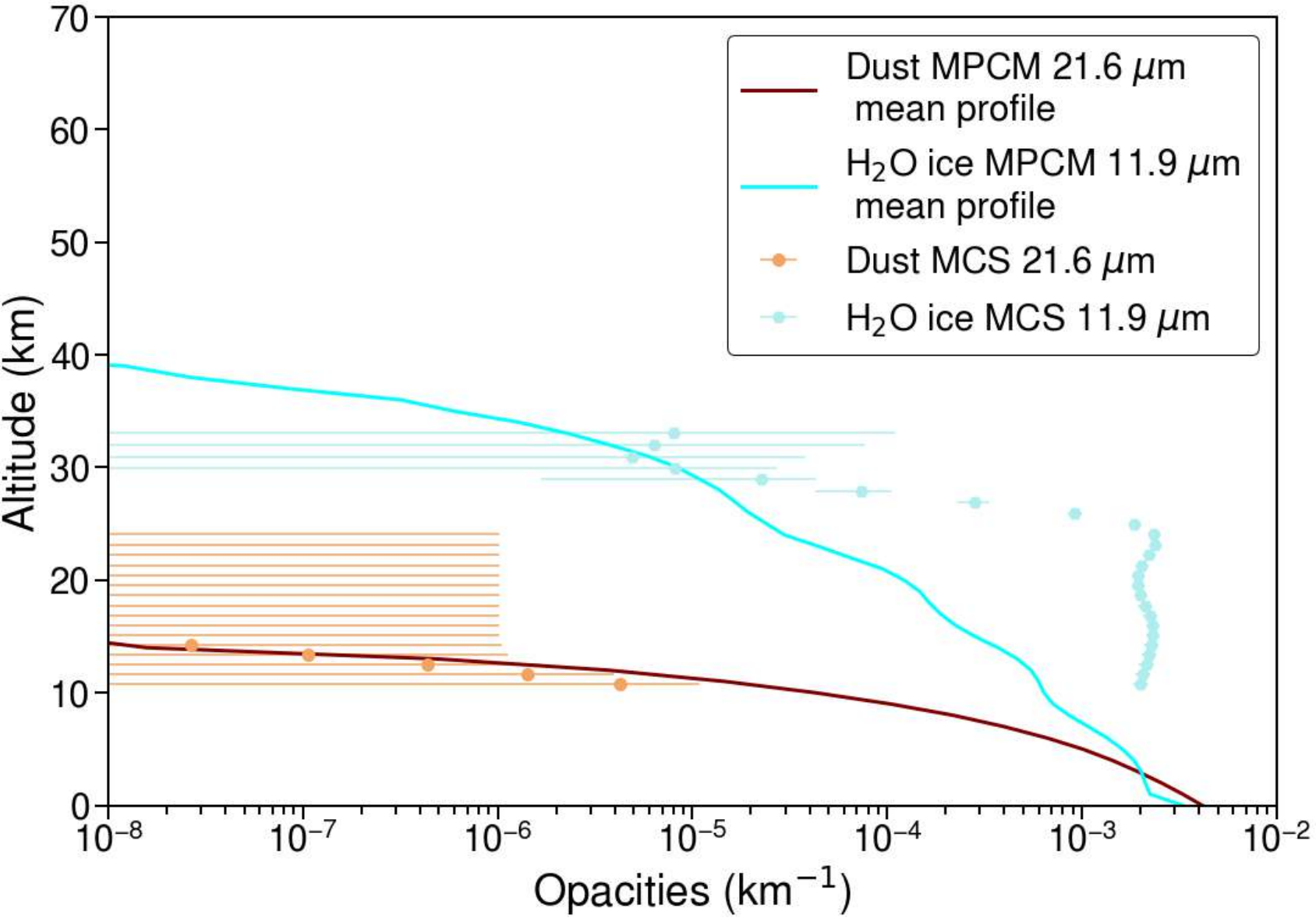}
            }
      \caption{Model production (left) and loss (middle) rates of HCl corresponding to the ACS HCl detection from Fig. \ref{Fig_HCl_vap_biases_north}, where modelled HCl is lower than observed by ACS. The right panel shows model (solid line) and MCS (closed circles) water ice and dust opacity profiles. The MCS data are taken as close as possible in space and time to the ACS MIR HCl observation. Model profiles are averaged over the 5$^{\circ}$ L$_\text{S}$ interval containing the observation; standard deviations are omitted here for clarity.}
         \label{Fig_prod_loss_biases_north}
\end{figure*}

For the first period, which corresponds to the beginning of MY 35, all large biases are found above 45$^{\circ}$ latitude. More precisely, most detections in the northern hemisphere occurred above 70$^{\circ}$N. HCl and water vapour VMRs profiles for one of these points are shown in Fig. \ref{Fig_HCl_vap_biases_north}. Production and loss rates, as well as water ice and dust opacity profiles are presented in Fig. \ref{Fig_prod_loss_biases_north}. 
As illustrated in Fig. \ref{Fig_HCl_vap_biases_north}, the atmosphere is generally very dry at these latitudes, with water vapour VMRs typically lower than 25 ppmv. In addition, water ice clouds are found over a large altitude range, typically from the surface to 30-40 km (see Fig. \ref{Fig_prod_loss_biases_north}), where HCl is often detected. Finally, the dust abundance decreases rapidly with altitude. 

As shown in Fig. \ref{Fig_prod_loss_biases_north}, the large vertical extent of water ice clouds significantly reduces HCl abundances. This reduction occurs due to uptake of HCl on water ice (via reaction h3), which is the main loss process in regions with high water ice abundance (see Fig. \ref{Fig_prod_loss_general_results} for another example).

In addition to the net chemical loss of HCl caused by comparatively large loss and small production rates, we find that the influence of horizontal transport is generally small for these HCl observations (see Appendix \ref{appendix_transport} for more details). As a result, the imbalance between chemical production and loss terms is not compensated by any inflow of HCl from elsewhere. 
As seen in Fig. \ref{Fig_MAXhcl_vs_Ls}, the observations with strong negative biases at the beginning of MY 35 occur between L$_{\text{S}}$=140$^{\circ}$ and L$_{\text{S}}$=220$^{\circ}$. This period encompasses the northern autumn equinox, when the inversion of the Hadley cell occurs, which explains the low transport fluxes. However, we find that in our model, low HCl VMRs are generally correlated with low horizontal transport fluxes at altitudes where the influence of atmospheric transport is not negligible compared to atmospheric photochemistry. 

Changes in atmospheric temperature (Fig. \ref{Fig_HCl_vap_biases_north}) also play a key role in changes in the uptake rates of HCl on both dust (reaction h2) and water ice (reaction h3). The uptake coefficients of these reactions exhibit negative temperature dependencies above 136.4 and 150.3 K, respectively, due to the use of an MPM-type formula \citep{taysum_observed_2024}. 
Consequently, the uptake of HCl on dust and water ice is most efficient at temperatures typically found at high latitudes during this period. As a result, conditions at high latitudes at this time of the year do not support the net production of large HCl VMRs: chlorine production through h1 is inhibited due to low water vapour and dust content, while the loss of HCl is enhanced by the large amount of water ice and low temperatures. This suggests that the current uptake coefficients are not well suited to reproduce HCl abundances under these conditions.

This also explains the larger biases arising after L$_{\text{S}}$=330$^{\circ}$: the atmosphere cools as dust settles down, leading to a reduction in atmospheric water vapour and an increase in water ice clouds. These conditions resemble those observed at the beginning of MY 35, resulting in lower HCl VMRs. Based on our analysis, we conclude that the rapid disappearance of HCl at the end of the dust season is due to the decrease of atmospheric dust and water vapour after the southern autumn months, when there is a large and rapid reduction in chlorine production. At the same time, atmospheric temperatures decrease and water ice clouds increase, resulting in rapid destruction of HCl.

\begin{figure*}[!h]
    \centering
   \resizebox{\hsize}{!}
            {
            \includegraphics{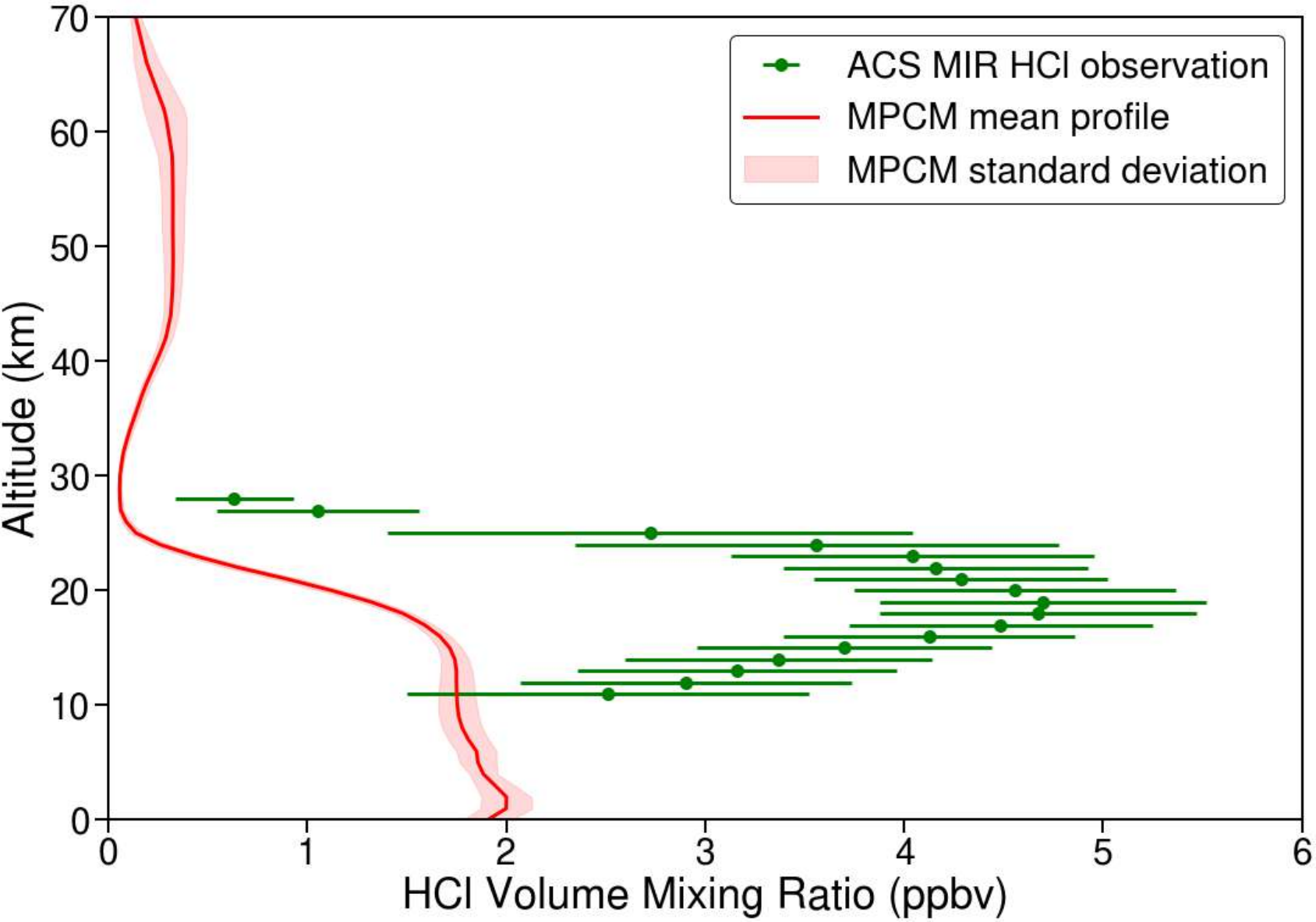}
            \includegraphics{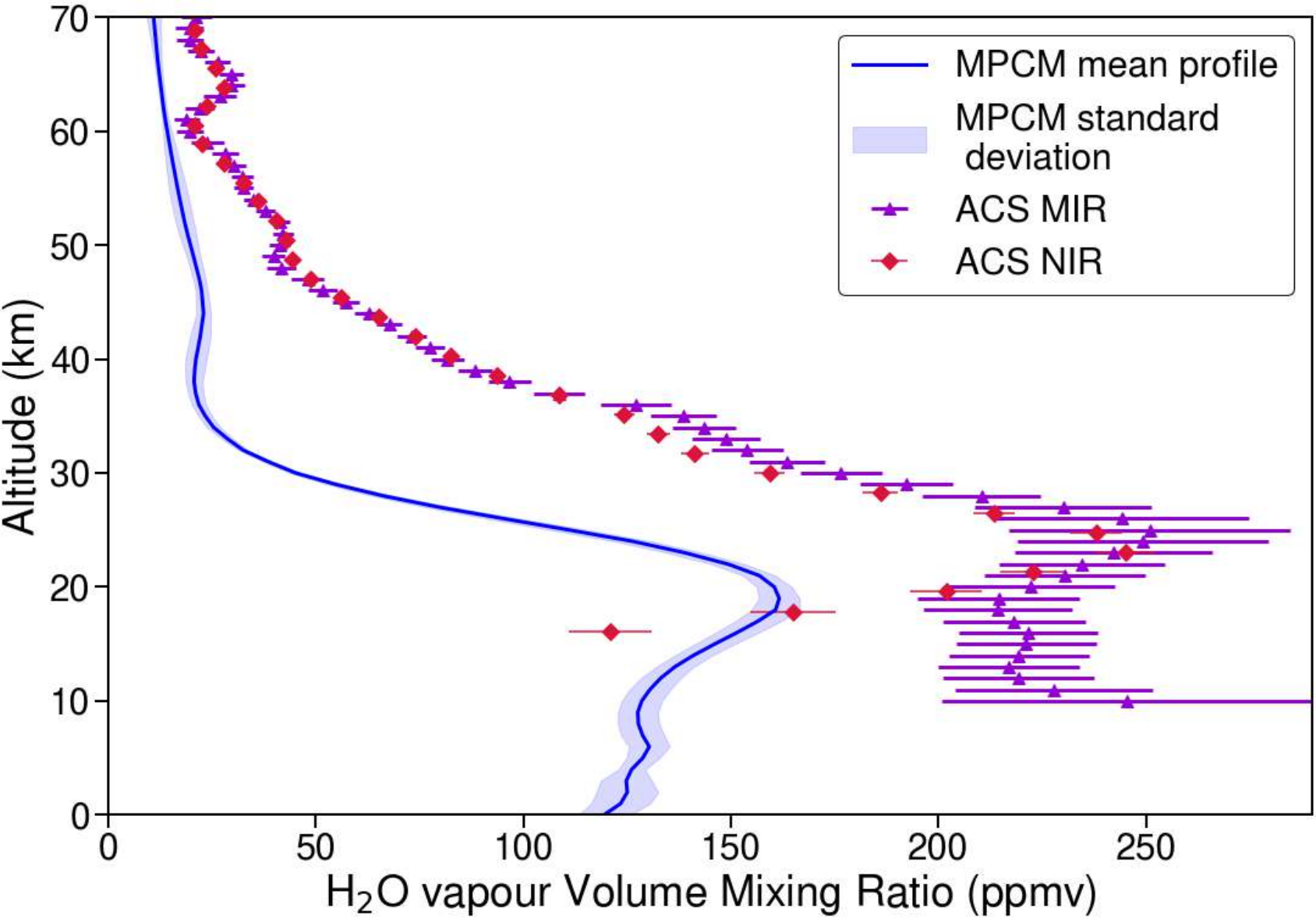}
            \includegraphics{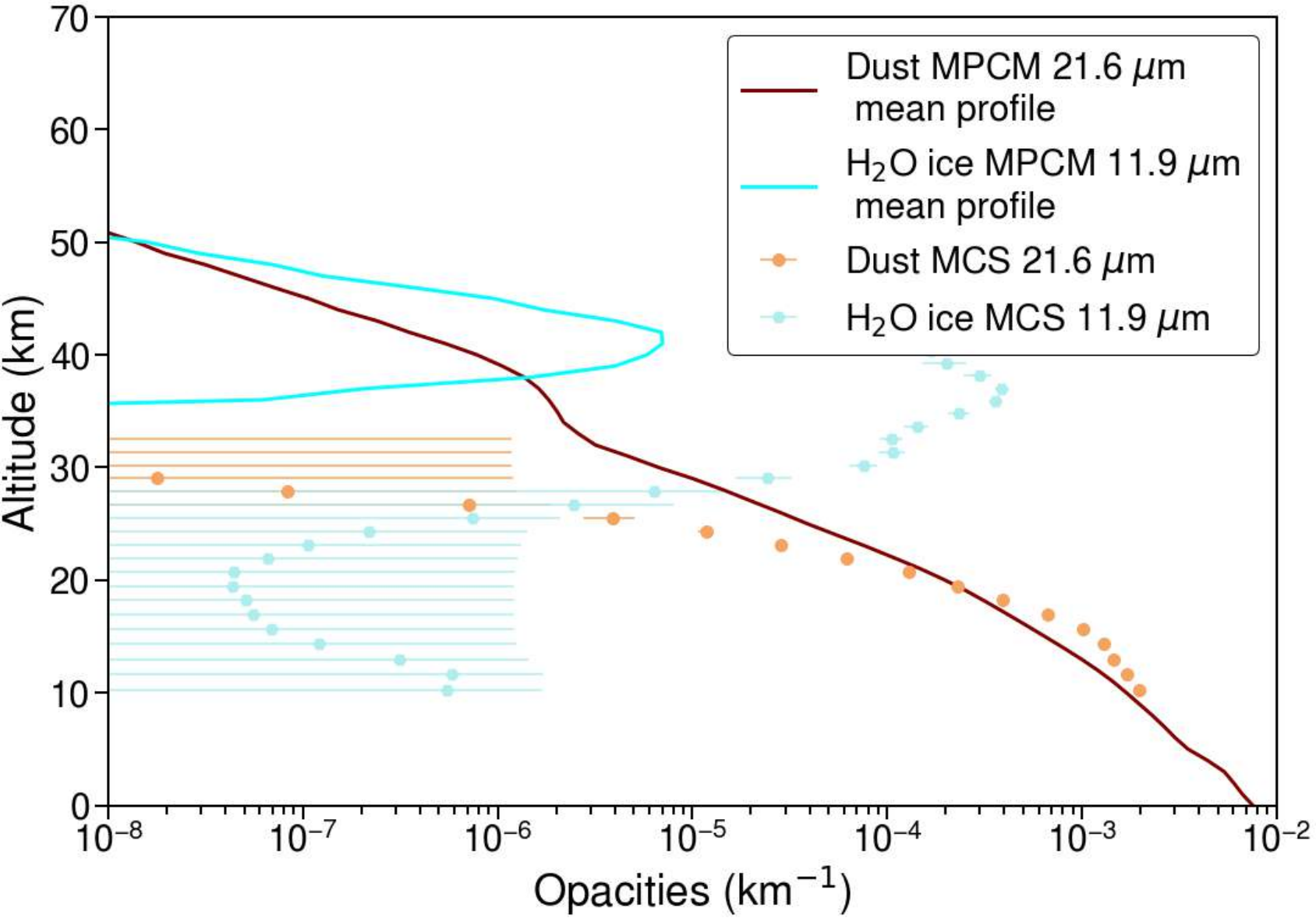}
            }
      \caption{Comparison between modelled and observed HCl VMRs (left),  water vapour VMRs (middle), and water ice and dust opacities (right) for a representative case where the modelled HCl VMR profile is lower than the observed profile from ACS MIR due to a negative bias in the water vapour abundance. Model profiles are averaged over the 5$^{\circ}$ L$_{\text{S}}$ interval containing the ACS observation. The standard deviation represents the variability of the computed profiles on this interval. These plots correspond to the ACS MIR occultation 013069\_N2\_E\_P1\_12\_F, located at latitude 56.83$^{\circ}$S, longitude 86.67$^{\circ}$E, 304.16$^{\circ}$ L$_{\text{S}}$, LT = 20.34 hrs in MY 35.}
         \label{Fig_HCl_vap_biases_mid}
\end{figure*}

We also observe large negative model biases in the L$_{\text{S}}$=275--310$^{\circ}$ range. We find that these biases result from two main causes: a strong negative bias in water vapour, the presence of a water ice cloud, or a combination of the two (e.g. Fig. \ref{Fig_HCl_vap_biases_mid}).
Therefore, all of these model biases are linked to the dust cycle, which influences the water vapour and water ice content in the atmosphere, as well as the atmospheric temperature. This is also consistent with the decrease in model biases between L$_{\text{S}}$=310--330$^{\circ}$ in both years, corresponding to the typical appearance of the late-season storm when surface mineral dust is lofted into the atmosphere, resulting in environmental conditions that elevate the production of atomic chlorine and reduce the loss of HCl. 

\subsubsection{Study of model positive biases}
\label{subs_study_pos_biases}

\begin{figure*}[!h]
    \centering
   \resizebox{\hsize}{!}
            {
            \includegraphics{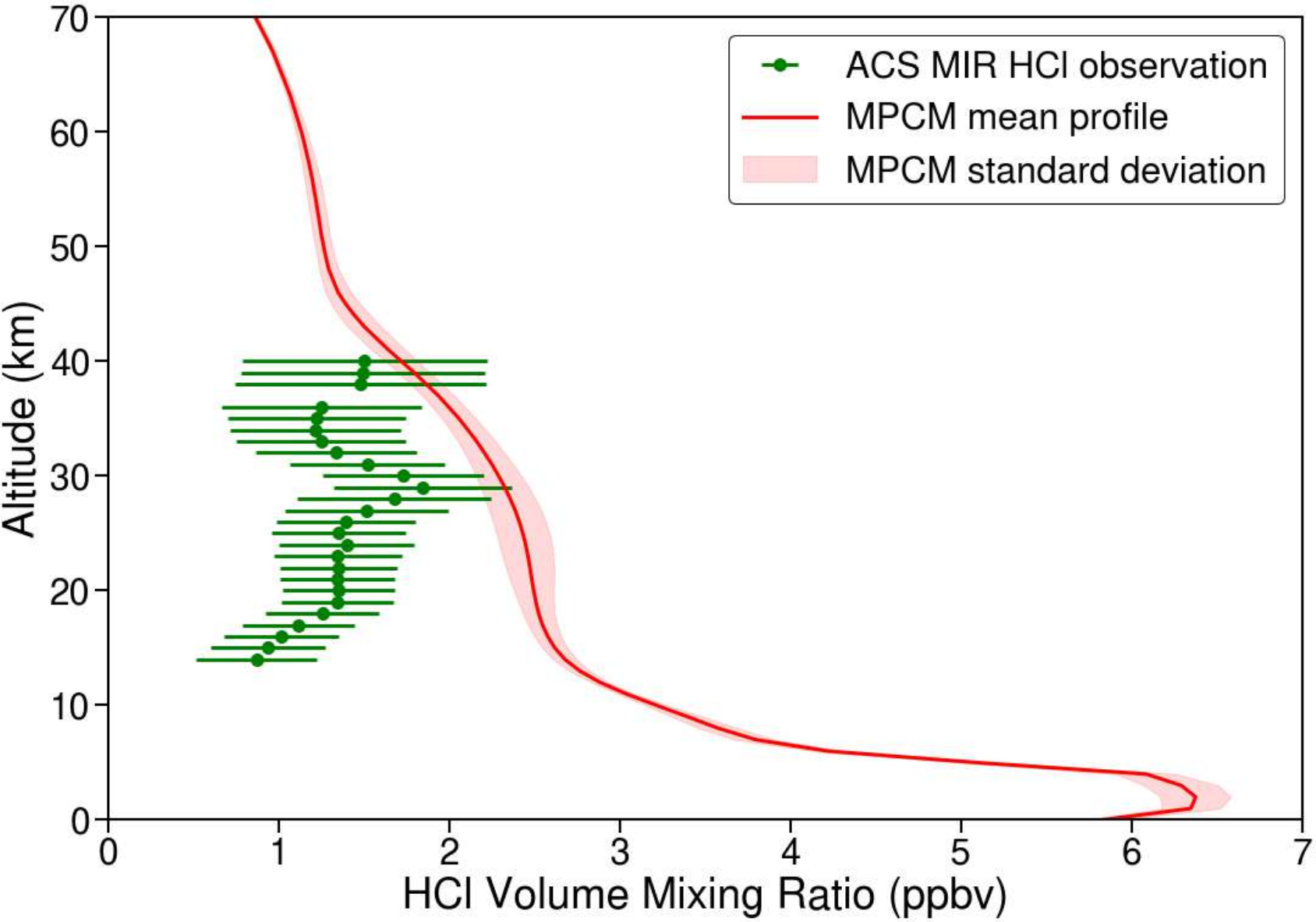}
            \includegraphics{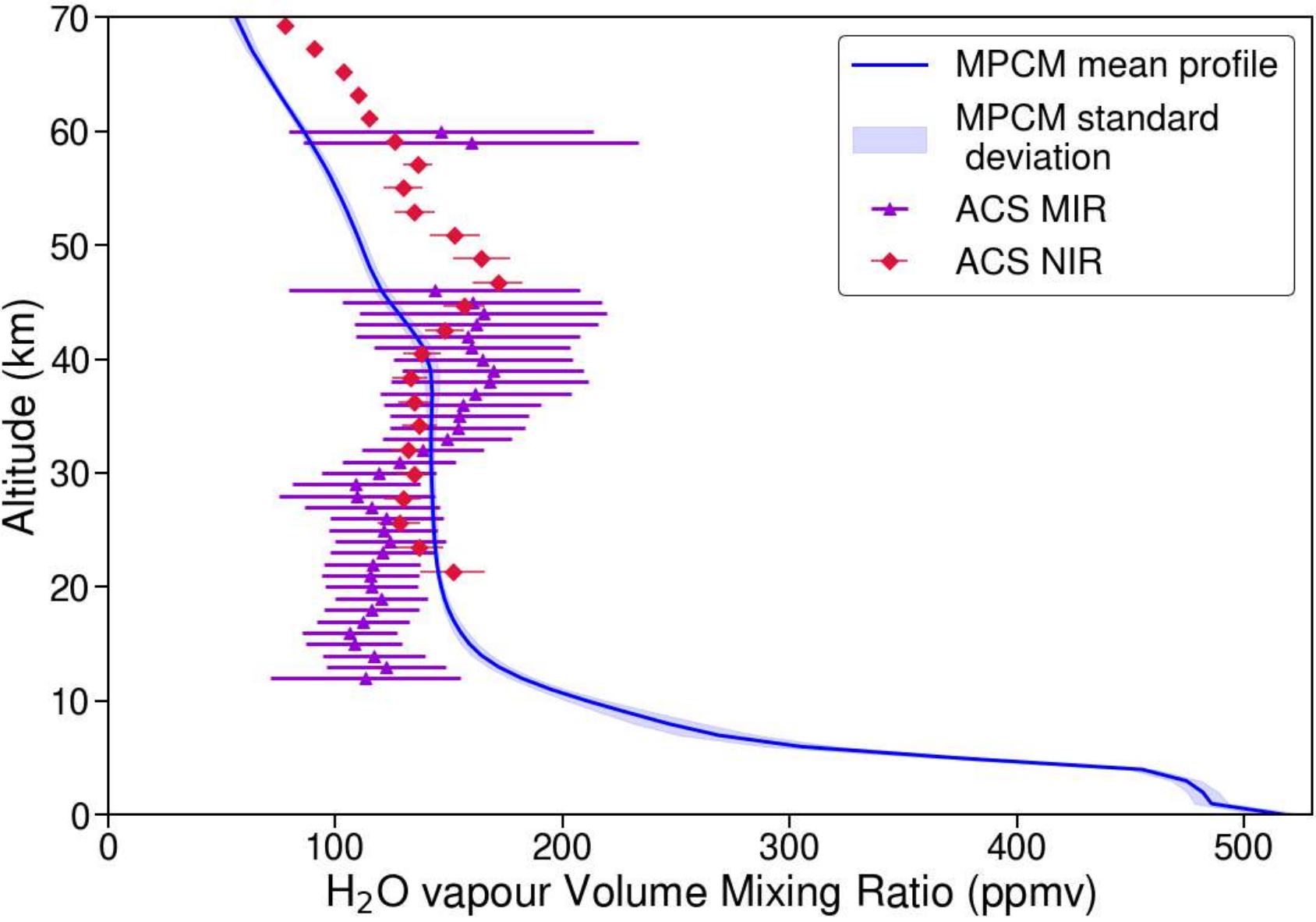}
            \includegraphics{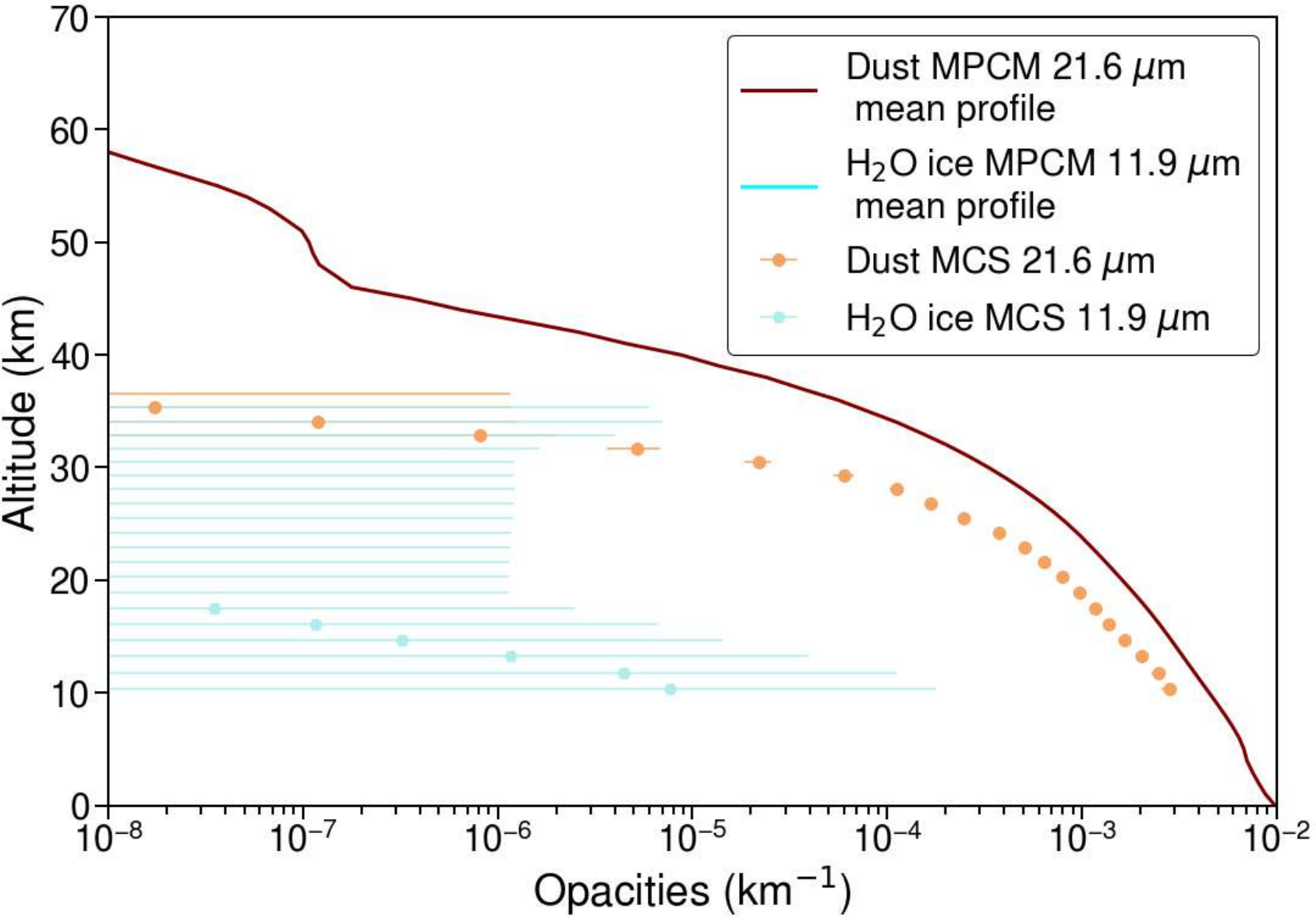}
            }
      \caption{Comparison between modelled and observed HCl VMRs (left), water vapour VMRs (middle), and water ice and dust opacities (right) for a representative case where the modelled HCl VMR profile is higher than the observed profile from ACS MIR. Model profiles are averaged over the 5$^{\circ}$ L$_{\text{S}}$ interval containing the ACS observation. The standard deviation represents the variability of the computed profiles over this interval. These plots correspond to the ACS MIR occultation 003782\_N2\_E\_P1\_11\_F, located at latitude 61.05$^{\circ}$S, longitude 76.76$^{\circ}$E, 258.89$^{\circ}$ L$_{\text{S}}$, LT = 21.73 hrs in MY 34.}
    \label{Fig_HCl_vap_biases_south}
\end{figure*}

In Fig. \ref{Fig_MAXhcl_vs_Ls}, we find that some profiles show that the model has a positive bias against ACS MIR retrievals. An example is shown in Fig. \ref{Fig_HCl_vap_biases_south}. For this case, water vapour and dust are abundant at these latitudes, 
particularly towards the surface. The production of atomic chlorine is strongly enhanced (reaction h1), while HCl losses are low due to the low abundance of atmospheric water ice (reaction h3). Additionally, temperature is high near the surface, which reduces the efficiency of reaction h2, further reducing the loss of HCl. 

These conditions are common at southern latitudes, and result in HCl peaks ($\sim 10^1$ ppbv) at altitudes below 10~km, as shown in Sect. \ref{subs_climatology} and illustrated by Fig. \ref{App_fig_FullFig_67}. These HCl peaks unfortunately lie outside the altitude range accessible by ACS MIR. However, similar HCl peaks at L$_{\text{S}}$=273$^{\circ}$ and L$_{\text{S}}$=306$^{\circ}$ were detected from ground-based telescopic observations \citep{aoki_global_2024} during MY 35, collected by the iSHELL high-resolution near-infrared echelle spectrometer \citep{rayner_ishell_2022} at the NASA Infrared Telescope Facility (IRTF). These measurements showed a strong correlation between HCl and water vapour, with HCl VMRs increasing strongly near the surface, where large water vapour VMRs are reached following the sublimation of the polar cap. This study also reported larger HCl VMRs towards high southern latitudes. Our findings are therefore consistent with these ground-based observations from \citet{aoki_global_2024}. A caveat is that the standard deviation of water vapour VMRs close to the surface in our model under these conditions is generally large, which also results in large standard deviations for HCl.

\subsection{Non-detections}
\label{subs_non_detect}

To evaluate our model, we also compared our HCl vertical profiles to ACS MIR measurements where this species could not be confirmed, thereafter referred to as non-detections. At these locations, an HCl profile was retrieved but
the uncertainties at each tangent height were too large relative to the retrieved abundance
to qualify as a definitive detection. For most of these retrievals, HCl VMRs are low and fitting uncertainties are large. However, we can still compare model results to these profiles to assess model consistency. We compare the model HCl with every ACS non-detection during MY 34 and MY 35, as well as model and observed profiles of water vapour, temperature, water ice, and dust opacities from ACS MIR, ACS NIR, and MCS. These figures are available on \href{https://zenodo.org/records/14717873}{Zenodo}. A figure similar to Fig. \ref{Fig_MAXhcl_vs_Ls} comparing model results to ACS MIR minimum uncertainties for these non-detections is also provided in Appendix (Fig. \ref{App_Fig_MAXhcl_vs_Ls_NonDet}).

Overall, we find that in most cases the model lies within the estimated uncertainties of the non-detection retrieved profiles during MY 34 and 35. If we compute the relative difference between the retrieved profiles and model outputs, considering the standard deviation in model outputs and the uncertainties in the retrieved values, yields a median relative difference of -54\% and a mean relative difference of 29\%. This is consistent with our model generally exhibiting a negative bias for HCl VMRs. 
Retrievals with low HCl VMRs typically correspond to low water vapour abundances, high water ice abundances, or low dust content. This result provides additional confidence in our model's ability to reproduce the chemical processes that describe HCl and the broader chlorine cycle in the Martian atmosphere. For a small number of retrievals, we find that model HCl VMRs exceed ACS MIR data. For almost all these profiles, we observe a positive model bias for the water vapour or dust abundance, or a negative bias for the water ice abundance. This implies that the observed HCl biases are not solely due to the chlorine chemistry implemented in the model but are also strongly influenced by model errors in the water and dust cycles. 

\subsection{Correlations between HCl and other atmospheric quantities}\label{subs_correlations}

\begin{figure*}[!h]
    \centering
   \resizebox{0.9\hsize}{!}
            {
            \includegraphics{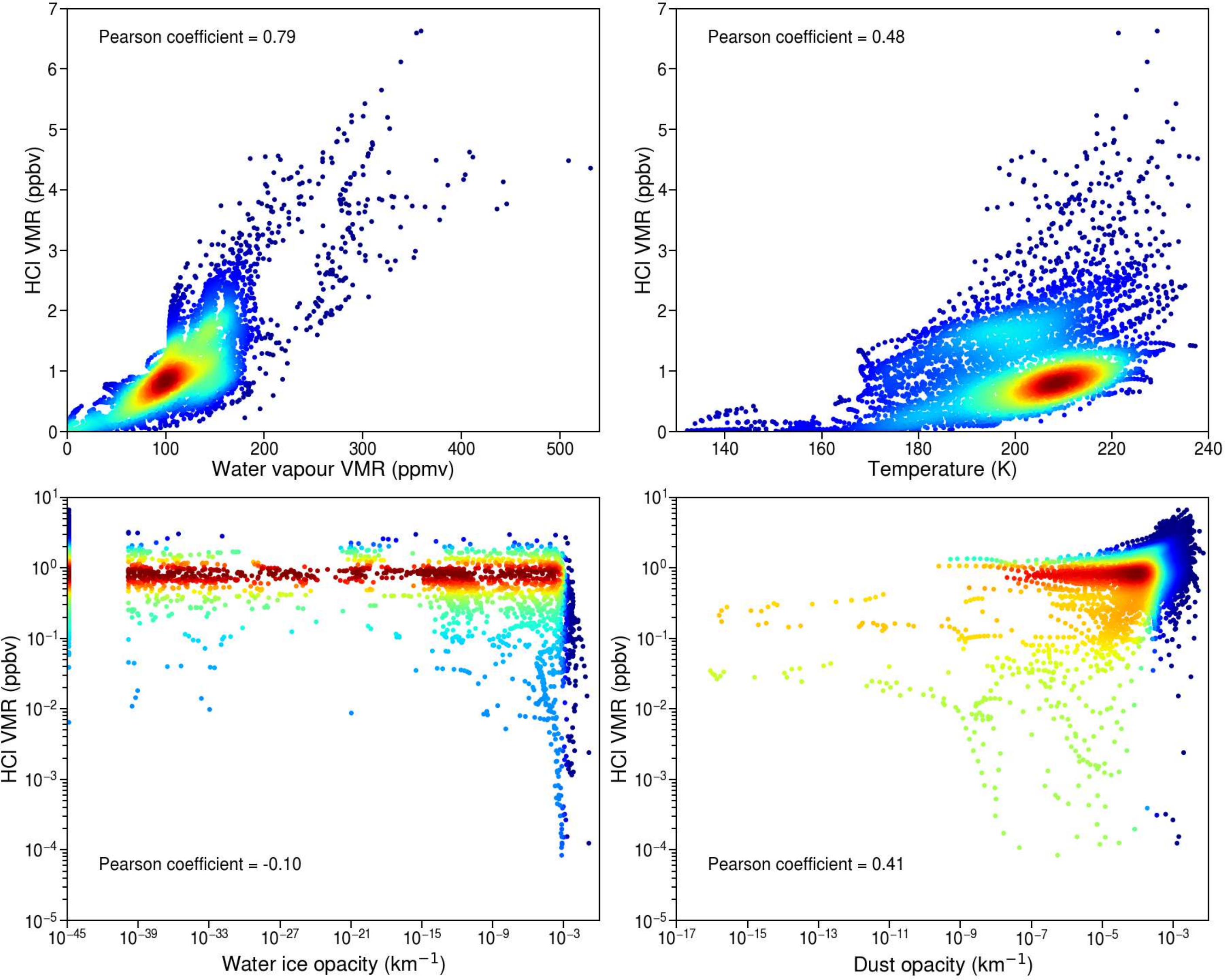}
            }
      \caption{Scatter plots of model HCl VMRs versus (top left) water vapour VMRs, (top right) temperature, (bottom left) water ice opacity, and (bottom right) dust opacity, sampled at each ACS MIR occultation where HCl was retrieved during MY 34 and MY 35. Only data from the altitudes where HCl was successfully retrieved are considered. Points with zero ice opacity are arbitrarily assigned an opacity of 1 10$^{-45}$ km$^{-1}$. The colour scale indicates point density, increasing from blue to red.}
         \label{Fig_stats_correl_Pearson_detect_MY3435}
\end{figure*}

We computed correlations between HCl VMRs and other key atmospheric quantities: water vapour VMRs, water ice opacity, dust opacity, and temperature. Similar computations were performed by \citet{olsen_relationships_2024-1} using ACS data so we chose to replicate their method to enable a direct comparison between obtained values. 
We therefore used all the points from the model profiles within the altitude range where HCl was detected by ACS MIR to calculate the Pearson correlation coefficients between HCl and the other quantities.
Our calculations considered data from MYs 34 and 35 simultaneously and the results are shown in Fig. \ref{Fig_stats_correl_Pearson_detect_MY3435}. Table \ref{Table_compar_Pears_coeffs} compares the correlation coefficients obtained from our model to those reported by \cite{olsen_relationships_2024-1}.

\begin{table}[!h]
\centering
    \caption{Pearson correlation coefficients between HCl and key atmospheric quantities.}
    \label{Table_compar_Pears_coeffs}
    \begin{tabular}{lcc}
        \hline \hline
        Quantity          & \begin{tabular}[c]{@{}c@{}}\citet{olsen_relationships_2024-1}\end{tabular} & \begin{tabular}[c]{@{}c@{}}This work\end{tabular} \\ \hline
        Water vapour VMRs & 0.63                                                                                                                            & 0.79                                                                                \\
        Temperature       & -0.2                                                                                                                            & 0.48                                                                                \\
        Water ice opacity & -0.09                                                                                                                           & -0.10                                                                               \\
        Dust opacity      & 0.18                                                                                                                            & 0.41                                                                                \\ \hline
    \end{tabular}
\end{table} 

We find good agreement between the correlation coefficients for HCl and water vapour and between HCl and water ice. The HCl–water ice relationship confirms that the correlation coefficient is close to zero at the observation locations, despite water ice being a very effective sink of HCl.  
The HCl–water vapour correlation can be linked to the production process of atomic chlorine, as described in Sect. \ref{subsect_vert_prof_comp}: atomic chlorine is produced through reaction h1, which involves water vapour, and then reacts with H$_2$ and HO$_2$, produced from water photolysis, to form HCl. Heterogeneous chemistry also explains the positive correlation between HCl and dust opacity. Because dust is involved in both the production and loss of HCl, through reactions h1 and h2, respectively, the correlation is lower than that for water vapour but remains positive. Our model correlation is, however, stronger than that derived by \citet{olsen_relationships_2024-1}. This could be a consequence of computing our coefficients at the exact same time and location as HCl, whereas \citet{olsen_relationships_2024-1} used MCS data co-located within a window of 0.25$^{\circ}$ L$_{\text{S}}$ and 500~km.
As a consistency check, we also computed the correlations between HCl, water ice, and dust opacities while restricting opacity values to those larger than 10$^{-7}$ km$^{-1}$, consistent with the range accessible to MCS. Removing opacity values lower than 10$^{-7}$ km$^{-1}$ did not significantly change the correlation coefficients obtained from model calculations: we find -0.13 and 0.38 for water ice and dust opacity, respectively, compared to -0.10 and 0.41 when considering the full range of values (see Table \ref{Table_compar_Pears_coeffs}).

The positive correlation with temperature can also be linked to positive correlations with dust and water vapour: dust helps to heat the atmosphere, which increases the amount of water vapour that can be stored through the Clausius-Clapeyron relationship \citep{olsen_relationships_2024}. 
The higher correlation coefficient found in our model compared to \citet{olsen_relationships_2024-1} results from instrumental limitations of ACS. As stated in \citet{olsen_relationships_2024-1}, while temperature is retrieved from strong CO$_2$ lines, HCl retrievals use weaker lines that are often close to instrumental noise, which negatively impacts the correlation between these two quantities. Therefore, since correlations between water vapour and HCl and between water vapour and temperature are seen in ACS data, a positive correlation between HCl and temperature should also be found, as is the case in our model, which is not affected by instrumental limitations.
As detailed in Sect. \ref{subsect_vert_prof_comp}, heterogeneous processes are key for both production and loss of HCl, and reactions h2 and h3 have negative temperature dependences above 136.4 and 150.3 K, respectively, due to the use of an MPM-type formula \citep{taysum_observed_2024}. 
In contrast, reaction h1, which produces atomic chlorine, is proportional to the square root of temperature, as shown in Sect. \ref{subsect_chem_scheme}. 
Therefore, HCl destruction by reactions h2 and h3 becomes less efficient as temperature increases, while production of atomic chlorine through reaction h1 is enhanced, leading to a strong correlation between HCl and temperature. 

Correlation coefficients can also be compared to values derived from ground-based observations by \citet{aoki_global_2024}. For their two observation campaigns, they find large correlation coefficients of 0.67 and 0.85 between HCl and water vapour. Similar calculations between HCl and dust opacity led to correlation coefficients of -0.23 and -0.51. 
To compare our results to \citet{aoki_global_2024}, we computed the correlation between HCl and water vapour column densities and between HCl column densities and dust column opacity for all HCl detections from ACS MIR. Using this approach, we find correlation coefficients of 0.90 for HCl–water vapour and 0.17 for HCl– dust. These results are consistent with those reported by \citet{aoki_global_2024}. We note, however, that the HCl–dust coefficient is higher, although it remains low and therefore consistent with an absence of correlation. The higher values could result from the use of MCS data that are not exactly collocated, as discussed previously. 

To provide a more comprehensive view of the correlations between HCl and the other atmospheric quantities, we also computed Spearman correlation coefficients.
The purpose of these calculations was to assess whether the observed correlations are impacted by the analysis method used. Instead of considering values for each altitude where HCl was observed by ACS MIR and using data for all detections simultaneously to compute the correlation coefficient, we computed a Spearman correlation coefficient for each HCl detection independently. These calculations use the full vertical profile of the different quantities and therefore do not restrict the data to the altitude range where HCl is detected by ACS. Using Spearman coefficients also allows us to consider a broader set of relationships between HCl and the other quantities, as we do not restrict the calculations to linear relationships, unlike the Pearson correlation coefficient.
These coefficients are computed for each HCl detection in both MY 34 and MY 35 and are presented in Fig. \ref{Fig_stats_correl_MY3435}. 
Using this method, we find that HCl is strongly correlated with water vapour (median value = 0.76) and temperature (median value = 0.69), correlated with dust (median = 0.37) and anticorrelated with water ice (median = -0.61). 
The anticorrelation between HCl and water ice is much stronger, while other positive correlations are consistent with the results presented earlier. 
For dust, despite the median value being similar to that obtained with the Pearson analysis, the distribution is broad and covers all possible values between -1.0 and 1.0, with the mode of the distribution at 1.0. This confirms that the lower correlation compared to water vapour arises from dust acting both as a source and a sink of HCl, with a slight advantage of the source. 
The correlation with temperature is even more pronounced with these results, confirming that using more data points and not being limited by instrumental constraints highlights the HCl-temperature correlation.  

\subsection{The atmospheric lifetime of HCl on Mars}\label{subs_hcl_lifetime}

Based on model knowledge of HCl loss rates, we now compute the column integrated chemical lifetime of HCl $\tau_{\text{c}}$:
\begin{equation}
    \tau_{\text{c}} = \frac{\text{\{HCl}\}}{L^{\int }} \times \frac{1}{88775}~\text{sols,}
\end{equation}
where \{HCl\} is the column density of HCl in cm$^{-2}$ and $L^{\int}$ is the column-integrated loss rate of HCl in cm$^{-2}$.s$^{-1}$. The factor $\frac{1}{88775}$ converts the calculated lifetimes to sols. 

\begin{figure*}[!h]
   \resizebox{\hsize}{!}
            {
            \includegraphics{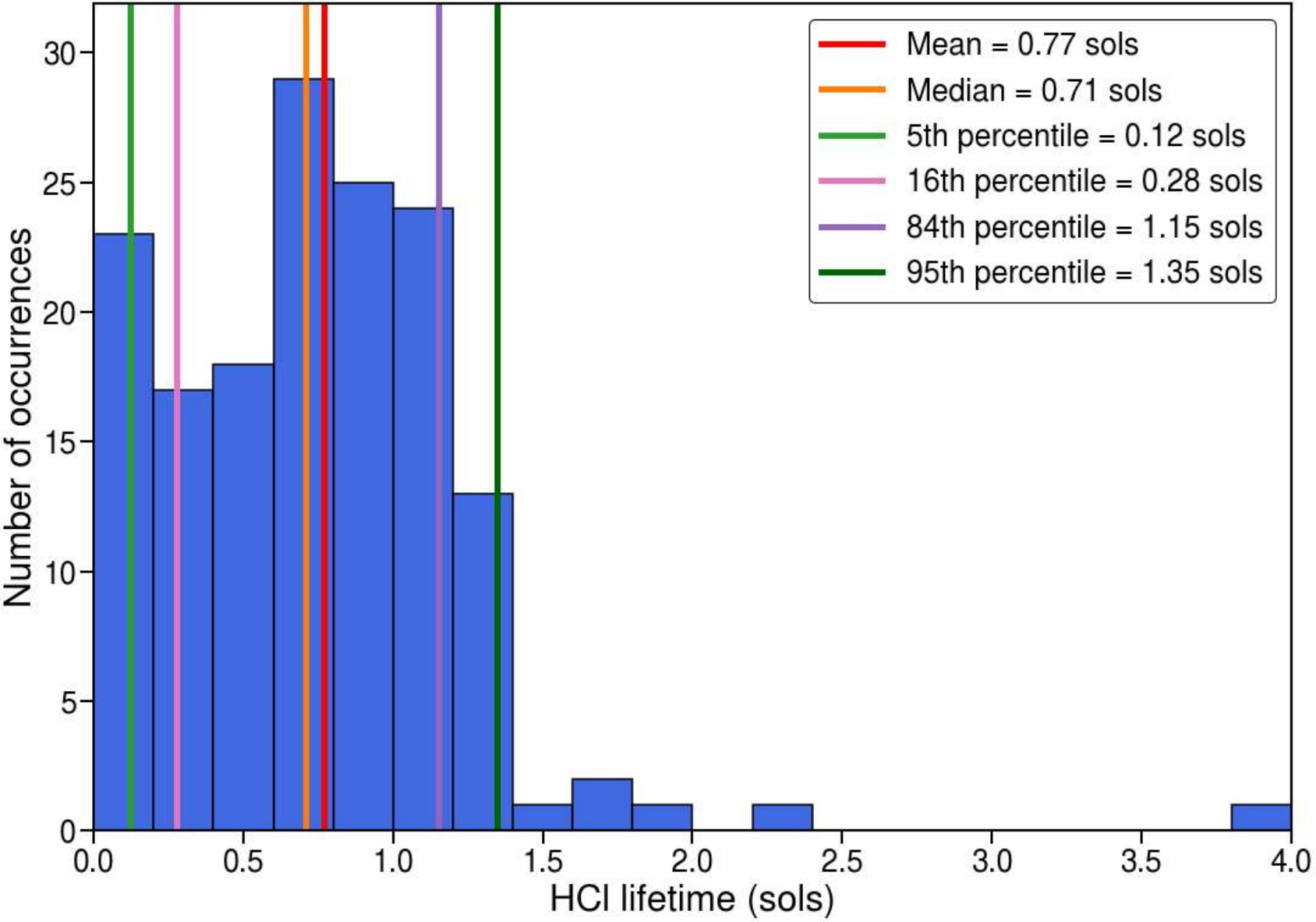}
            \includegraphics{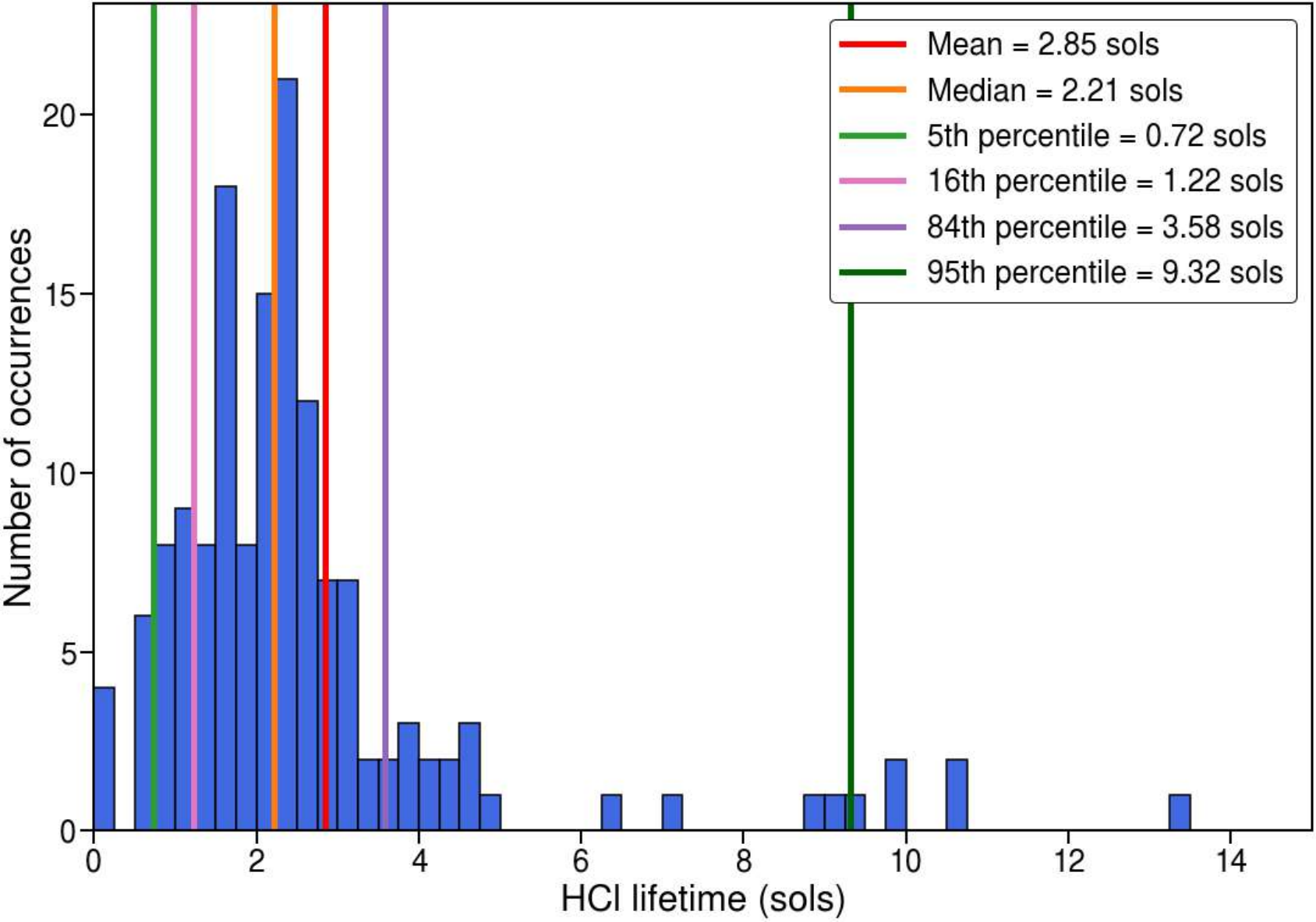}
            }
      \caption{Chemical lifetime of HCl columns at each ACS observation point where HCl was detected during MY 34 (left) and MY 35 (right).}
         \label{fig_HCl_lifetime}
\end{figure*}

Figure \ref{fig_HCl_lifetime} shows histograms of HCl chemical lifetimes computed at each point where HCl was detected by ACS in MY 34 and MY 35. We find that the median lifetime of HCl is only a few sols, which is consistent with results from the 1D model of \citet{taysum_observed_2024}. 
This is mainly due to heterogeneous uptake, which constitutes the main loss process for HCl. Our model lifetime is significantly lower than the upper limit of 73 sols reported by \citet{krasnopolsky_photochemistry_2022} which is required to explain the rapid disappearance of HCl at the end of the dust season. 
We find that the longer lifetimes found in MY 35 are due to a significant decrease in the loss rates related to reactions h2 and h3 compared to MY 34. This decrease is linked to differences in the water vapour and dust content in the atmosphere between MY 34 and 35. In the former, more water vapour was found in the latter part of the year, during the global dust storm. The latter part of MY 35 therefore had less water ice clouds and dust, leading to significantly lower h2 and h3 rates. This also reduces the production of atomic chlorine through reaction h1, but we find that the global loss rate of HCl decreases more than the global production rate between MY 34 and MY 35. Consequently, the lifetime of HCl increases, which may facilitate reaching ppbv-level VMRs even in the absence of a global dust storm in MY 35.

\subsection{Aphelion detections of HCl} \label{subs_aphel_detect}

The ACS data collected during the first half of MY 35 highlighted two aphelion HCl detections in the northern hemisphere \citep{olsen_seasonal_2021, olsen_relationships_2024-1}. These two detections occur in the Arcadia region, at (L$_{\text{S}}=109^{\circ}$, 66$^{\circ}$N, 117$^{\circ}$W) and (L$_{\text{S}}=115^{\circ}$, 46$^{\circ}$N, 97$^{\circ}$W) \citep{olsen_seasonal_2021} and were the only detections occurring during the aphelion season in MY 35. Another single detection at  L$_{\text{S}}=105^{\circ}$ was made during MY 36, which motivated a more detailed investigation during MY 37 \citep{olsen_relationships_2024-1}. The atmospheric conditions at the locations where HCl was detected are not unique \citep{olsen_relationships_2024-1} which raises questions about whether these isolated detections can be reproduced using chemistry involving dust and water vapour.

\begin{figure*}[!h]
   \resizebox{\hsize}{!}
            {
            \includegraphics{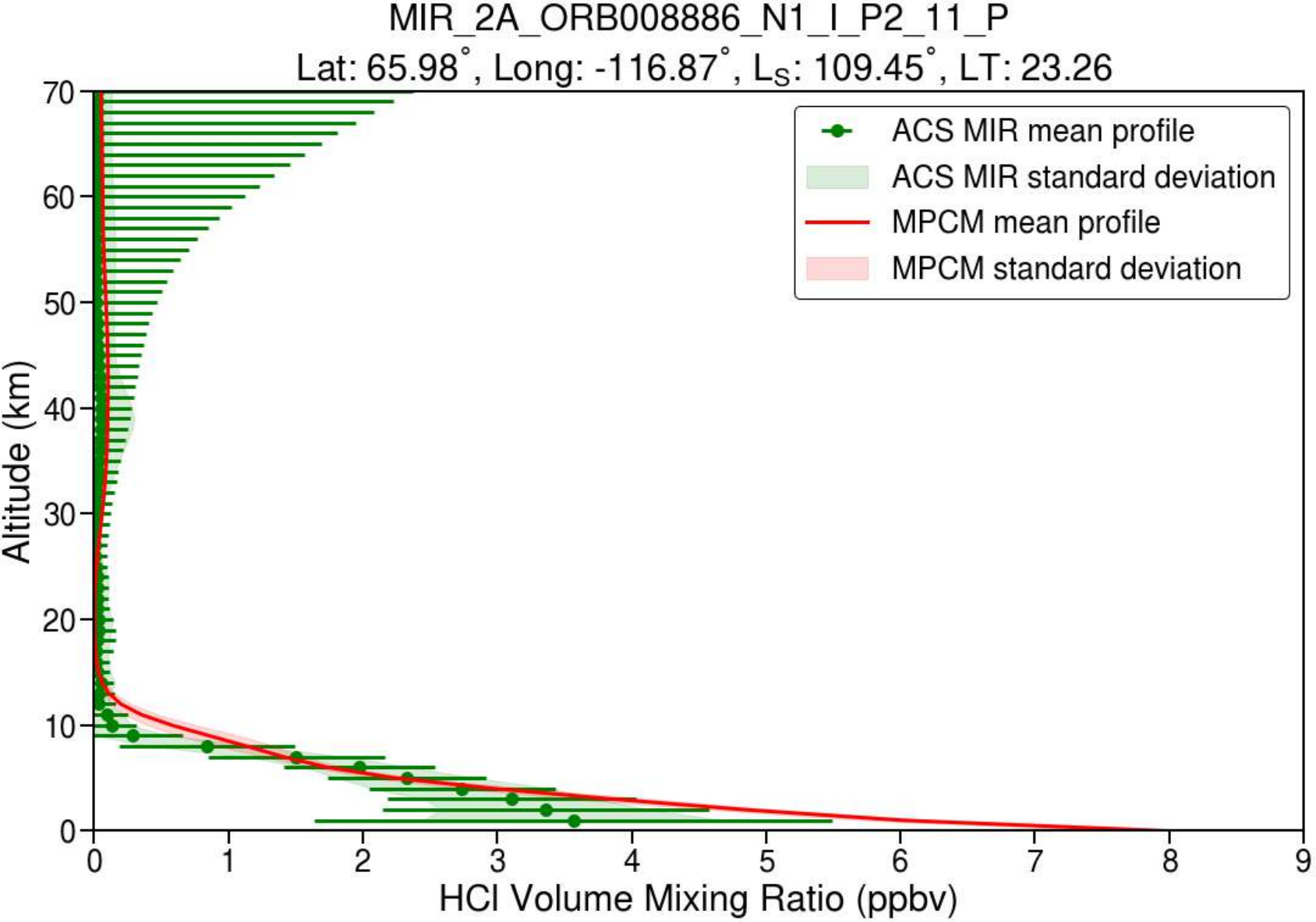}
            \includegraphics{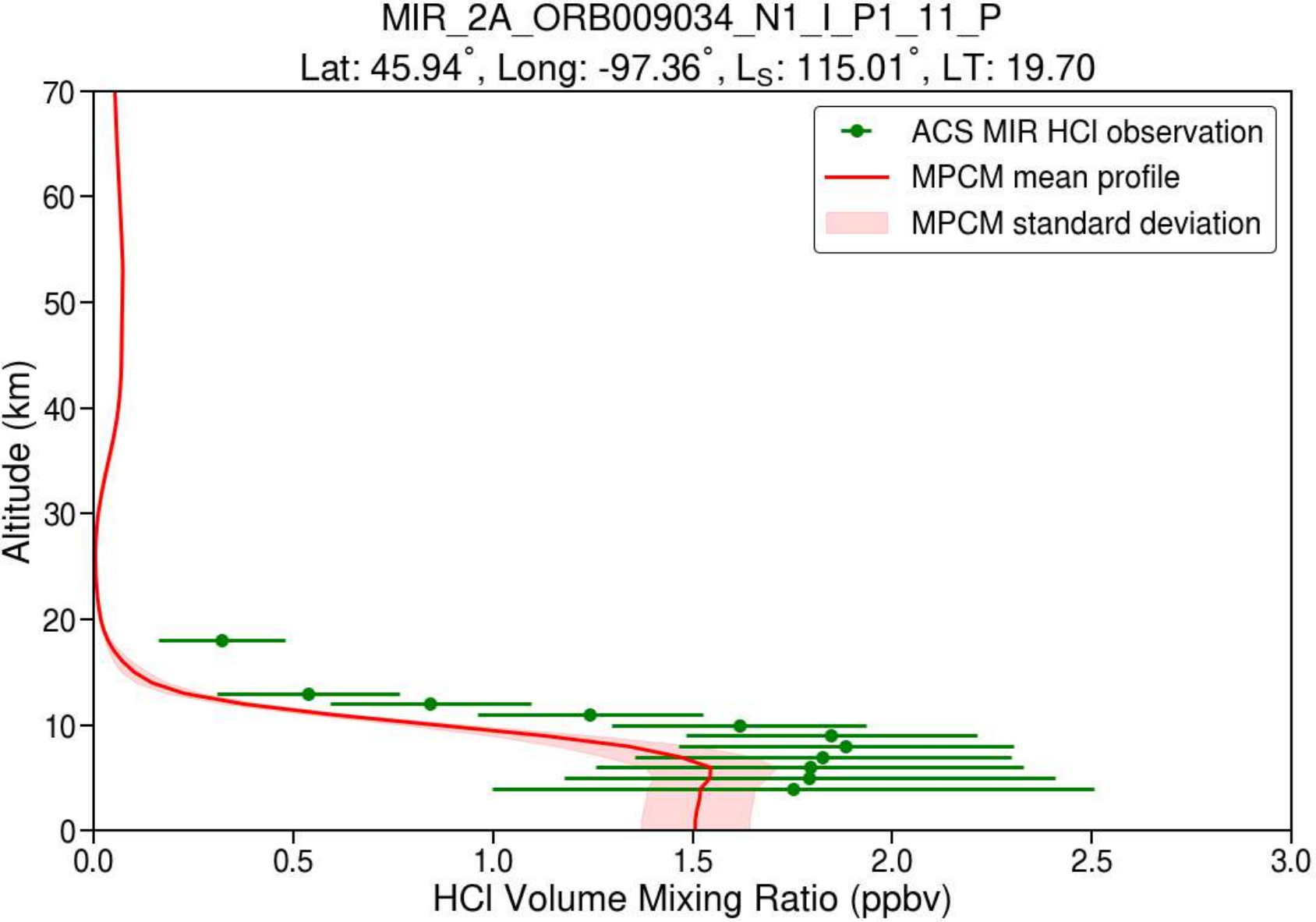}
            }
      \caption{Mean vertical profiles of HCl for the two aphelion detections of MY 35. Error bars indicate the uncertainty of the ACS retrieved value, while the standard deviation reflects the variance among the retrieved profiles used to compute the mean. The model profile represents the mean HCl profile averaged over the 5$^{\circ}$ L$_{\text{S}}$ period containing the time of the ACS detection. The standard deviation indicates the 1-$\sigma$ variation of this profile over this period.}
         \label{fig_HCl_aphel_detect}
\end{figure*}

Figure \ref{fig_HCl_aphel_detect} shows the model and ACS MIR vertical profiles of HCl at these two locations. We find that the model reproduces the observed HCl vertical profiles in both cases. At both locations, we find that dust and water vapour are abundant and there is an absence of water ice. However, we acknowledge that for the L$_{\text{S}}=109^{\circ}$ observation, the model exhibits a large negative model bias for water ice opacity compared to nearby MCS data. Using the MCS water ice profile, model HCl levels would be lower. From our analysis in Sect. \ref{subsect_vert_prof_comp}, we find that all the favourable conditions exist for significant production of HCl and a low loss rate. In such cases, production dominates the loss of HCl below 30~km, with production increasing towards the surface along with the abundances of water vapour and dust. As a result, the HCl profile strongly mimics the shape of the water vapour profile.  

If we study ACS non-detections of HCl around this time and region, we find that our model produces large amounts of HCl close to the surface for most ACS MIR observations. In addition, examining the spatial distribution of HCl at the time it was detected shows that it is not confined to the area in which it was observed by ACS (see Fig. \ref{App_fig_aphel_det_VMRs_maps}). 
This indicates that, with the chlorine chemical scheme used in our model, HCl is not confined to the Arcadia region. However, for almost all these points, the model water vapour exhibits a large positive bias compared to ACS MIR and ACS NIR observations. 
Figure \ref{Fig_climato_hcl_vap} shows that this region and time of year correspond to the peak of sublimation in the water cycle, which produces large water vapour VMRs. As noted previously for high southern latitudes during the dust season, under these conditions our model produces large amounts of HCl close to the surface. HCl peaks at 10-20 ppbv below 10--20~km, with a vertical extent determined by the amount of water ice at these altitudes. Because these peaks are close to the surface, they are typically not observed by ACS. This, coupled with positive model bias for water vapour, makes it difficult to compare the model and ACS data and to understand why HCl is unexpectedly detected at these two locations. From our initial model results, the production of large HCl VMRs does not seem to be as isolated as suggested by ACS observations.

\section{Concluding remarks}
\label{sect_conclu}

In this work, we implemented the chlorine photochemical scheme developed by \citet{taysum_observed_2024} in the MPCM. 
The uptake coefficients of the most important heterogeneous reactions were slightly modified following sensitivity studies, to allow results consistent with TGO observations. 
We ran simulations for both MY 34 and 35, using initial conditions from the MCD v6.1. 
As in the 1D model \citep{taysum_observed_2024}, this scheme correctly reproduces most of the ACS MIR HCl observations. 
This species is mainly produced by gas-phase reactions between H$_2$, HO$_2$ and atomic chlorine, which is produced by a heterogeneous reaction involving airborne dust, water vapour, and UV radiation. 
HCl is then mainly destroyed by uptake on water ice when clouds are present, uptake on dust in the first 50 km above the surface, and by gas-phase reactions with H and OH, as well as photolysis at higher altitudes. 
Heterogeneous reactions dominate the loss of HCl and reduce its lifetime to only a couple of sols at the locations where HCl was detected by ACS MIR in MY 34 and MY 35. 
This short lifetime is consistent with transient observations of this species and could explain its rapid disappearance at the end of the dust season. 

Using this model, we are able to reproduce ACS MIR detections of HCl during MY 34 and MY 35. The HCl VMRs could reach ppbv levels for most of the year, from L$_{\text{S}}\approx 50^{\circ}$ and L$_{\text{S}} \approx$ 340$^{\circ}$ in both years. Peak values occur during summer in each hemisphere, with variations correlated with the water sublimation cycle.
This shows that using a heterogeneous reaction to produce atomic chlorine allows production of HCl at ppbv level both in a year of 'classic' dust activity (MY 35) and in a year with a global dust storm (MY 34).
We find that model biases for HCl vertical profiles are closely related to water vapour biases, reflected by a large correlation coefficient of 0.79. This correlation arises from the role of water vapour in the production of atomic chlorine and H$_2$ and HO$_2$.
At the same time, we find that HCl correlates with temperature (correlation coefficient = 0.48), and dust (correlation coefficient = 0.41). 
The temperature correlation results from heterogeneous reactions, as the reaction producing atomic chlorine is proportional to temperature, whereas those responsible for HCl loss are inversely proportional.
The correlation with dust is positive, despite dust being involved in both production and loss of HCl.
We link these positive correlations to dust activity, as dust helps to heat the atmosphere, allowing for a higher content of water vapour. 
We also find that HCl is anticorrelated with water ice, as the uptake of HCl on water ice is a very efficient sink. 
These results are consistent with observed correlations from \citet{olsen_relationships_2024-1}. 

Our model is also consistent with 70\% of HCl non-detections by ACS MIR. For the remaining 30\%, we find that the discrepancy between model and observations results from biases associated with water vapour, dust, or water ice abundances. For most non-detections, low HCl VMRs at these locations result from high amounts of water ice in the lower atmosphere, coupled with low amount of dust and water vapour. 
This combination inhibits the production of atomic chlorine and increases HCl loss, leading to low VMRs. 
This case also occurs at locations where HCl is detected, typically at high northern latitudes or at southern latitudes outside of the dust season, preventing reproduction of observations. 
Further studies on model parameters and the chemical network are needed to improve our results in these areas. 
In cases where our model produces HCl amounts that exceed the retrieved values, we find this is linked to positive biases in water vapour or dust, or to negative biases in water ice, or a combination of these biases. 

Our model shows elevated HCl VMRs in both northern and southern summer months, creating a quasi-symmetrical cycle. This contrasts with the asymmetrical cycle observed by the TGO instruments, although our model largely reproduces observed detections and non-detections of HCl. This discrepancy between observed and model hemispheric cycles of HCl results from the chemical reactions that govern the production of atomic chlorine and HCl, which strongly link HCl to water vapour. Further research into modelling the Martian water cycle is therefore crucial to determine if this difference suggests that key processes affecting HCl abundances, e.g. heterogeneous processes, are not fully described in this work.

We also show that with the chemistry used in this work, we could reproduce aphelion detections made during MY 35. In terms of detections made during the perihelion season, HCl is produced in areas where water vapour and dust are abundant and water ice is not. 
However, we were unable to fully investigate the reason for the few  detections in this area and time of year, as our model overestimates water vapour abundance near the water cycle sublimation peak, which leads to apparent overestimation of HCl at non-detection points near the two aphelion detection locations.

Our study relies on uncertain uptake coefficients to describe heterogeneous chemical reactions in the Martian atmosphere. Although we use the best experimental data available, which we have modified via sensitivity studies to reproduce TGO observations, there is an urgent need to reevaluate the underlying empirical data. Only through new laboratory data can we improve the confidence in our model to interpret HCl data collected by instruments aboard the TGO.

\section*{Data availability}
Figures comparing the vertical profiles of key atmospheric quantities (HCl, water vapour, temperature, ice, and dust opacities) from our model and the corresponding profiles measured by ACS MIR (HCl and water vapour), ACS NIR (water vapour and temperature), and MCS (ice and dust opacities) for all HCl detections and non-detections of ACS MIR during MY 34 and MY 35 are available on \href{https://zenodo.org/records/14717873}{Zenodo}.

\section*{Acknowledgements}
B. Benne and P. I. Palmer acknowledge funding from the UK Space Agency Aurora Science programme (ST/Y003284/1). B. Benne acknowledge E. Millour, M. Cohen and A. Mechineau for useful advice at various stages of this work. 
\newpage

\begin{appendix}

\section{Chlorine photochemical scheme}
\label{Appendix_cl_scheme}

Table \ref{tab_app_added_species} lists all the species added in the MPCM to model the chlorine photochemistry considered in this work. Table \ref{Tab_app_chlo_scheme} lists all the reactions of this specific part of the photochemical scheme, along with reaction rates and references for these rates, and Table \ref{tab_App_uptake_coeffs} the uptake coefficients used in heterogeneous reactions involving chlorinated species.

\begin{table}[h!]
    \begin{center}
        \begin{tabular}{lcl}
            \hline \hline
            Formula     &  Name                 &  \begin{tabular}[c]{@{}l@{}} Molar mass\\(g/mol)\end{tabular}  \\ \hline
            Cl          & Atomic chlorine      & 35.453             \\
            ClO         & Chlorine monoxide    & 51.452             \\
            Cl$_2$      & Molecular chlorine   & 70.906             \\
            OClO        & Chlorine dioxide     & 67.452             \\
            ClOO        & Chloroperoxyl        & 67.452             \\
            Cl$_2$O$_2$ & Chlorine peroxide    & 102.905            \\
            HCl         & Hydrogen chloride    & 36.451             \\
            HOCl        & Hypochlorous acid    & 52.460             \\
            ClCO        & Chloroformyl radical & 63.463             \\
            ClO$_3$     & Chlorine trioxide    & 83.451             \\
            ClO$_4$     & Perchlorate          & 99.451             \\
            HClO$_4$    & Perchloric acid      & 100.459            \\ \hline
        \end{tabular}
        \caption{List of chlorine species added in the MPCM in this work. \\ The molar mass of these species are taken from the NIST database (\url{https://webbook.nist.gov/}).}
        \label{tab_app_added_species}
    \end{center}
\end{table}

\begin{longtable}{cllc}
   \caption{Chlorine photochemical scheme included in the MPCM}
   \label{Tab_app_chlo_scheme}\\
    \hline 
    Label  & Reaction    & Reaction rate  & Reference \\ \hline
    \endfirsthead

    \multicolumn{4}{c}%
    {{\tablename\ \thetable{} - Continued}} \\
    \hline
    \endhead

    \hline
    \endfoot

    \\
    \multicolumn{4}{l}%
    {{Values between brackets are number densities in cm$^{-3}$. [M] is the number density of the third body for three-body reactions. }}\\
    \multicolumn{4}{l}%
    {{$S_{\text{dust}}$ and $S_{\text{ice}}$ are the surface area of dust and water ice, respectively, in cm$^{-1}$. }}\\
    \multicolumn{4}{l}%
    {{$v_{\text{th}}$(species) is the thermal speed of the considered species in cm/s, computed as $v_{\text{th}}(z)=\sqrt{\frac{8RT(z)}{\pi M}}$, with $R$ the ideal gas}} \\
    \multicolumn{4}{l}%
    {{constant, $T(z)$ the temperature at altitude level $z$ and $M$ the molar mass of the considered species in g/mol.}} \\ \\
    \multicolumn{4}{l}%
    {{[1] - \cite{heays_photodissociation_2017}, 
    [2] - \cite{burkholder_chemical_2019}, 
    [3] - \citet{taysum_observed_2024}, 
    [4] - \citet{keller_equatorial_2006},
     }} \\
    \multicolumn{4}{l}%
    {{[5] - \citet{boynton_evidence_2009},
    [6] - \citet{huynh_heterogeneous_2020},
    [7] - \citet{huynh_heterogeneous_2021},
    [8] - \citet{hynes_interaction_2001},
     }} \\
    \multicolumn{4}{l}%
    {{[9] - \citet{kippenberger_trapping_2019},
    [10] - \citet{sander_community_2019},
    [11] - \citet{atkinson_evaluated_2006-1},
    [12] - \citet{zhu_ab_2010}, 
     }} \\
    \multicolumn{4}{l}%
    {{[13] - \citet{berho_uv_1999},
    [14] - \citet{neufeld_chemistry_2009}, from the KIDA database \citep{wakelam_kinetic_2012}, }} \\
    \multicolumn{4}{l}%
    {{[15] - \citet{catling_atmospheric_2010},
    [16] - \citet{simonaitis_perchloric_1975}, 
    [17] - \citet{xu_ab_2003}, }} \\
    \multicolumn{4}{l}%
    {{[18] - \citet{s_zhu_ab_2001}, 
    [19] - \citet{xu_computational_2010} }} \\
    \endlastfoot

        ph1   & HCl + $h\nu$ $\longrightarrow$ H + Cl               &                                                                                                                                                                                                                  & [1]                                     \\
        ph2   & Cl$_2$ + $h\nu$ $\longrightarrow$ Cl + Cl           &                                                                                                                                                                                                                  & [2]                                            \\
        ph3   & ClO + $h\nu$ $\longrightarrow$ Cl + O($^1$D)        &                                                                                                                                                                                                                  & [2]                                           \\
        ph4   & ClO + $h\nu$ $\longrightarrow$ Cl + O               &                                                                                                                                                                                                                  & [2]                                            \\
        ph5   & HOCl + $h\nu$ $\longrightarrow$ Cl + OH             &                                                                                                                                                                                                                  & [2]                                            \\
        ph6   & ClOO + $h\nu$ $\longrightarrow$ Cl + O$_2$          &                                                                                                                                                                                                                  & [2]                                             \\
        ph7   & OClO + $h\nu$ $\longrightarrow$ ClO + O             &                                                                                                                                                                                                                  & [2]                                            \\ \hline
        h1    & 6 H$_2$O + dust $\longrightarrow$ Cl + 6 H$_2$O           & $\frac{1}{4}\times v_{\text{th}}(\text{H$_2$O})\times \gamma_1 \times 0.0049 \times S_{\text{dust}}$                                                                                                                                    & [3] based on [4]                                                             \\
        h2    & 2 HCl + dust $\longrightarrow$ 2 H               & $\frac{1}{4}\times v_{\text{th}}(\text{HCl})\times \gamma_2 \times 0.04 \times S_{\text{dust}}$                                                                                                                                           & [3], based on [6] and [7]                                            \\
        h3    & HCl + water ice $\longrightarrow$ H + Cl or H                & $\frac{1}{4}\times v_{\text{th}}(\text{HCl})\times \gamma_3 \times S_{\text{ice}}$                                                                                                                                                        & [3] based on [8] and [9] \\
        h4    & Cl$_2$ + dust $\longrightarrow$ $\varnothing$            & $\frac{1}{4}\times v_{\text{th}}(\text{Cl$_2$})\times \gamma_4 \times 10^{-2} \times S_{\text{dust}}$                                                                                                                                     & [3] based on [2]                                           \\
        h5    & ClO + water ice $\longrightarrow$ O                & $\frac{1}{4}\times v_{\text{th}}(\text{ClO})\times \gamma_5 \times S_{\text{ice}}$                                                                                                                                                         & [3], based on [2]                                          \\ \hline
        cl1   & Cl + O$_3$ $\longrightarrow$ ClO + O$_2$            & 2.8$\times 10^{-11} \, \exp{\left ( \frac{-250}{T(z)} \right )}$                                                                                                                                                                      & [10]                                                                   \\
        cl2   & ClO + O $\longrightarrow$ Cl + O$_2$                & 2.5$\times 10^{-11} \, \exp{\left ( \frac{110}{T(z)} \right )}$                                                                                                                                                                       & [10]                                                                   \\
        cl3   & ClO + ClO $\longrightarrow$ Cl$_2$ + O$_2$          & 1.0$\times 10^{-12} \, \exp{\left ( \frac{-1590}{T(z)} \right )}$                                                                                                                                                                     & [10]                                                                   \\
        cl4   & ClO + ClO $\longrightarrow$ 2 Cl + O$_2$            & 3.0$\times 10^{-11} \, \exp{\left ( \frac{-2450}{T(z)} \right )}$                                                                                                                                                                     & [10]                                                                   \\
        cl5   & ClO + ClO $\longrightarrow$ Cl + OClO               & 3.5$\times 10^{-13} \, \exp{\left ( \frac{-1370}{T(z)} \right )}$                                                                                                                                                                     & [10]                                                                   \\
        cl6   & ClO + ClO + M $\longrightarrow$ Cl$_2$O$_2$ + M     & \begin{tabular}[c]{@{}l@{}}$k_0 = 1.9\times 10^{-32} \, \left ( \frac{298}{T(z)} \right )^{3.6}$\\ $k_{\infty} = 3.7\times 10^{-12} \, \left ( \frac{298}{T(z)} \right )^{1.6}$\end{tabular}                                          & [10]                                                                   \\
        cl7   & Cl$_2$O$_2$ $\longrightarrow$ ClO + ClO             & $\frac{\text{cl6}}{2.16\times 10^{-27} \, \exp{\left ( \frac{8537}{T(z)} \right )}}$                                                                                                                                                  & [10]                                                                   \\
        cl8   & Cl + H$_2$ $\longrightarrow$ HCl + H                & 3.9$\times 10^{-11} \, \exp{\left ( \frac{-2310}{T(z)} \right )}$                                                                                                                                                                     & [10]                                                                   \\
        cl9   & Cl + HO$_2$ $\longrightarrow$ HCl + O$_2$           & 4.4$\times 10^{-11}$ - 7.5$\times 10^{-11} \, \exp{\left ( \frac{-620}{T(z)} \right )}$                                                                                                                                               & [10]                                                                   \\
        cl10  & Cl + HO$_2$ $\longrightarrow$ ClO + OH              & 7.5$\times 10^{-11} \, \exp{\left ( \frac{-620}{T(z)} \right )}$                                                                                                                                                                      & [10]                                                                   \\
        cl11  & Cl + H$_2$O$_2$ $\longrightarrow$ HCl + HO$_2$      & 1.1$\times 10^{-11} \, \exp{\left ( \frac{-980}{T(z)} \right )}$                                                                                                                                                                      & [10]                                                                   \\
        cl12  & ClO + OH $\longrightarrow$ Cl + HO$_2$              & 7.3$\times 10^{-12} \, \exp{\left ( \frac{300}{T(z)} \right )} \times 0.94$                                                                                                                                                           & [10]                                                                   \\
        cl13  & ClO + OH $\longrightarrow$ HCl + O$_2$              & 7.3$\times 10^{-12} \, \exp{\left ( \frac{300}{T(z)} \right )} \times 0.06$                                                                                                                                                           & [10]                                                                   \\
        cl14  & ClO + HO$_2$ $\longrightarrow$ HOCl + O$_2$         & 2.2$\times 10^{-12} \, \exp{\left ( \frac{340}{T(z)} \right )}$                                                                                                                                                                       & [10]                                                                   \\
        cl15  & HCl + OH $\longrightarrow$ Cl + H$_2$O              & 1.7$\times 10^{-12} \, \exp{\left ( \frac{-230}{T(z)} \right )}$                                                                                                                                                                      & [10]                                                                   \\
        cl16  & HOCl + OH $\longrightarrow$ ClO + H$_2$O            & 3.0$\times 10^{-12} \, \exp{\left ( \frac{-500}{T(z)} \right )}$                                                                                                                                                                      & [10]                                                                   \\
        cl17  & CO + Cl + M $\longrightarrow$ ClCO + M              & 2.5\, {[}M{]} \, 1.3$\times 10^{-33} \, \left ( \frac{300}{T(z)} \right )^{3.8}$                                                                                                                                                      & [11]                                                           \\
        cl18  & ClOO + Cl $\longrightarrow$ ClO + ClO               & 1.2$\times 10^{-11}$                                                                                                                                                                                                                  & [2]                                                                \\
        cl19  & ClOO + Cl $\longrightarrow$ O$_2$ + Cl$_2$          & 2.3$\times 10^{-10}$                                                                                                                                                                                                                  & [2]                                                                \\
        cl20  & ClOO + M $\longrightarrow$ Cl + O$_2$ + M               & 2.8$\times 10^{-10} \, \exp{\left ( \frac{-1820}{T(z)} \right )}$ \, {[}M{]}                                                                                                                                                          & [11]                                                           \\
        cl21  & HClO$_4$ + OH $\longrightarrow$ ClO$_4$ + H$_2$O    & $\frac{2.8\times 10^{-10}}{T(z)^{2.99}} \, \exp{\left ( \frac{1664}{T(z)} \right )}$                                                                                                                                                  & [12]                                                                \\
        cl22  & Cl + O$_2$ + M $\longrightarrow$ ClOO + M           & 2.5\, {[}M{]} \, 1.4$\times 10^{-33} \, \left ( \frac{300}{T(z)} \right )^{3.9}$                                                                                                                                                      & [11]                                                           \\
        cl23  & Cl + Cl$_2$O$_2$ $\longrightarrow$ Cl$_2$ + ClOO    & 1.0$\times 10^{-10}$                                                                                                                                                                                                                  & [11]                                                           \\
        cl24  & ClCO $\longrightarrow$ CO + Cl                      & {[}M{]} \, 4.1$\times 10^{-10} \, \exp{\left ( \frac{-2960}{T(z)} \right )}$                                                                                                                                                          & [11]                                                           \\
        cl25  & Cl$_2$ + O($^1$D) $\longrightarrow$ Cl$_2$ + O      & {[}Cl$_2${]} \, $\frac{1}{4} \times 2.7\times 10^{-10}$                                                                                                                                                                               & [2]                                                                \\
        cl26  & Cl$_2$ + O($^1$D) $\longrightarrow$ Cl + ClO        & $\frac{3}{4} \times 2.7\times 10^{-10}$                                                                                                                                                                                               & [2]                                                                \\
        cl27  & Cl$_2$ + OH $\longrightarrow$ HOCl + Cl             & 2.6$\times 10^{-12} \, \exp{\left ( \frac{-1100}{T(z)} \right )}$                                                                                                                                                                     & [2]                                                                \\
        cl28  & Cl$_2$ + H $\longrightarrow$ HCl + Cl               & 8.0$\times 10^{-11} \, \exp{\left ( \frac{-416}{T(z)} \right )}$                                                                                                                                                                      & [13]                                                              \\
        cl29  & HCl + O($^1$D) $\longrightarrow$ HCl + O            & 0.12 $\times 1.5\times 10^{-10}$ \, {[}HCl{]}                                                                                                                                                                                         & [2]                                                                \\
        cl30  & HCl + O($^1$D) $\longrightarrow$ ClO + H            & 0.22 $\times 1.5\times 10^{-10}$                                                                                                                                                                                                      & [2]                                                                \\
        cl31  & HCl + O($^1$D) $\longrightarrow$ Cl + OH            & 0.66 $\times 1.5\times 10^{-10}$                                                                                                                                                                                                      & [2]                                                                \\
        cl32  & HCl + O $\longrightarrow$ OH + Cl                   & 1.0$\times 10^{-11} \, \exp{\left ( \frac{-3300}{T(z)} \right )}$                                                                                                                                                                     & [2]                                                                \\
        cl33  & HCl + H $\longrightarrow$ H$_2$ + Cl                & 1.49$\times 10^{-11} \, \exp{\left ( \frac{-1763}{T(z)} \right )}$                                                                                                                                                                    & [14]                                                                    \\
        cl34  & HOCl + O $\longrightarrow$ ClO + OH                 & 1.7$\times 10^{-13}$                                                                                                                                                                                                                  & [11]                                                           \\
        cl35  & Cl + O$_3$ + M $\longrightarrow$ ClO$_3$ + M        & 2.5 \, {[}M{]} \, $\times 1.0\times 10^{-31}$                                                                                                                                                                                         & [15], based on [16]                                                            \\
        cl36  & ClO + ClO$_3$ $\longrightarrow$ ClOO + OClO         & 1.85$\times 10^{-18} \, \exp{\left ( \frac{-2417}{T(z)} \right )} \, T(z)^{2.28}$                                                                                                                                                     & [15], based on [17]                                                            \\
        cl37  & ClO + ClO$_3$ $\longrightarrow$ OClO + OClO         & 1.42$\times 10^{-18} \, \exp{\left ( \frac{-2870}{T(z)} \right )} \, T(z)^{2.11}$                                                                                                                                                     & [15], based on [17]                                                            \\
        cl38  & ClO + ClO$_3$ + M $\longrightarrow$ Cl$_2$O$_4$ + M & \begin{tabular}[c]{@{}l@{}}$k_0 = \frac{1.43\times 10^{-1}}{T(z)^{10.19}} \, \exp{\left ( \frac{-1597}{T(z)} \right )}$\\ $k_{\infty} = 1.43\times 10^{-10} \, \exp{\left ( \frac{-82}{T(z)} \right )} \, T(z)^{0.094}$\end{tabular} & [17]                                                            \\
        cl39  & ClO$_3$ + OH $\longrightarrow$ HClO$_4$             & 6.67$\times 10^{-13}$                                                                                                                                                                                                                 & [15], based on [16]                                                            \\
        cl40  & ClO$_3$ + OH + M $\longrightarrow$ HClO$_4$ + M     & \begin{tabular}[c]{@{}l@{}}$k_0 = \frac{1.94\times 10^{36}}{T(z)^{15.3}} \, \exp{\left ( \frac{-5542}{T(z)} \right )}$\\ $k_{\infty} = 3.2\times 10^{-10} \, \exp{\left ( \frac{-25}{T(z)} \right )} \, T(z)^{0.07}$\end{tabular}    & [15], based on [18]                                                            \\
        cl41  & ClO$_3$ + OH $\longrightarrow$ OClO + HO$_2$        & 2.1$\times 10^{-10} \, \exp{\left ( \frac{-18}{T(z)} \right )} \, T(z)^{0.09}$                                                                                                                                                        & [15], based on [18]                                                            \\
        cl42  & OClO + O + M $\longrightarrow$ ClO$_3$ + M          & \begin{tabular}[c]{@{}l@{}}$k_0 = 3.0\times 10^{-31} \, \left ( \frac{298}{T(z)} \right )^{3.1}$\\ $k_{\infty} = 8.3\times 10^{-12} \, \left ( \frac{298}{T(z)} \right )$\end{tabular}                                                & [2]                                                            \\
        cl43  & OClO + O$_3$ $\longrightarrow$ ClO$_3$ + O$_2$      & 2.1$\times 10^{-12} \, \exp{\left ( \frac{-4700}{T(z)} \right )}$                                                                                                                                                                     & [2]                                                            \\
        cl44  & ClO$_4$ + Cl $\longrightarrow$ ClO$_3$ + ClO        & 8.05$\times 10^{-11} \, T(z)^{0.158} \exp{\left ( \frac{-49}{T(z)} \right )}$                                                                                                                             & [17]                                                         \\
        cl45  & ClO$_4$ + HOCl $\longrightarrow$ ClO + HClO$_4$     & 1.35$\times 10^{-18} \, T(z)^{1.73} \exp{\left ( \frac{1017}{T(z)} \right )}$                                                                                                                            & [19]                                            \\
        cl46  & HOCl+ Cl $\longrightarrow$ Cl$_2$ + OH              & 0.935$\times 3.4\times 10^{-12} \, \exp{\left ( \frac{-130}{T(z)} \right )}$                                                                                                                                                                    & [2] and [15]                                                 \\
        cl47  & HOCl+ Cl $\longrightarrow$ HCl + ClO                & 0.065$\times 3.4\times 10^{-12} \, \exp{\left ( \frac{-130}{T(z)} \right )}$                                                                                                                                                                     & [2] and [15]                                                 \\
        cl48  & OClO+ O $\longrightarrow$ ClO + O$_2$               & 2.4$\times 10^{-12} \, \exp{\left ( \frac{-960}{T(z)} \right )}$                                                                                                                                                                      & [11]                                                           \\
        cl49  & OClO+ OH $\longrightarrow$ HOCl + O$_2$             & 1.4$\times 10^{-12} \, \exp{\left ( \frac{600}{T(z)} \right )}$                                                                                                                                                                       & [2]                                                                \\
        cl50  & OClO+ Cl $\longrightarrow$ ClO + ClO                & 3.4$\times 10^{-11} \, \exp{\left ( \frac{160}{T(z)} \right )}$                                                                                                                                                                       & [2]                                                                \\ \hline
\end{longtable}

\vspace{0.5cm}

\begin{table}[h!]
\centering
    \begin{tabular}{cllc}
    \hline \hline
    Label & Reaction                                   & Uptake coefficient                                                                                                                                                                                                 & Reference                                                                                                             \\ \hline
    $\gamma_1$    & 6 H$_2$O + dust $\longrightarrow$ Cl + 6 H$_2$O  & \begin{tabular}[c]{@{}l@{}}$ 0.1 \times 6.3\times 10^{-2} \times j_{\text{ClO}}(z)$ \end{tabular} & \begin{tabular}[c]{@{}c@{}} [3] based on [20] and [21] \end{tabular}             \\
    $\gamma_2$    & 2 HCl + dust $\longrightarrow$ 2 H      & MPM(0.2, 2.03$\times 10^{3}$, 12.30$\times 10^{3}$)                                                                                                                                                                 & [3] based on [6] and [7]                                                                                      \\
    $\gamma_3$    & HCl + water ice $\longrightarrow$ H + Cl or H & MPM(0.09, 8.33$\times 10^{8}$, 30.50$\times 10^{3}) \times (1-\theta_{\text{Lang}})$                                                                                                                                 & \begin{tabular}[c]{@{}c@{}} [3] based on [8] and [9]\end{tabular} \\
    $\gamma_4$    & Cl$_2$ + dust $\longrightarrow$ $\varnothing$   & 1$\times 10^{-3}$                                                                                                                                                                                                  & {[}2{]} (upper limit)                                                                                                 \\
    $\gamma_5$    & ClO + water ice $\longrightarrow$ O & 1$\times 10^{-4}$                                                                                                                                                                                                  & {[}2{]} (upper limit)                                                                                                 \\ \hline
    \end{tabular}
    \caption{Uptake coefficients for heterogeneous reactions involving chlorinated species. \\ The MPM function is MPM(A, $\frac{A_{\text{des}}}{A_{\text{R}}}$, $\Delta E$) = A / [1 + $\frac{A_{\text{des}}}{A_{\text{R}}} \times \exp{\left ( -\Delta E / RT(z) \right )}$ ], with $A_{\text{des}}$ the pre-exponential factor of the adsorbed molecule (s$^{-1}$), $A_{\text{R}}$ the pre-exponential factor of the adsorbed molecule ionisation (s$^{-1}$), and $\Delta E$ the difference in the activation energies from the Arrhenius equations controlling the adsorption of molecules onto the solid and the ionisation of the adsorbed molecule onto the solid surface (J.mol$^{-1}$). $R$ is the ideal gas constant and $T(z)$ the atmospheric temperature at altitude $z$ \citep{berland_surface_1997,taysum_observed_2024}.  \\$\theta_{\text{Lang}}$ is the water ice surface coverage as defined by \citet{kippenberger_trapping_2019}: $\theta_{\text{Lang}} = \frac{K_{\text{Lang}}[\text{HCl}]}{1+K_{\text{Lang}}[\text{HCl}]}$, with $K_{\text{Lang}}=9.6\times 10^{-11}$ cm$^3$.mol$^{-1}$ and [HCl] the number density of HCl (mol.cm$^{-3}$). \\ \, [20] - \citet{zhang_reaction_2022}, [21] - \citet{seisel_water_2005}. For other references, see footnote of Table \ref{Tab_app_chlo_scheme}}
    \label{tab_App_uptake_coeffs}
\end{table}
\newpage

\section{Supplementary figures}
\label{Appendix_complementary_figures}

\subsection{HCl maximum VMRs versus altitude, latitude, longitude and local time}

Here, we present scatter plots showing the repartition of the HCl maximum VMRs at each output time step. For this, we use outputs averaged over 5$^{\circ}$ L$_{\text{S}}$ intervals, with 12 points per sol. The total number of points considered is therefore 72 $\times$ 12 = 864 points per MY. 
Results are first shown for MY 34, then for MY 35.

\begin{figure*}[h!]
\centering
   \resizebox{\hsize}{!}
            {
            \includegraphics{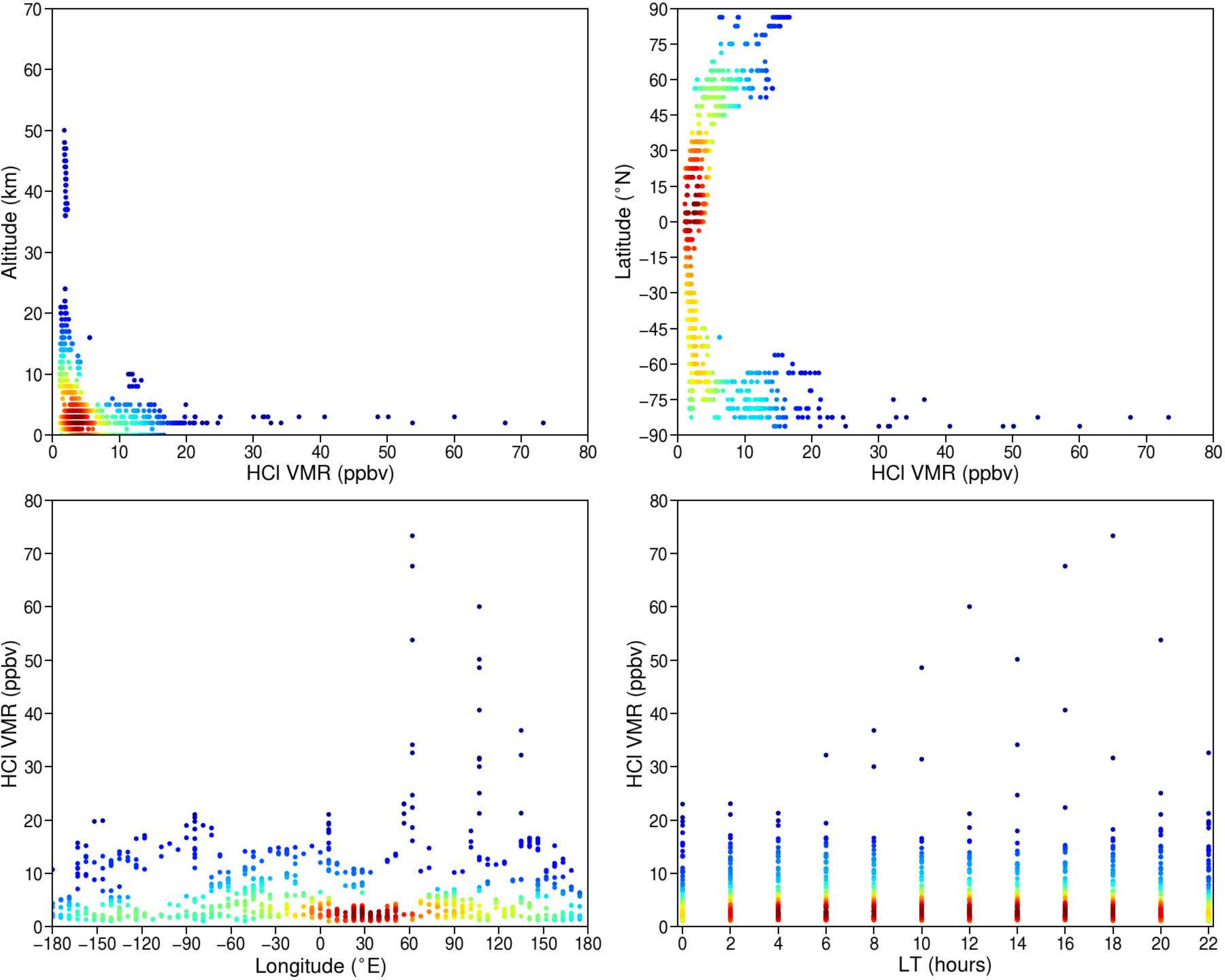}
            }
      \caption{Repartition of HCl maximum VMRs in altitude (top left), latitude (top right), longitude (bottom left) and local time (bottom right) for MY 34. The colour scale indicates the density of points, with increasing density going from blue to red.}
         \label{app_Fig_scatt_altlatlonlt_MY34}
\end{figure*}

\begin{figure*}[h!]
\centering
   \resizebox{\hsize}{!}
            {
            \includegraphics{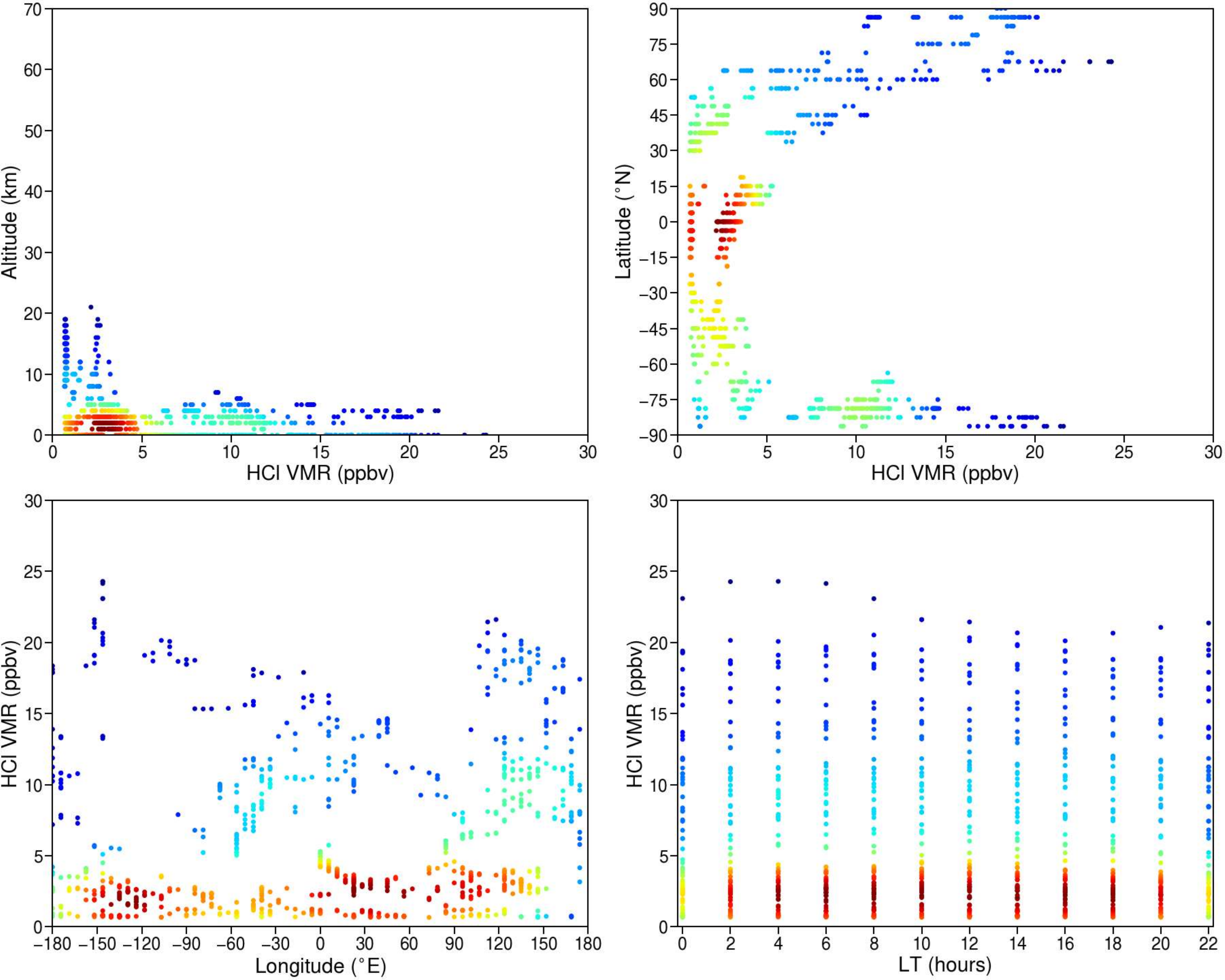}
            }
      \caption{Repartition of HCl maximum VMRs in altitude (top left), latitude (top right), longitude (bottom left) and local time (bottom right) for MY 35. The colour scale indicates the density of points, with increasing density going from blue to red.}
         \label{app_Fig_scatt_altlatlonlt_MY35}
\end{figure*}
\newpage

~
\newpage

\subsection{Relative differences between model and observations for HCl and water vapour VMRs}

We present histograms showing the relative difference between modelled and observed VMRs of HCl and water vapour. For each solar occultation where HCl was detected by ACS MIR, we compute the relative difference with model results for each point of the ACS MIR profile. For water vapour, we do similar computations for occultations where HCl is successfully detected, but use both ACS MIR and ACS NIR data. These calculations are done for every HCl detection of MY 34 and MY 35. 
In the figure, top panels show the relative difference between mean values from the model and observations, ignoring the standard deviations on these values. Model mean values are computed over a 5$^{\circ}$ L$_{\text{S}}$ interval containing the ACS detection. On the contrary, bottom panels show relative differences considering both standard deviations in model results and uncertainties in ACS data.

\begin{figure*}[h!]
\centering
   \resizebox{\hsize}{!}
            {
            \includegraphics{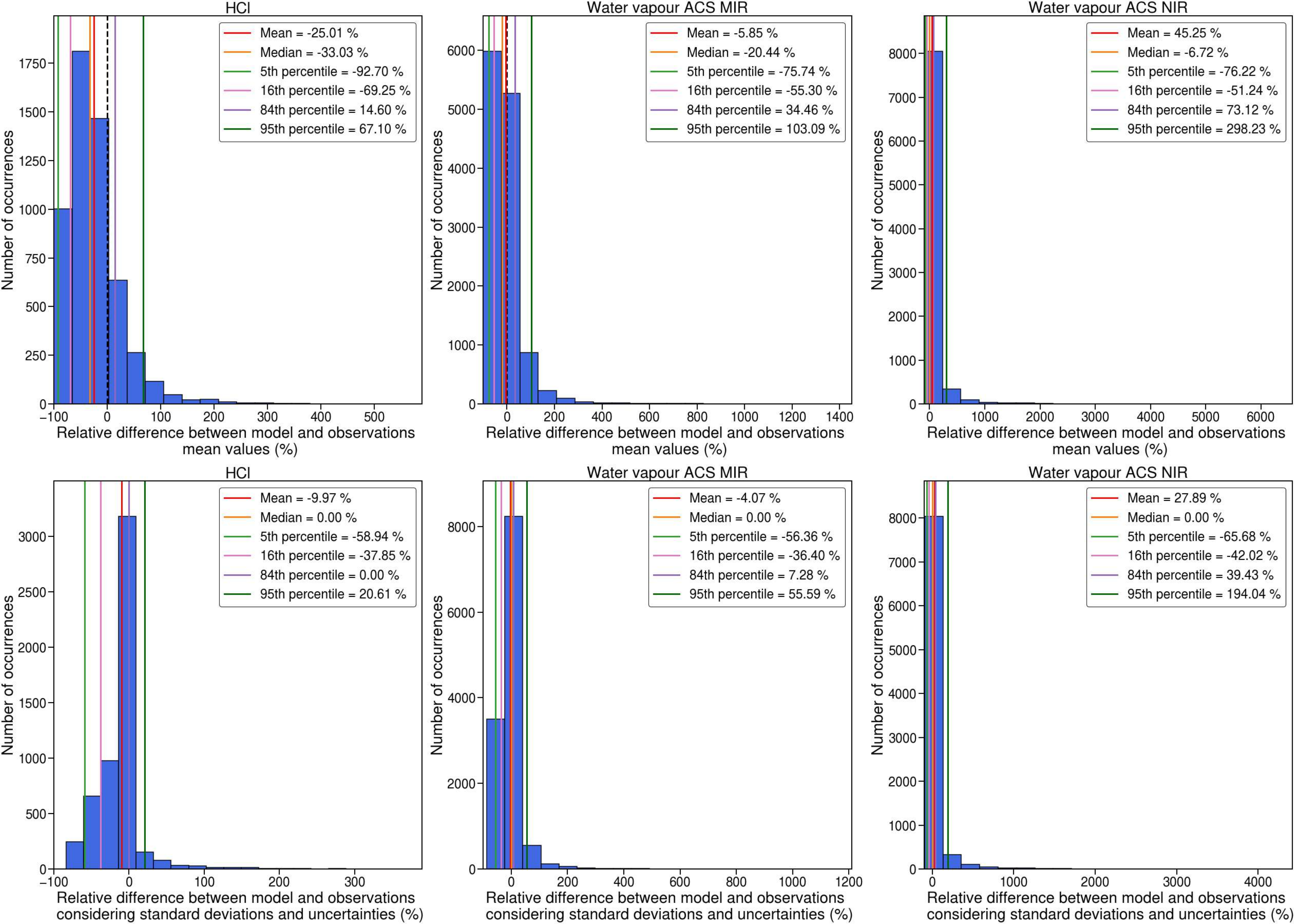}
            }
      \caption{Relative difference between modelled and observed HCl and water vapour VMRs. These are computed for each point of the retrieved profiles from ACS MIR/NIR observations where HCl was detected in MYs 34 and 35. Top panels present the relative difference between mean values, while bottom panels considers the standard deviations in model results and observations uncertainties in the computations.}
    \label{app_Fig_rel_diff_MY3435}
\end{figure*}
\newpage

\subsection{Vertical profiles of HCl VMRs, water vapour, and water ice and dust opacities from a location with a near-surface peak of HCl}

Here, we give complementary results for Sect. \ref{subs_study_pos_biases}. As developed in this section, our model produces peaks of HCl at high-southern latitudes during the dust season, when large amounts of dust and water vapour are available near the surface. The following figure illustrates the typical behaviour of HCl, water vapour and dust opacity when a peak of $\sim$ 10 ppbv HCl is found in our model.

\begin{figure*}[h!]
\centering
   \resizebox{\hsize}{!}
            {
            \includegraphics{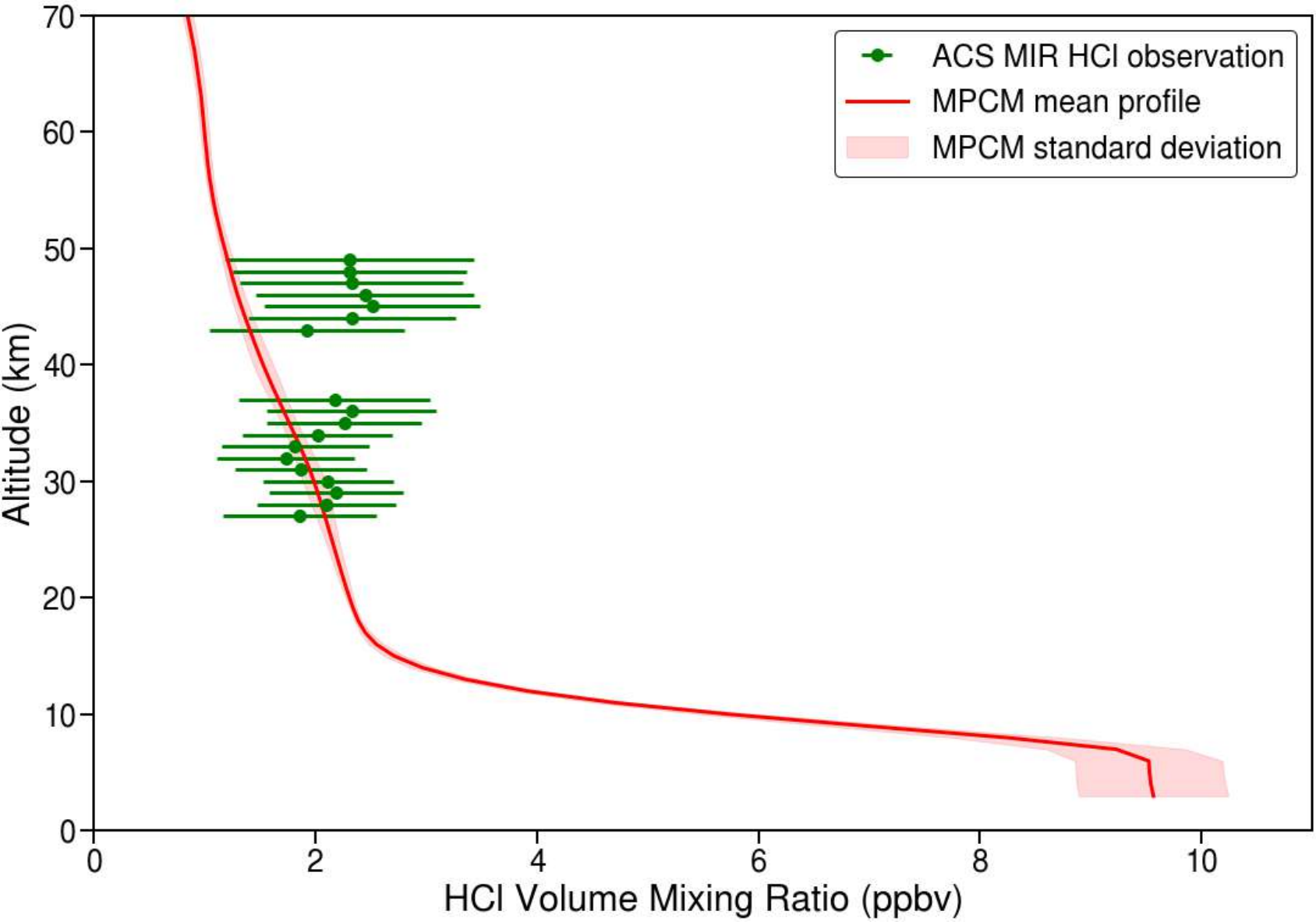}
            \includegraphics{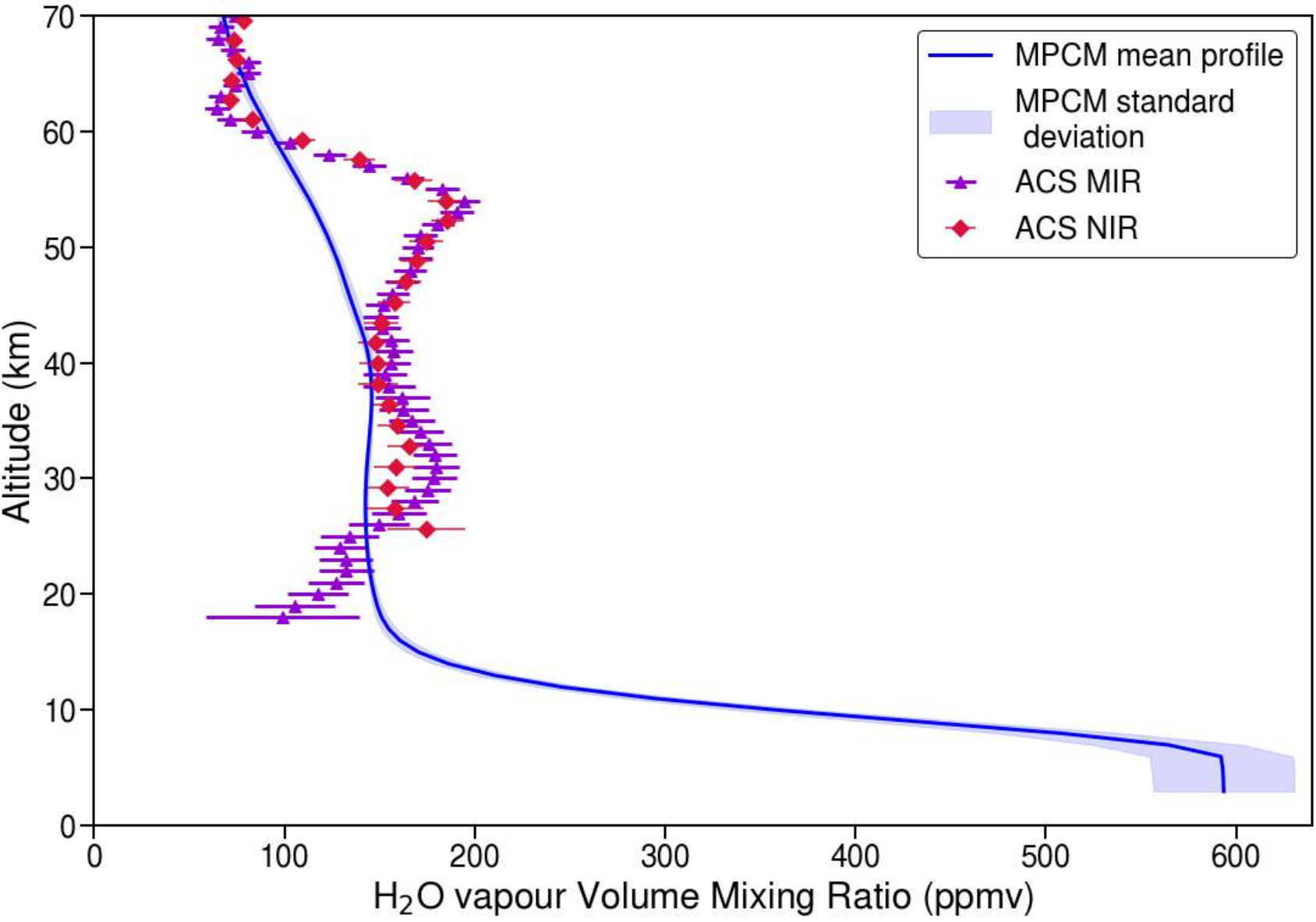}
            \includegraphics{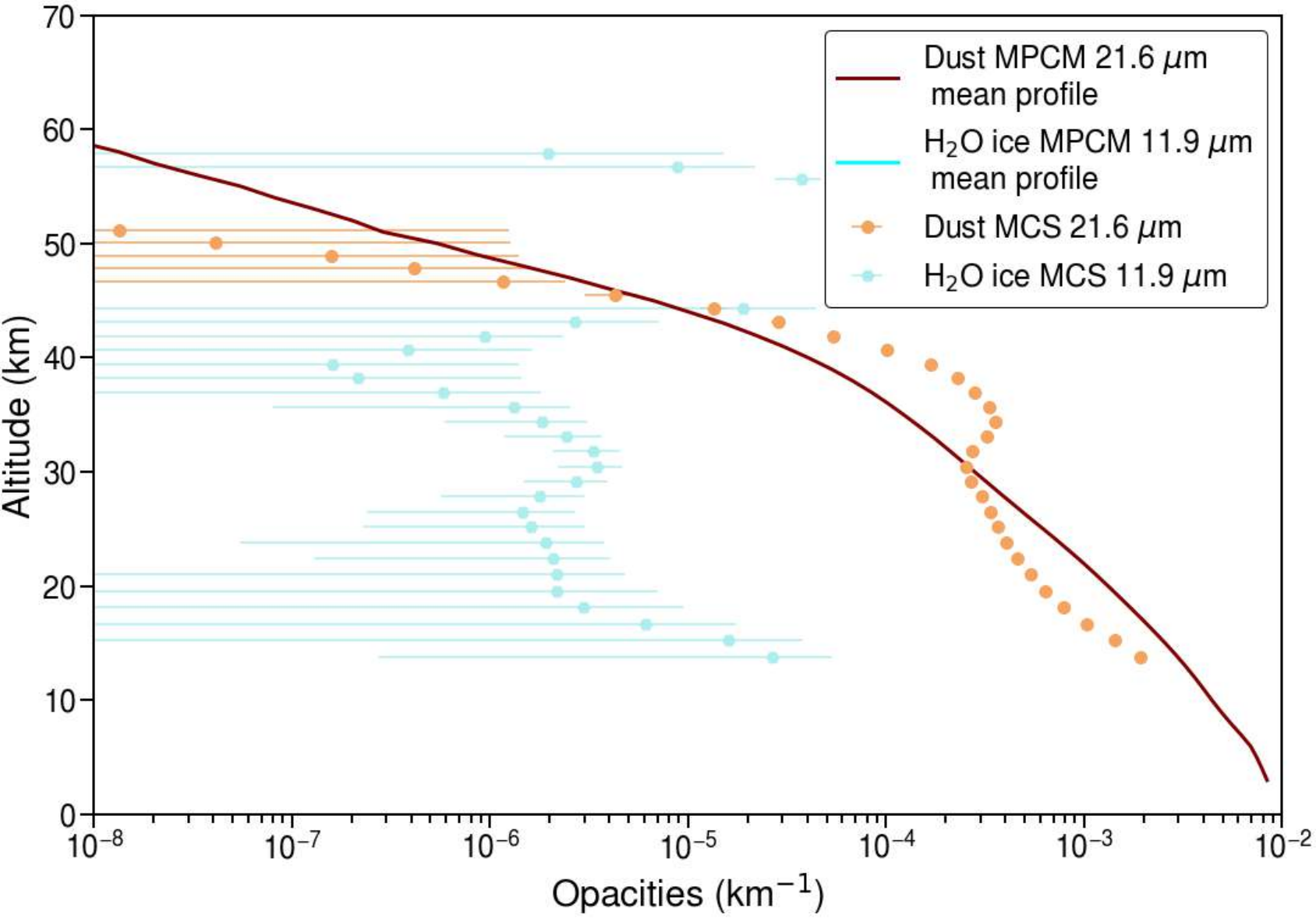}
            }
      \caption{Vertical profiles of HCl VMRs (left), water vapour VMRs (middle), and water ice and dust opacities (right) computed with our model (solid lines) and compared to observations. Model profiles are averaged over the 5$^{\circ}$ L$_{\text{S}}$ interval containing the ACS observation. The standard deviation represents the variability of the computed profiles within this interval. The plots correspond to the ACS MIR occultation 003814\_N2\_E\_P1\_12\_F, located at latitude 54.05$^{\circ}$S, longitude 136.46$^{\circ}$W, 260.55$^{\circ}$ L$_{\text{S}}$, LT = 20.64 hrs in MY 34.}
         \label{App_fig_FullFig_67}
\end{figure*}

\subsection{Comparison of model HCl VMRs with ACS MIR data for non-detections}

In the following figure, we compare the HCl VMRs computed in our model to the minimum uncertainty obtained in the ACS MIR retrievals at locations where HCl could not definitively be detected in both MY 34 and MY 35. 
Despite HCl not being properly confirmed, this comparison allows to further validate our model by showing its consistency with most non-detections, as it produces low HCl VMRs at most locations. 
In most cases, when the model produces positive HCl biases, we also find a positive bias in water vapour or dust opacity, or a negative bias in water ice opacity compared to observations. 

\begin{figure*}[h!]
   \resizebox{\hsize}{!}
            {
            \includegraphics{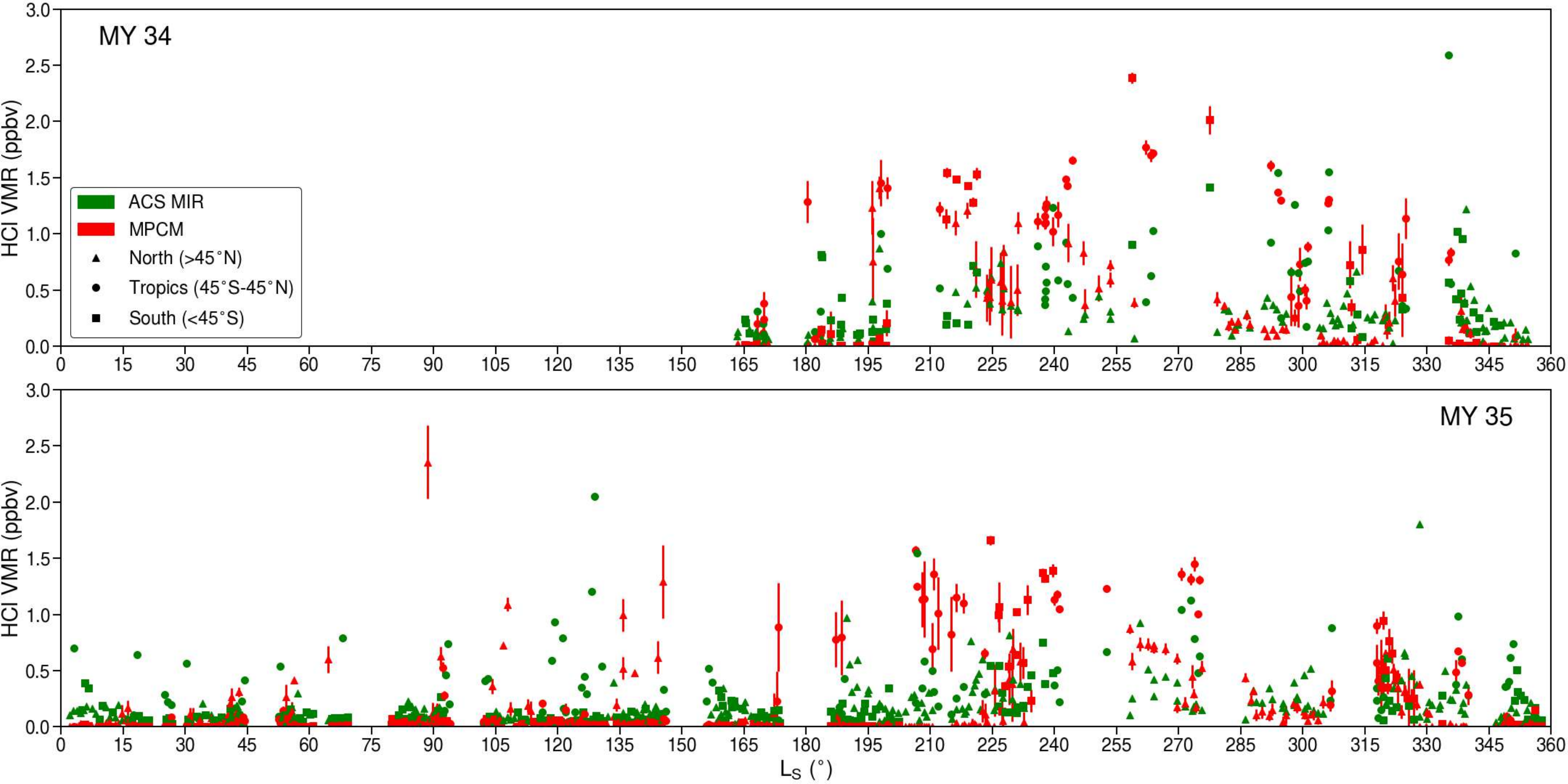}
            }
      \caption{HCl VMR from our model (red) and minimum retrieved ACS MIR uncertainty (green) at every location where HCl could not be confirmed (i.e. non-detection) during MYs 34 (top) and 35 (bottom). Triangles correspond to observations made at northern latitudes ($>$45$^{\circ}$N), circles at tropical latitudes (45$^{\circ}$S-45$^{\circ}$N), and squares to southern latitudes ($<$45$^{\circ}$S). Error bars display the standard deviation of modelled HCl within the 5$^{\circ}$L$_{\text{S}}$ interval containing the observation.}
    \label{App_Fig_MAXhcl_vs_Ls_NonDet}
\end{figure*}

\subsection{Spearman coefficients}

In addition to the computation of Pearson correlation coefficients, we performed computations of the Spearman coefficients between the HCl VMRs and other key atmospheric quantities: water vapour VMRs, temperature, water ice opacity and dust opacity. 
These computations are done to check that the correlations between these quantities are not affected by the analysis method used, as Pearson and Spearman coefficients differ on multiple points. (1) For Spearman computations, we compute correlations between the considered quantities by considering their full vertical profile. We therefore obtain one correlation coefficient per HCl detection, instead of a single coefficient based on all the values retrieved at each altitude where HCl is detected by ACS MIR. We therefore did not restrict the computations to the altitude range where HCl is detected by ACS MIR, as was done in the Pearson calculations. (2) While Pearson coefficients are used to study if there is a linear relationship between the studied quantities, this is not the case for the Spearman correlation, which is a rank correlation. This allows us to consider a broader range of relationships between HCl and the other quantities. 
The histograms showed in this section therefore display the Spearman coefficients computed for each ACS MIR HCl detection of MY 34 and MY 35.

\begin{figure*}[h!]
\centering
   \resizebox{\hsize}{!}
            {
            \includegraphics{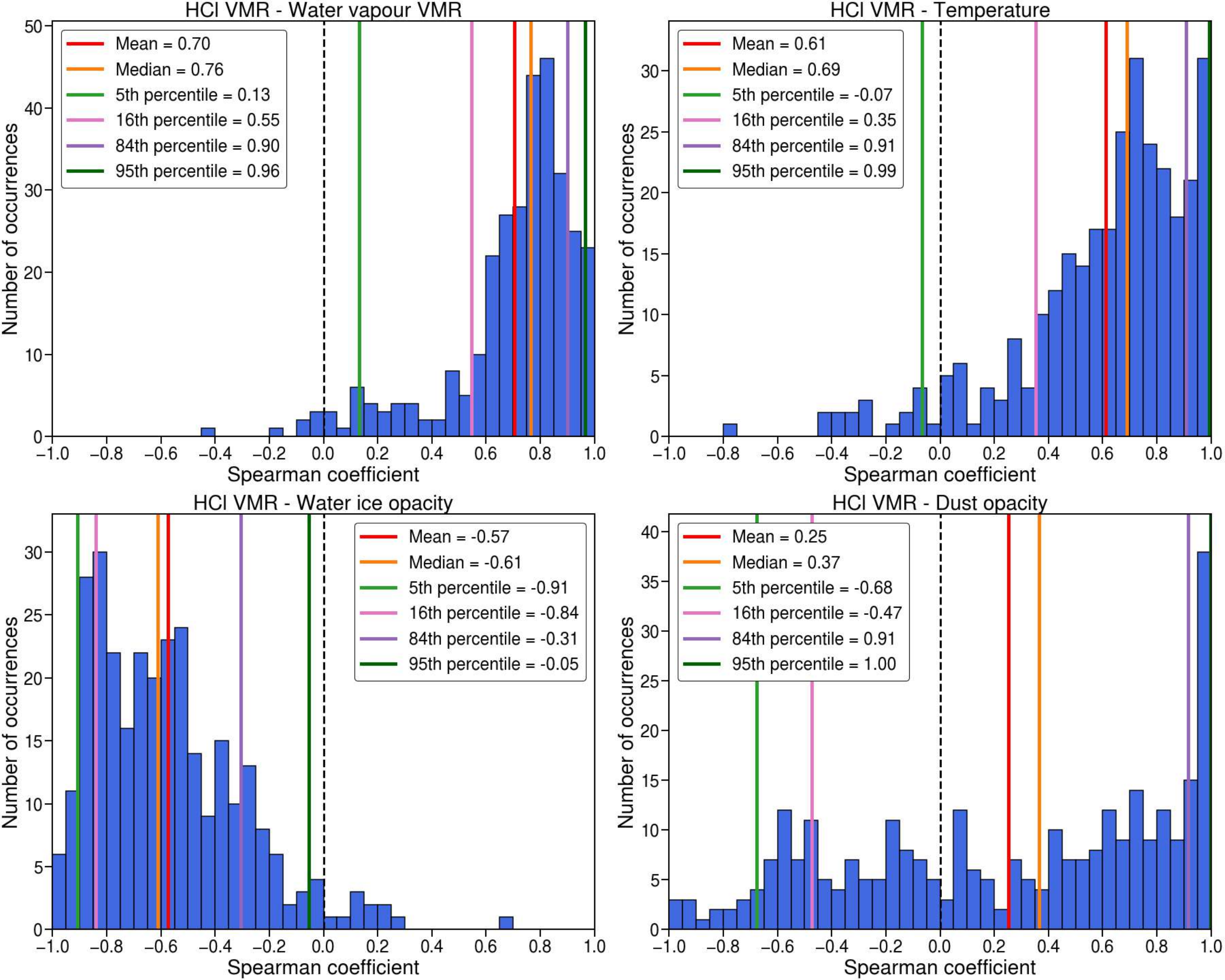}
            }
      \caption{Spearman correlation coefficients between vertical profiles of HCl VMRs and (top left) water vapour VMRs, (top right) temperature, (bottom left) water ice opacity, and (bottom right) dust opacity. The coefficients are computed for each HCl detection of MYs 34 and 35 by ACS MIR.}
         \label{Fig_stats_correl_MY3435}
\end{figure*}
\newpage

\subsection{Spatial distribution of HCl at the time of the aphelion detections}

In this section, we present maps of the maximum HCl VMRs reached at the time of the MY 35 aphelion detections that showed the presence of HCl in the Arcadia region. These maps show that HCl is not confined to this region, and that higher VMRs are reached at higher northern latitudes. These results directly depend on the chlorine chemical scheme used in our model and to the modelled water cycle, which tends to overestimate the water vapour content at high northern latitudes at that time of the year.

\begin{figure*}[h!]
\centering
   \resizebox{\hsize}{!}
            {
            \includegraphics{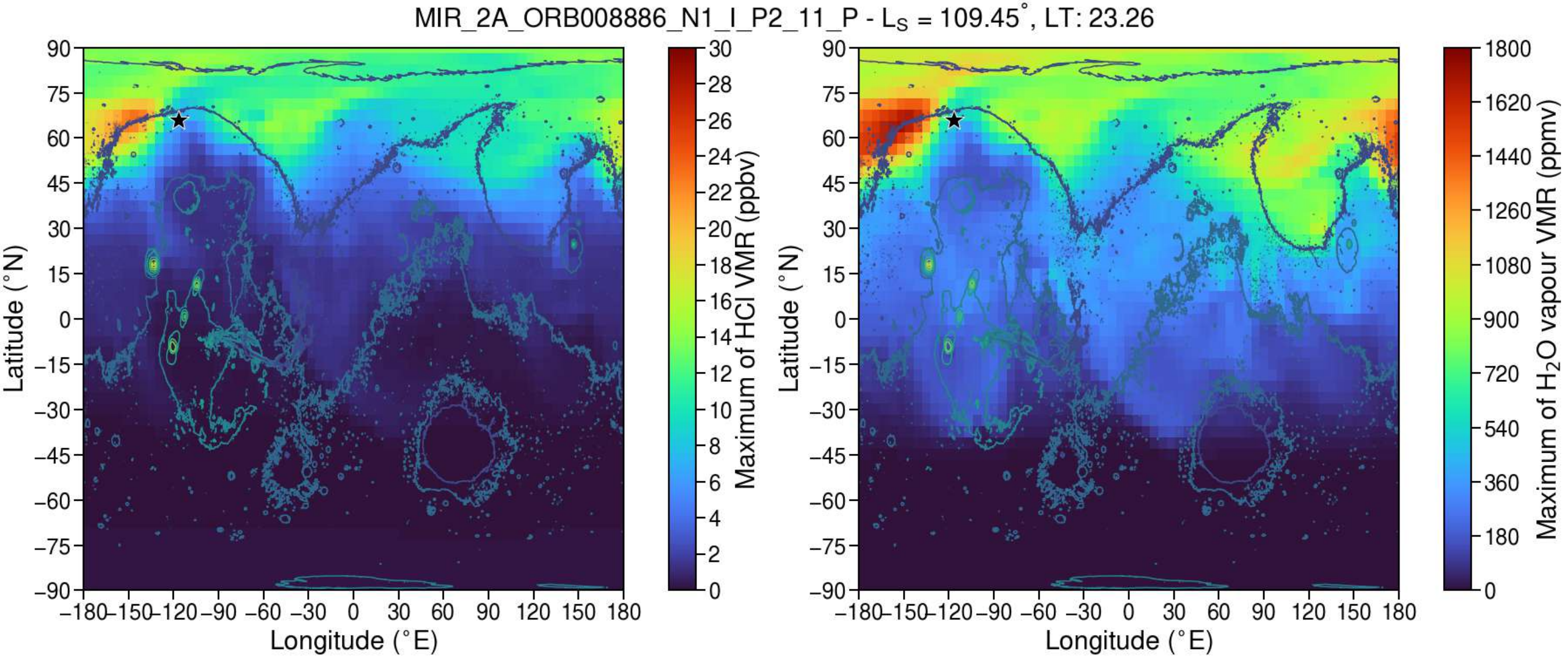}
            }
    \resizebox{\hsize}{!}
            {
            \includegraphics{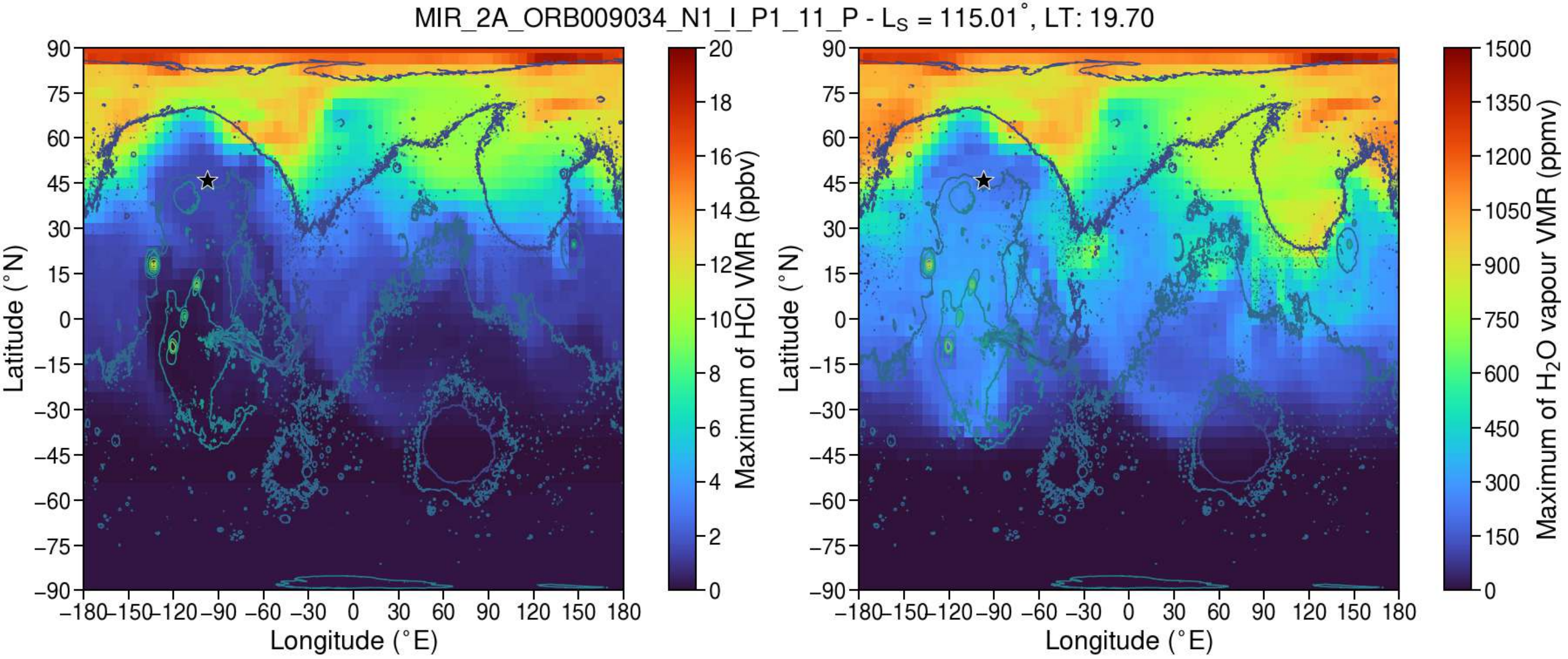}

            }
      \caption{Maps of the maximum HCl VMRs (left) and water vapour VMRs (right) in each atmospheric column considered in our model at the time of the two aphelion detections of MY 35. Top plots refer to the detection showed in Fig. \ref{fig_HCl_aphel_detect}(left), and bottom plots to Fig. \ref{fig_HCl_aphel_detect}(right). For each detection, the location where HCl was detected by ACS MIR is indicated by a black star.}
         \label{App_fig_aphel_det_VMRs_maps}
\end{figure*}

\section{Horizontal transport of HCl at northern latitudes}
\label{appendix_transport}

To investigate the impact of atmospheric transport on HCl at northern latitudes ($>$ 45$^{\circ}$N), we compute the HCl mass meridional ($\Psi_{\text{m}}$) and zonal ($\Psi_{\text{z}}$) streamfunctions. They are computed as in \citet{braam_stratospheric_2023}:
\begin{align*}
    \Psi_{\text{m}} &= \frac{2 \pi R_{\text{M}} \cos{\phi}}{g} \int_0^P \bar{v}~\text{d}P \times \text{mmr(HCl)} \\
    \Psi_{\text{z}} &= \frac{2 \pi R_{\text{M}} }{g} \int_0^P \bar{u}~\text{d}P \times \text{mmr(HCl)},
\end{align*}
\noindent where $R_{\text{M}}$ is the radius of Mars, 
$\phi$ is the latitude, $g$ is the gravitational acceleration, 
$\bar{v}$ and $\bar{u}$ are the zonal and meridional mean of the northward and eastward wind components, respectively, $P$ is the atmospheric pressure, and mmr(HCl) is the mass mixing ratio of HCl. These give the mass flux of HCl (kg/s) at a considered location. 

To better assess the transport from equatorial regions to northern latitudes, we perform a Helmholtz decomposition of the horizontal wind ($\vec{u}$, $\vec{v}$) to isolate the divergent components ($\vec{u'}$, $\vec{v'}$), as detailed in \citet{hammond_rotational_2021}, and use these components in our computations of $\Psi_{\text{m}}$ and $\Psi_{\text{z}}$. We then consider that the net transport flux in a given cell of the atmosphere is equal to the difference between the incoming fluxes from adjacent cells and the outgoing flux from the considered cell. 
\newline

To understand when horizontal transport can influence the HCl vertical profiles, we compute the dynamical lifetimes associated with meridional ($\tau_{\text{v'}}$) and zonal ($\tau_{\text{u'}}$) transport \citep{braam_stratospheric_2023} and compare these lifetimes with the one derived from HCl photochemistry ($\tau_{\text{c}}$):

\begin{align*}
    \tau_{\text{v'}} &= \frac{\pi R_{\text{M}}}{v'}  \\
    \tau_{\text{u'}} &= \frac{2\pi R_{\text{M}}}{u'}  \\
    \tau_{\text{c}}  &= \frac{n(\text{HCl})}{L},
\end{align*}
\noindent where $n(\text{HCl})$ is the number density of HCl (cm$^{-3}$) and $L$ is the total loss rate of HCl (cm$^{-3}$.s$^{-1}$).
\newline 

\begin{figure*}[h!]
\centering
   \resizebox{\hsize}{!}
            {
            \includegraphics{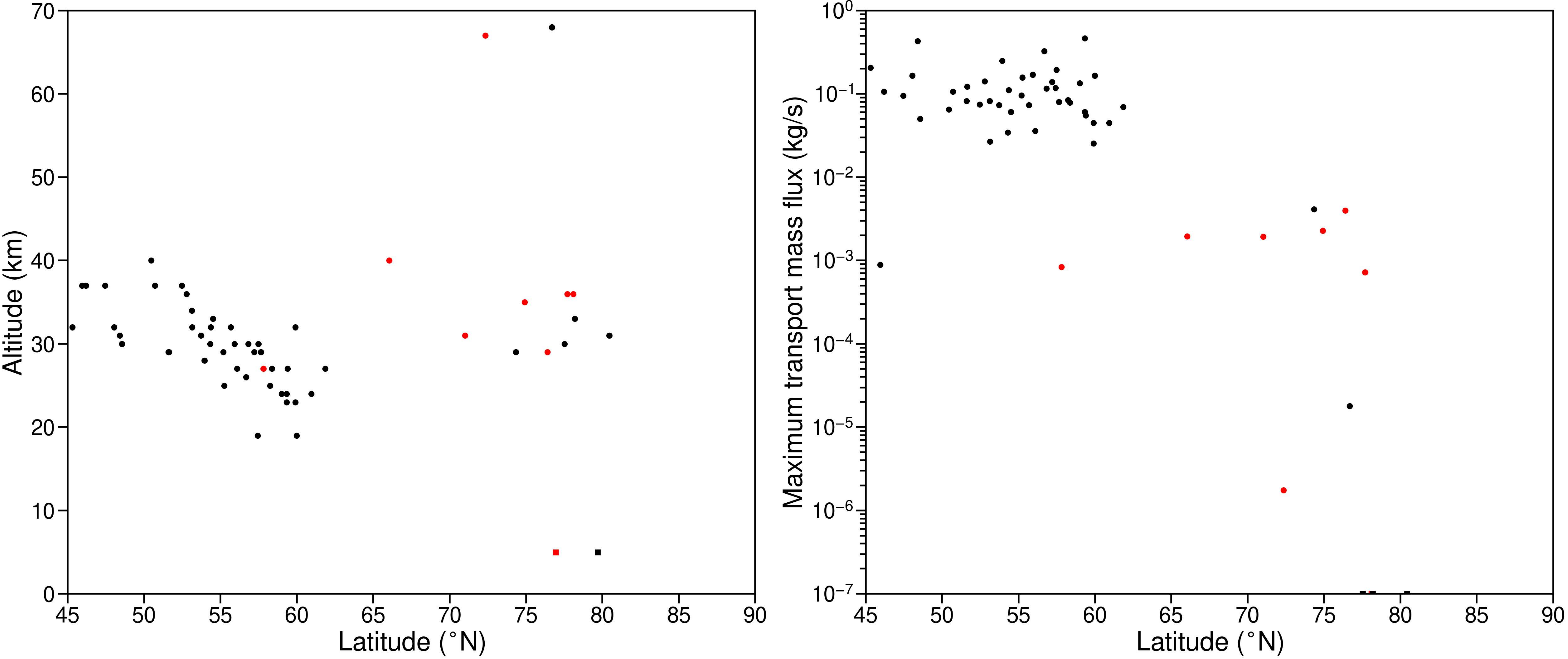}
            }
      \caption{Horizontal transport modelled properties for northern latitudes ($>$ 45$^{\circ}$N) ACS MIR HCl detections of MY 35. Left: lowest altitude where the horizontal transport lifetime of HCl becomes lower than the photochemical lifetime (i.e. where horizontal transport can compete with chemistry) in the atmospheric column corresponding to the ACS MIR observation. Squares located at 5 km are used when the horizontal transport lifetime is larger than the photochemical lifetime in the entire column. Right: maximum net flux of HCl resulting from horizontal transport in the atmospheric column where the HCl observation is performed. Squares located at 10$^{-7}$ kg/s are used when the maximum net mass flux in lower or equal to zero. In both panels, red dots indicate the observations where we find the largest negative biases between model results and ACS MIR observations in MY 35, corresponding to observations between L$_\text{S}$=140-220$^{\circ}$.}
         \label{App_fig_scat_transp_MY35_det}
\end{figure*}

Results from these computations for all the northern latitudes ($>$ 45$^{\circ}$N) HCl detections of MY 35 are shown in Fig. \ref{App_fig_scat_transp_MY35_det}. We focus on these observations as we find the largest negative model biases occur at these latitudes between L$_\text{S}$=140-220$^{\circ}$ of that year. Results for these particular observations are indicated in red in Fig. \ref{App_fig_scat_transp_MY35_det}. The left panel shows that at these latitudes, horizontal transport can only compete with chemistry above $\approx$30 km, which corresponds to the upper limit of the altitude range where extensive water ice clouds are produced in our model. The right panel shows that the maximum HCl mass flux resulting from atmospheric transport appears to decrease by more than one order of magnitude above $\approx$ 60$^{\circ}$N, and that all the observations where we find large negative biases correspond to cases with low horizontal transport. This shows that these biases are due to both strong photochemical loss (as described in Sect.\ref{subsect_vert_prof_comp}), as well as the lack of HCl supply from HCl rich regions by horizontal transport. These low fluxes at this time of the year (L$_\text{S}$=140-220$^{\circ}$) can be linked to the inversion of the Hadley cell, which takes place around the equinox at L$_\text{S}$=180$^{\circ}$. During this period and at these latitudes, the HCl meridional streamfunction switches from positive values (northward mass flux) to negative values (southward mass flux), therefore significantly reducing the HCl flux in the meantime.

\end{appendix}

\newpage
\bibliographystyle{abbrvnat}
\bibliography{bib_Mars_20240702}

\end{document}